\documentclass{aa}  
\usepackage{tabularx}

\usepackage{graphicx}
\usepackage{natbib}
\usepackage{tabularx}
\usepackage{adjustbox}
\usepackage{rotating}
\usepackage{subcaption}
\usepackage{txfonts}
\usepackage{array}
\usepackage[utf8]{inputenc}
\usepackage[T1]{fontenc}
\usepackage{lmodern} 
\usepackage{threeparttable}
\usepackage{newtxtext,newtxmath}
\usepackage{adjustbox}
\newcommand{\HI}{H\,{\sc i}}
\newcommand{\HII}{H\,{\sc ii}}

\newcommand{\kms}{~km\,s$^{-1}$}
\usepackage[autostyle]{csquotes}
\usepackage{textcomp}
\usepackage{rotating}
\usepackage{float}
\MakeRobust{\overrightarrow}
\usepackage{graphicx}
\usepackage{tikz}
\usetikzlibrary{tikzmark, calc}
\usepackage{hyperref}
\usepackage{enumitem}
\usepackage{subcaption}
\usepackage{amssymb,amsmath}
\usepackage{pifont}
\usepackage{multirow}
\usepackage{graphicx}

\usepackage[compatibility=false]{caption}
\hypersetup{
    colorlinks=true,
    linkcolor=blue,
    filecolor=magenta,      
    urlcolor=blue,
    citecolor=blue,
    pdftitle={Overleaf Example},
    pdfpagemode=FullScreen,
    }
\usepackage{calc} 

\usepackage{booktabs}  

\usepackage{caption}[=v1]
\begin{document} 

\titlerunning{\HI\ Distribution and Kinematics of ESO444-G084 and [KKS2000]23}
\authorrunning{Namumba et al.}
 \title{Investigating the \HI\ distribution and kinematics of ESO444-G084 and [KKS2000]23: New insights from the MHONGOOSE survey}

\author{B.\ Namumba\inst{1,2}\textsuperscript {\ding{72}},
R.\ Ianjamasimanana\inst{1},
B.\ S.\ Koribalski\inst{3,4},
A.\ Bosma\inst{5},
E. \ Athanassoula\inst{5},
C.\ Carignan\inst{6,7,8},
G.\ I.\ G.\ J\'ozsa\inst{9,10},
P.\ Kamphuis\inst{11},
R.\ P.\ Deane\inst{2,12},
S.\ P.\ Sikhosana\inst{13,14},
L.\ Verdes-Montenegro\inst{1},
A.\ Sorgho\inst{1},
X.\ Ndaliso\inst{2},
P.\ Amram\inst{5},
E.\ Brinks \inst{15},
L.\ Chemin\inst{16,17},
F.\ Combes\inst{18},
W.\ J.\ G.\ de Blok\inst{19,6,20},
N.\ Deg\inst{21},
J.\ English\inst{22},
J.\ Healy\inst{23,24},
S.\ Kurapati\inst{19},
A. \ Marasco\inst{25},
S.\ S.\ Mc Gaugh\inst{26},
K.\ A. \ Oman\inst{27,28},
K.\ Spekkens\inst{29},
S.\ Veronese\inst{19,20}
O.\ I.\ Wong\inst{30,31}}

\institute{
\begin{center}
(\textit{Affiliations are listed at the end of the paper})
\end{center}
\vspace{-1.5em}  
}

\date{
\begin{center}
Received 06 March 2025; accepted 01 June 2025
\end{center}
}

\abstract{We present the \HI\ distribution, kinematics, mass modeling, and gravitational stability of the dwarf irregular galaxies ESO444--G084 and [KKS2000]23 using high spatial, spectral, and column density sensitivity data from the MHONGOOSE survey obtained with MeerKAT. ESO444--G084 exhibits centrally concentrated \HI\ emission, while [KKS2000]23 has irregularly distributed high-density pockets. The total \HI\ fluxes measured down to column density thresholds of \(10^{19}\,\mathrm{cm}^{-2}\) and \(10^{18}\,\mathrm{cm}^{-2}\) are nearly the same, suggesting that the increase in the \HI\ diameter at lower column densities is primarily due to the larger beam size, and that no significant additional emission is detected. The total \HI\ masses are \((1.1 \pm 0.1) \times 10^8\) M\(_{\odot}\) for ESO444--G084 and \((6.1 \pm 0.3) \times 10^8\) M\(_{\odot}\) for [KKS2000]23. We derived rotation curves using 3D kinematic modeling tools PyFAT and TiRiFiC, which allow us to fully capture the gas kinematics. Both galaxies exhibit disk-like rotation, with ESO444--G084 showing a kinematic warp beyond \(\sim1.8\) kpc. Its relatively fast-rising rotation curve suggests a more centrally concentrated dark matter distribution, whereas [KKS2000]23’s more gradual rise indicates a more extended mass distribution. Mass modeling with an isothermal halo and \((M/L)^{*}_{3.4\,\mu\text{m}} = 0.20\) for ESO444--G084 and \(0.18\) for [KKS2000]23 yields physically consistent results. We further analyze disk stability using spatially resolved maps of the Toomre Q parameter and \(\Sigma_{\rm gas}/\Sigma_{\rm crit}\), linking these with recent star formation traced by H\(\alpha\) and FUV emission. ESO444--G084 supports localized star formation despite global stability, while [KKS2000]23 is gravitationally unstable yet lacks strong H\(\alpha\) emission, suggesting that turbulence, gas depletion, or past feedback suppresses star formation. The absence of detectable inflows or outflows implies that internal processes regulate star formation. This study highlights the interplay between \HI\ morphology, kinematics, dark matter distribution, and disk stability, demonstrating how internal mechanisms shape dwarf galaxy evolution.} 


   \maketitle

\section{Introduction}
Dwarf galaxies, despite their small size, serve as key laboratories for studying galaxy formation and evolution. Their relatively simple structures make them valuable testbeds for investigating star formation, feedback, and the interplay between baryons and dark matter \citep{1988ApJ...332L..33C, 2012AJ....144..134H}. Their extended neutral hydrogen (\HI) reservoirs allow for detailed kinematic studies beyond the stellar disk, providing a unique opportunity to probe their gravitational potential \citep{2001AJ....122.2396D, 2008AJ....136.2648D}.

An essential approach to studying the structure and dynamics of galaxies is through rotation curves (RCs), which trace the circular motion of gas or stars as a function of galactocentric radius. These curves provide fundamental insights into the mass distribution, highlighting the contributions of both baryonic and dark matter components in galaxies \citep{1978PhDT.......195B, 1998ApJ...506..125C}. The shape of a galaxy’s rotation curve, whether steeply rising, flat, or declining, reveals how baryonic matter and dark matter interact. By examining these curves, one can accurately determine a galaxy’s gravitational potential and overall dynamical behavior, leading to a deeper understanding of its mass distribution \citep{2001AJ....122.2396D}.

Among the different methods for studying galaxy kinematics, neutral hydrogen observations provide a reliable approach for deriving rotation curves and probing the distribution of dark matter. Although optical tracers, such as Balmer lines from ionized hydrogen (H$\alpha$), are crucial for constraining the inner regions of rotation curves due to their high spatial resolution, they are limited to the optical disk, leaving the dark matter-dominated outskirts unexplored \citep{2001AJ....122.2396D, 2004A&A...420..147B}. On the other hand, the \HI\ disc in galaxies extends much further into the outer regions, often well beyond the optical disk \citep{1985ApJ...294..494C}. In these outer regions, where dark matter dominates the gravitational potential \citep{1988ApJ...332L..33C}, \HI\ observations provide a direct means of studying the structure and kinematics of the dark matter halo.

Although rotation curves are a fundamental tool for probing mass distributions, accurately determining their shape remains challenging. Beam smearing, noncircular motions, and projection effects can distort observed velocity fields, introducing systematic uncertainties in mass modeling \citep{2008AJ....136.2761O,2016MNRAS.462.3628R}. Recent studies \citep{2015MNRAS.452.3650O,2023MNRAS.522.3318D} have demonstrated that $\Lambda$CDM based hydrodynamical simulations yield consistent circular velocity profiles for low mass galaxies, yet they do not capture the diversity seen in observed dwarf galaxy rotation curve shapes. In many cases, observed inner rotation speeds are lower than predicted, a discrepancy attributed either to various dynamical effects such as mergers, inflows, outflows, warps, and halo asymmetries or to systematic biases that affect the inferred kinematics.

Traditional 2D tilted ring models \citep{1989A&A...223...47B} have been widely used to study galaxy kinematics, but they often fail to fully correct for distortions, leading to rotation curve shapes that do not accurately reflect the true underlying circular velocities \citep{2008AJ....136.2648D,2003ApJ...583..732S}. To overcome these limitations, advanced 3D kinematic modeling techniques, such as TiRiFiC \citep{2007A&A...468..731J} and $^\mathrm{3D}$ Barolo \citep{2015MNRAS.451.3021D} have been developed. These methods utilize the complete data cube—including detailed line profiles and intensity distributions to correct for beam smearing. They also incorporate extra parametric components to explicitly model noncircular motions, including radial flows and warps. Although these techniques still assume a locally axisymmetric disk, they represent a significant improvement over 2D approaches in constraining the kinematics of galaxies.

Understanding the kinematics of dwarf galaxies is also critical for assessing their gravitational stability and star formation processes. The stability of a galaxy’s disk is often characterized using the Toomre \( Q \) parameter, which measures the susceptibility of the gas to collapse and form stars \citep{1964ApJ...139.1217T}. Dwarf galaxies typically exhibit low gas surface densities, making them marginally stable or even gravitationally unstable in certain regions \citep{2015ApJ...805..145E}. Since \HI\ serves as both a tracer of galaxy kinematics and a reservoir for future star formation, linking disk stability analyses with rotation curve modeling provides a more complete picture of how gas dynamics, gravitational instabilities, and dark matter influence the evolution of dwarf galaxies \citep{2012AJ....143....1E,2018NewA...63...38G}. 

The MeerKAT \HI\ Observations of Nearby Galactic Objects: Observing Southern Emitters (MHONGOOSE) survey \citep{2024A&A...688A.109D}, one of the MeerKAT Large Survey Projects, provides an unprecedented opportunity to study the kinematics and mass distribution of dwarf galaxies. Building upon previous surveys such as THINGS \citep{2008AJ....136.2563W}, LITTLE-THINGS \citep{2012AJ....144..134H}, and LVHIS \citep{2018MNRAS.478.1611K}, MHONGOOSE offers significantly improved velocity resolution, making it ideal for capturing fine-scale kinematic structures and resolving asymmetries at sub-kiloparsec scales. Furthermore, its high sensitivity to low-column-density gas (\(N_{\mathrm{HI}} < 10^{19}\) cm\(^{-2}\)), allows the detection of extended \HI\ structures well beyond the optical disk. Preliminary results from the MHONGOOSE survey have already demonstrated the power of high-sensitivity \HI\ observations in detecting extended gas distributions and resolving kinematic asymmetries at sub-kiloparsec scales \citep{2025A&A...693A..97V,2024A&A...687A.254H,2020A&A...643A.147D}. These advances are crucial for distinguishing between different dark matter halo profiles and probing the outer regions of dwarf galaxies, where dark matter dominates even more significantly, making its intrinsic properties more apparent. 

This study focuses on ESO444--G084 and [KKS2000]23, two gas-rich, isolated dwarf galaxies with well-resolved \HI\ kinematics and extended gas distributions, making them ideal candidates for detailed mass modeling. Their isolation minimizes environmental effects, enabling a clearer analysis of internal kinematics, disk stability, and dark matter distribution. By examining their \HI\ morphology, kinematics, rotation curves, and mass models, we investigate how gas dynamics shape their evolution and influence star formation. The structure of this paper is as follows: Section \ref{sec:sample} describes the sample selection; Section \ref{sec:datareduction} details the observations and data reduction process; Section \ref{sec:distribution} presents the \HI\ morphology and distribution; Section \ref{sec:kinematics} discusses kinematics and rotation curve derivation; Section \ref{sec:massmodel} explores mass modeling; Section \ref{sec:starformation} details the gravitational instability and star formation and Sections \ref{sec:discussion} and \ref{sec:conclude} provide discussion and conclusions, respectively.

\section{The galaxy sample} \label{sec:sample}
The MHONGOOSE sample consists of 30 nearby star-forming main sequence galaxies, providing a representative set for studying typical star formation processes \citep{2024A&A...688A.109D}. The sample covers a broad range of stellar and \HI\ masses ($\sim$ 10$^{6}$ to 10$^{11}$ M$_{\odot}$) and includes both disk galaxies and dwarf systems. The galaxies are situated at distances of 3--23 Mpc, spanning a range of inclination angles from nearly face-on to edge-on, offering a diverse selection for kinematic and structural analysis. The survey achieves a 3$\sigma$ column density sensitivities down to $\sim$10$^{17}$ cm$^{-2}$ at 90$^{\prime \prime}$ resolution and $\sim$10$^{19}$ cm$^{-2}$ at 7$^{\prime \prime}$ resolution, with a velocity resolution of 1.4 \kms. This study focuses on two MHONGOOSE dwarf galaxies, ESO444--G084 and [KKS2000]23, which were the first dwarfs in the MHONGOOSE sample to be observed with the full 50-hour integration time. Additionally, these galaxies were prioritized due to their well-resolved \HI\ kinematics, extended gas distributions, and their importance for detailed mass modeling. A comprehensive overview of the MHONGOOSE sample selection and survey methodology is provided in \citet{2024A&A...688A.109D}.

In addition to the MeerKAT \HI\ observations, this study makes use of publicly available, multi-wavelength ancillary data to provide a more comprehensive view of the galaxy properties. Far-ultraviolet (FUV) images were obtained from the GALEX data release \citep{2005ApJ...619L...1M} and optical images were taken from the DECaLS survey \citep{2019AJ....157..168D}. Mid-infrared data from the Wide-field Infrared Survey Explorer (WISE) were provided through private communication. H$\alpha$ images for both galaxies were accessed via the NASA/IPAC Extragalactic Database (NED)\footnote{\url{https://ned.ipac.caltech.edu/}}, with ESO444--G084 data from \citet{2009ApJ...703..517D} and from [KKS2000]23 \citet{2006ApJS..165..307M}. These multi-wavelength datasets, all openly available online, are used to trace star formation activity, stellar structure, and ionized gas, complementing the high-resolution \HI\ kinematics provided by the MHONGOOSE survey.

\subsection{ESO444--G084 (HIPASS J1337--28)}
ESO444--G084 is a dwarf irregular (dIrr) galaxy located at a distance of 4.6 $\pm$ 0.4 Mpc, based on the Tip of the Red Giant Branch (TRGB) method \citep{2002A&A...385...21K}. It is one of several dwarf companions surrounding the massive spiral galaxy M 83. Classified as an Im galaxy \citep{1991rc3..book.....D}, ESO444--G084 has a total B-band magnitude of 15.06 \citep{2002A&A...385...21K}. The galaxy's \HI\ distribution was mapped by \citet{2000AJ....120.3027C} using the Australia Telescope Compact Array (ATCA), revealing a symmetric extended \HI\ disk with significant warping in its velocity field. The heliocentric velocity was measured at 588 \kms. The \HI\ flux density was found to be 21.1 $\pm$ 3.2 Jy \kms, which corresponds to an \HI\ mass of 1.1 $\times$ 10$^{8}$ M$_{\odot}$ at the adopted distance of 4.6 Mpc \citep{2002A&A...385...21K}. These results are consistent with those of \citet{2018MNRAS.478.1611K}, who studied ESO444--G084 as part of the Local Volume HI Survey (LVHIS). The rotation curve derived by \citet{2000AJ....120.3027C} shows a maximum velocity of $\sim$ 60 \kms at a radius of 3.2 kpc, and a very high dark-to-luminous mass ratio, with a dark matter fraction over 90$\%$ within the last measured point. A summary of the key properties of ESO444--G084 is provided in Table~\ref{table1}.

\subsection{[KKS2000]23 (HIPASS J1106--14)}
Located at a distance of 13.9 Mpc \citep{2016AJ....152...50T}, the dwarf galaxy [KKS2000]23 was first identified by \citet{KKS2000} in a search for low surface brightness (LSB) objects. A chain of \HII\ regions in its northern region was later detected through H$\alpha$ imaging by \citet{Whting2002}. The first \HI\ spectrum was obtained by \citet{Huchtmeier2001} using the Effelsberg 100-m Radio Telescope, measuring an \HI\ flux density of 11.9 Jy \kms\ and a systemic velocity of 1041 \kms. These values were confirmed by \citet{Ryan-Weber2002} and \citet{Koribalski2004}, with an estimated \HI\ mass of $7.1 \times 10^{8}$~M$_{\odot}$. Observations using the Green Bank Telescope (GBT) by \citet{2019MNRAS.482.1248S} revealed an \HI\ extent of 15.1 arcmin and an \HI\ flux of $11.1 \pm 0.4$ Jy \kms. Previous studies classified [KKS2000]23 as having one of the lowest H$\alpha$-based SFRs in its sample \citep{2006ApJS..165..307M}. However, it is not the most passive galaxy in the SINGG survey, emphasizing that H$\alpha$ probes recent star formation on short timescales (1–3 Myr), whereas FUV traces star formation over longer periods (100 Myr). A summary of the key properties of [KKS2000]23 is provided in Table~\ref{table1}.
\begin{table}[ht]
\centering
\begin{threeparttable}
\caption{Basic Properties of ESO444--G084 and [KKS2000]23.}
\label{table1}
\setlength{\tabcolsep}{3pt}
\renewcommand{\arraystretch}{1.2}
\setlength{\extrarowheight}{3pt} 

\begin{tabular}{lcc} 
\hline\hline  
Parameter & ESO444--G084 & [KKS2000]23 \\                        
\hline 
Right ascension (J2000) & 13:37:19.9\tnote{a} & 11:06:12.0\tnote{b} \\
Declination (J2000)      & --28:02:42.0\tnote{a} & --14:24:25.7\tnote{b} \\
Morphology             & I\tnote{c} & I\tnote{d} \\
Distance (Mpc)         & 4.6\tnote{e} & 13.9\tnote{f}  \\
B magnitude            & 15.06\tnote{e} & 15.80\tnote{d} \\
V$_{\rm hel}$ (km\,s$^{-1}$) & 586.7\tnote{g} & 1035.9\tnote{g} \\
Total \HI\ mass (M$_{\odot}$) & 1.0 $\times$ 10$^{8}$\tnote{g} & 5.5 $\times$ 10$^{8}$\tnote{g} \\
\HI\ effective \\ radius (at 1 M$_{\odot}$pc$^{-2}$) & 4$\farcm 9$ (this work) & 4$\farcm 0$ (this work) \\
Stellar mass (M$_{\odot}$) & 4.9 $\times$ 10$^{6}$\tnote{g} & 3.2 $\times$ 10$^{7}$\tnote{g} \\
\hline 
\end{tabular}
\begin{tablenotes}
\footnotesize
\item \parbox[t]{1.0\linewidth}{References: (a) \citet{2003MNRAS.339..652K}; (b) \citet{2007AJ....133..715W}; (c) \citet{1991rc3..book.....D}; (d) \citet{2001A&A...377..801H}; (e) \citet{2002A&A...385...21K}; (f) \citet{2016AJ....152...50T}; (g) \citet{2024A&A...688A.109D}}
\end{tablenotes}

\end{threeparttable}
\end{table}
\section{MeerKAT observations and data reduction}\label{sec:datareduction}
The 21-cm \HI\ observations were carried out using the MeerKAT telescope \citep{2016mks..confE...1J,2018ApJ...856..180C} as part of the MHONGOOSE survey \citep{2024A&A...688A.109D}. Each target was observed for a total integration time of 50 hours, achieving a final spectral resolution of ~1.4 \kms\ after averaging over two channels. Data reduction was performed using the CARACal pipeline, the detailed procedure for which is described in \citet{2024A&A...688A.109D}.

For each galaxy, six data cubes were generated, covering spatial resolutions from 7\(\arcsec\) to 90\(\arcsec\). The corresponding 3\(\sigma\) column density sensitivity ranges from \(5.9 \times 10^{19}\) cm\(^{-2}\) to \(5.8 \times 10^{17}\) cm\(^{-2}\) over a velocity range of 16 km s\(^{-1}\). The observational parameters for ESO444--G084 and [KKS2000]23 are summarized in Table~\ref{table2}.  
\begin{table}
\centering
\caption{Parameters of the MeerKAT observations and properties of the highest and lowest resolution data cubes.}
\footnotesize
\setlength{\tabcolsep}{5pt}
\renewcommand{\arraystretch}{0.95}
\begin{tabular}{ll}
\hline \hline
Number of antennas & 58 to 64 \\
Total integration (target) & 50 h on source \\
FWHM of primary beam & $\sim$1$^{\circ}$ \\
Calibrated channel width & 6.4 kHz (1.4\kms) \\
\hline
\multicolumn{2}{l}{ESO444-G084} \\
\quad Robust = 0, taper = 0 & \\
\quad Pixel ($^{\prime\prime}$) & 2 \\
\quad Beam ($^{\prime\prime}$) & $8.0 \times 7.3$ \\
\quad Noise per channel (mJy/beam) & 0.22 \\
\quad 3$\sigma$, 16\kms\ $N_{\textsc{HI}}$ (cm$^{-2}$) & $5.9 \times 10^{19}$ \\
\quad Robust = 1, taper = 90 & \\
\quad Pixel ($^{\prime\prime}$) & 30 \\
\quad Beam ($^{\prime\prime}$) & $95.2 \times 91.3$ \\
\quad Noise per channel (mJy/beam) & 0.32 \\
\quad 3$\sigma$, 16\kms\ $N_{\textsc{HI}}$ (cm$^{-2}$) & $5.8 \times 10^{17}$ \\
\hline
\multicolumn{2}{l}{[KKS2000]23} \\
\quad Robust = 0, taper = 0 & \\
\quad Pixel ($^{\prime\prime}$) & 2 \\
\quad Beam ($^{\prime\prime}$) & $8.4 \times 6.9$ \\
\quad Noise per channel (mJy/beam) & 0.22 \\
\quad 3$\sigma$, 16\kms\ $N_{\textsc{HI}}$ (cm$^{-2}$) & $5.9 \times 10^{19}$ \\
\quad Robust = 1, taper = 90 & \\
\quad Pixel ($^{\prime\prime}$) & 30 \\
\quad Beam ($^{\prime\prime}$) & $94.1 \times 91.2$ \\
\quad Noise per channel (mJy/beam) & 0.32 \\
\quad 3$\sigma$, 16\kms\ $N_{\textsc{HI}}$ (cm$^{-2}$) & $5.8 \times 10^{17}$ \\
\hline \hline
\end{tabular}
\label{table2}
\end{table}
\section[HI morphology and distribution]{\HI\ morphology and distribution} \label{sec:distribution} 
\begin{figure*} 
\centering
   \advance\leftskip0cm
   \includegraphics[width=8.5 cm]{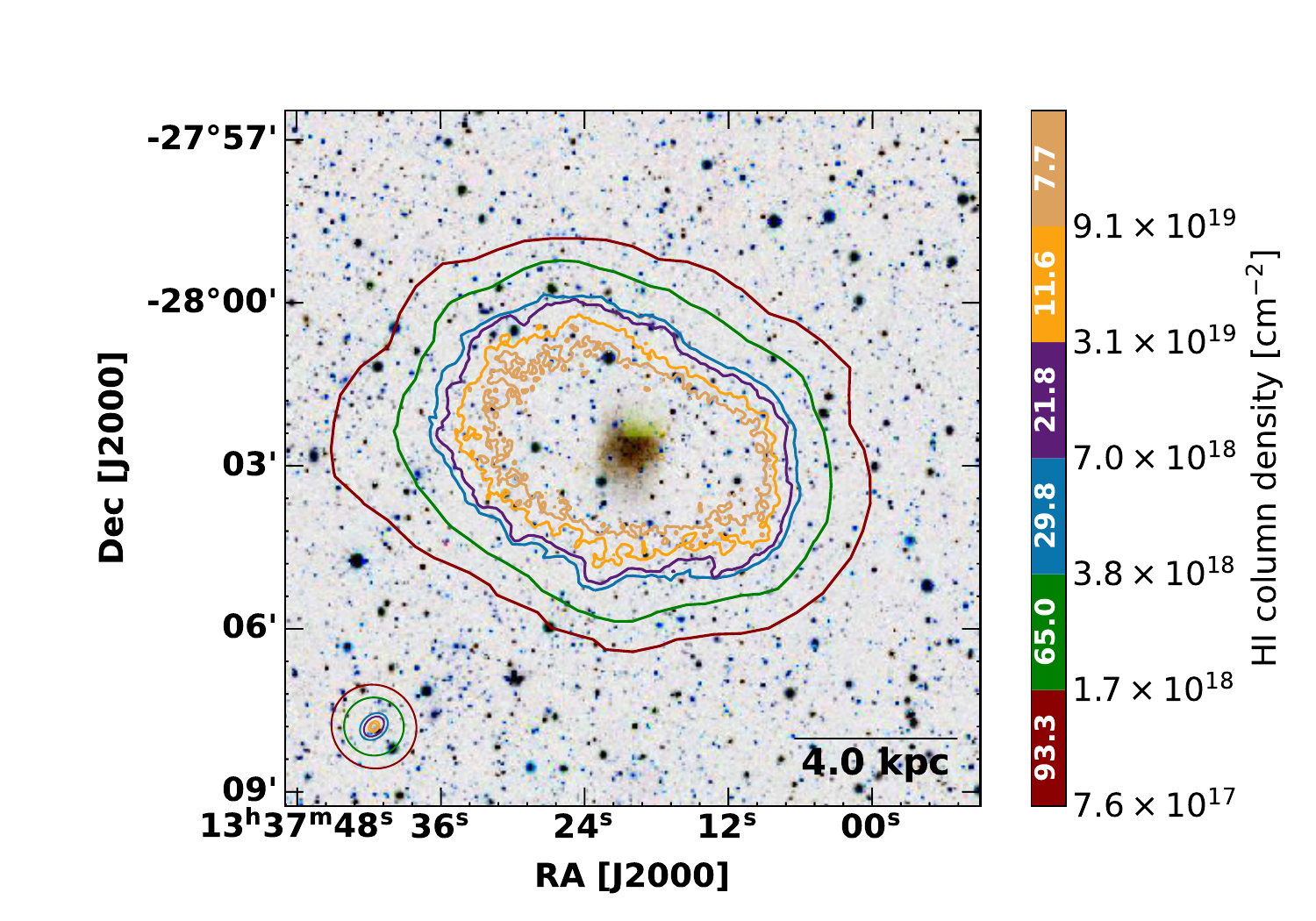}
     \includegraphics[width=8.5cm]{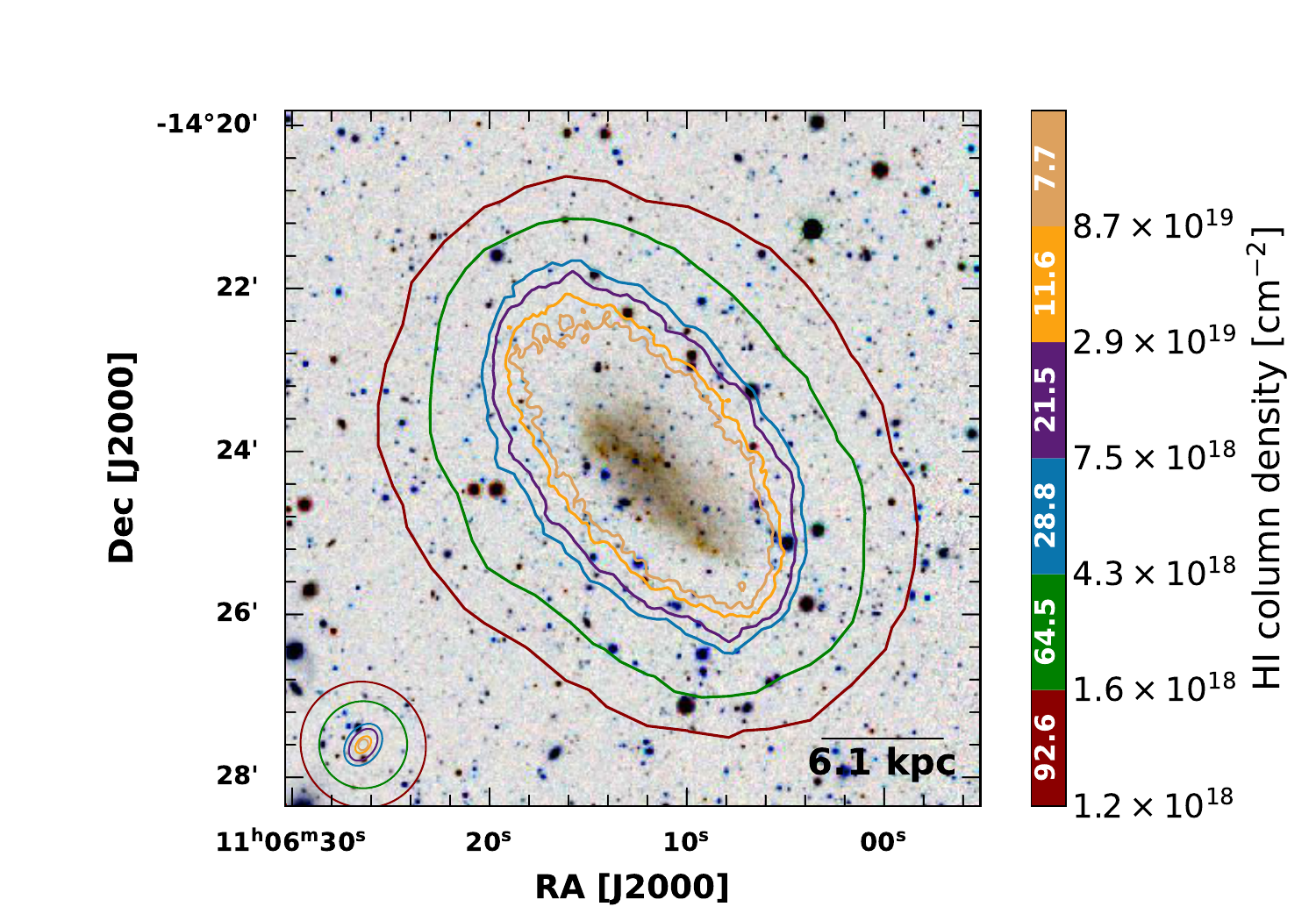}
\caption{MeerKAT \HI\ column density contours from different \HI\ resolution cubes of ESO444--G084 (left) and [KKS2000]23 (right) are overlaid on DECaLS grayscale images in the \textit{gri} bands. Each contour level represents the \HI\ column density at S/N = 3 for each spatial resolution. The color bar values indicate the average beam size corresponding to each spatial resolution. The ellipses in the lower-left corner of each map denote the beam sizes at different resolutions, while the black horizontal line represents the scale in kpc.}
\label{fig:mom0}
\end{figure*}

The results in this section are highlighted in Figs.~\ref{fig:mom0}, \ref{fig:mom0high}, \ref{fig:mom0high1} and \ref{fig:profile}. Fig.~\ref{fig:mom0} presents the \HI\ column density maps of ESO444--G084 and [KKS2000]23, derived from data cubes at different spatial resolutions. The lowest \HI\ contour at each resolution, corresponding to S/N = 3, is shown. Contours are color-coded and overlaid on deep grayscale images composed of \textit{gri} bands from the Dark Energy Camera Legacy Survey (DECaLS; \citealt{2019AJ....157..168D}) \footnote{\url{https://www.legacysurvey.org/decamls/}}. The \HI\ contours expand smoothly with increasing sensitivity. However, we find that the increase in \HI\ diameter from low (\(\sim 10^{19}\) cm\(^{-2}\)) to high (\(\sim 10^{18}\) cm\(^{-2}\)) column densities does not significantly impact the total fluxes of the galaxies (see Table~\ref{table11}). This suggests that the increase in \HI\ diameter at lower column densities is primarily due to the larger beam size, with no indication of significant additional emission.

\subsection{ESO444--G084}
The highest-resolution \HI\ column density map of ESO444--G084, presented in Fig.~\ref{fig:mom0high}, highlights a smooth gas distribution with high column density regions concentrated near the center of the galaxy. Distinct spiral arm structures are visible in the \HI\ morphology. The far-ultraviolet (FUV) emission from GALEX \citep{2005ApJ...619L...1M} as shown in Fig.~\ref{fig:mom0high} overlaps with the high-density \HI\ regions. Additionally, Fig.~\ref{fig:mom0high} includes velocity and velocity dispersion maps derived from the kinematic analysis, which will be discussed further in Section~\ref{sec:kinematics}.

Using the highest-resolution maps, we quantified the \HI\ effective diameter by applying a surface density threshold of \(1 \, \mathrm{M}_{\odot} \text{ pc}^{-2}\),  which is equivalent to a column density threshold of \(1.24 \times 10^{20}\) cm\(^{-2}\). Along the major axis, the effective \HI\ diameter was measured as 6.5 kpc (4$\farcm$9). Following the methodology outlined in \citet{2024ApJ...974..247C}, we estimated the optical diameter of ESO444–G084 using the concept of a “stellar edge,” rather than relying on a fixed isophotal threshold such as the traditional $D_{25}$. This approach defines the outer limit of the stellar disk based on features in the surface brightness and color profiles—specifically, a flattening or break in the light profile and a significant change in the $g - r$ color, which may indicate a transition in stellar populations or a truncation of the stellar disk. We constructed smoothed $g$- and $r$-band surface brightness profiles from DECaLS DR10 imaging, using elliptical annuli aligned with the galaxy’s kinematic parameters. A clear inflection point was identified near a semi-major axis where the $g - r$ color profile shows a distinct reddening and the surface brightness profiles begin to flatten. Based on this, we adopt a stellar edge diameter of $D_{\text{stellar}} = 0.94$ kpc for ESO444–G084. This yields an \HI-to-optical size ratio of \(\sim 6.9\).

The primary beam-corrected global \HI\ profile of ESO444--G084 is shown in Fig.~\ref{fig:profile}. The profile shows a symmetric shape. The total \HI\ flux integrated from the profile is \(21.2 \pm 0.2\) Jy\kms, which is consistent with the single-dish value of \(21.1 \pm 3.2\) Jy\kms \citep{Koribalski2004,2024A&A...688A.109D}. At a distance of 4.6 Mpc, this flux corresponds to an \HI\ mass of \((1.1 \pm 0.1) \times 10^{8} \mathrm{M}_{\odot}\). The profile widths at the 20\% and 50\% flux levels, determined using the \textsc{GIPSY} task \texttt{profglob}, are \(76.3 \pm 1.18\) \kms\ and \(55.8 \pm 2.1\) \kms, respectively. A summary of the MeerKAT \HI\ properties of ESO444--G084 is provided in Table~\ref{tablehi}.

\subsection{[KKS2000]23}
\begin{figure*}
\centering
   \advance\leftskip0cm
   \includegraphics[height=14cm,keepaspectratio]{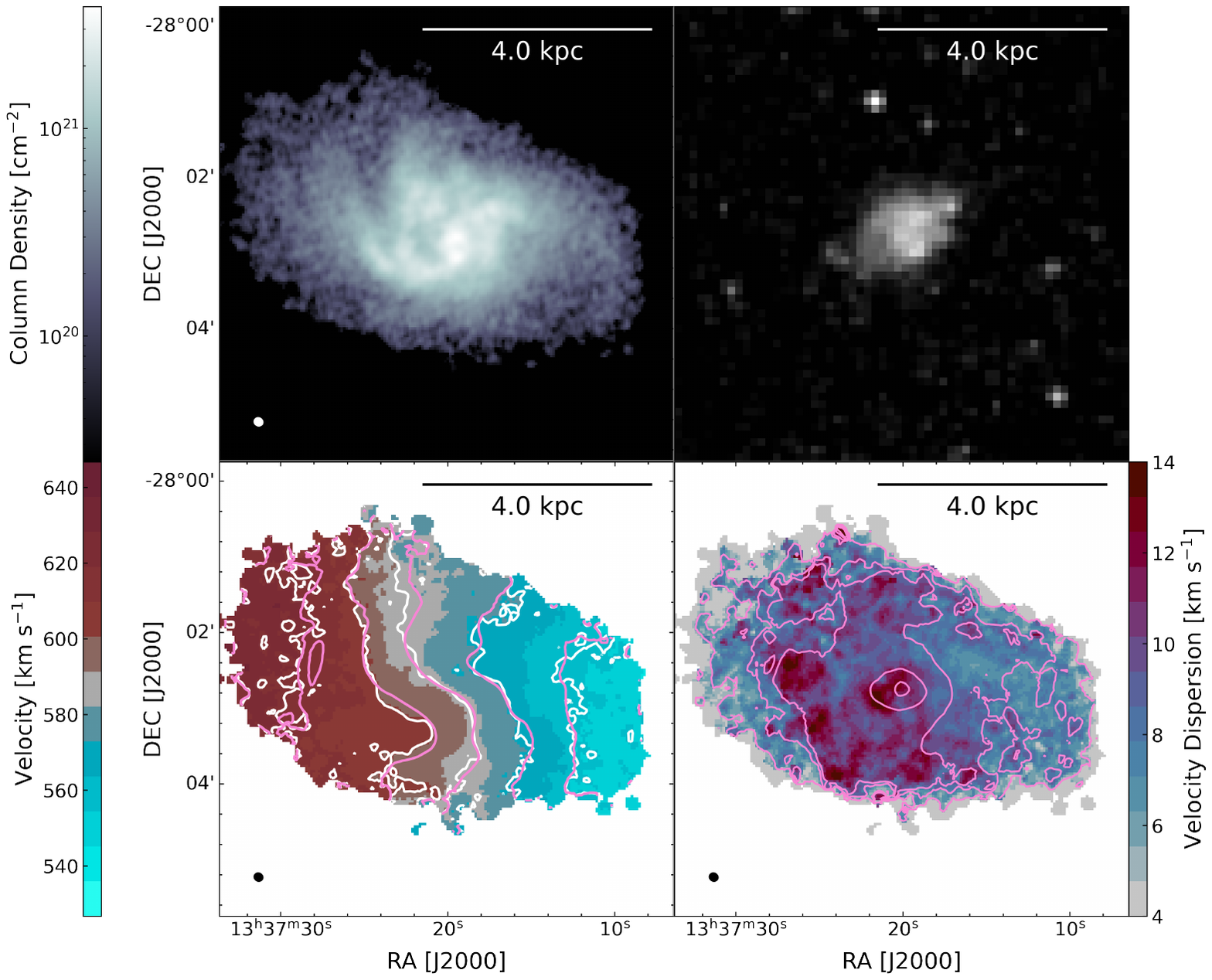}
\caption{High-resolution maps of ESO444--G084. \textit{Top left:} \HI\ column density maps from the highest resolution cubes. \textit{Top right:} The GALEX FUV images. \textit{Bottom left:} Velocity field models (pink contours) and observed (white contours) overlaid on velocity field maps (observed). \textit{Bottom right:} Dispersion map models (pink contours) overlaid on the data. The velocity field and dispersion map models were derived from the kinematic analysis described in Section \ref{sec:kinematics}. Contours for the velocity field are set at 526.7, 541.7, 556.7, 571.7, 586.7, 601.7, 616.7, 631.7, and 646.7 km\,s$^{-1}$, and contours for the velocity dispersion are set at 5, 7, 8, 10, and 12 km\,s$^{-1}$. The ellipses in the lower right corner of each map represent the beam sizes.} 
\label{fig:mom0high}
\end{figure*}

\begin{figure*}

\centering
   \advance\leftskip0cm
        \includegraphics[height=14cm,keepaspectratio]  {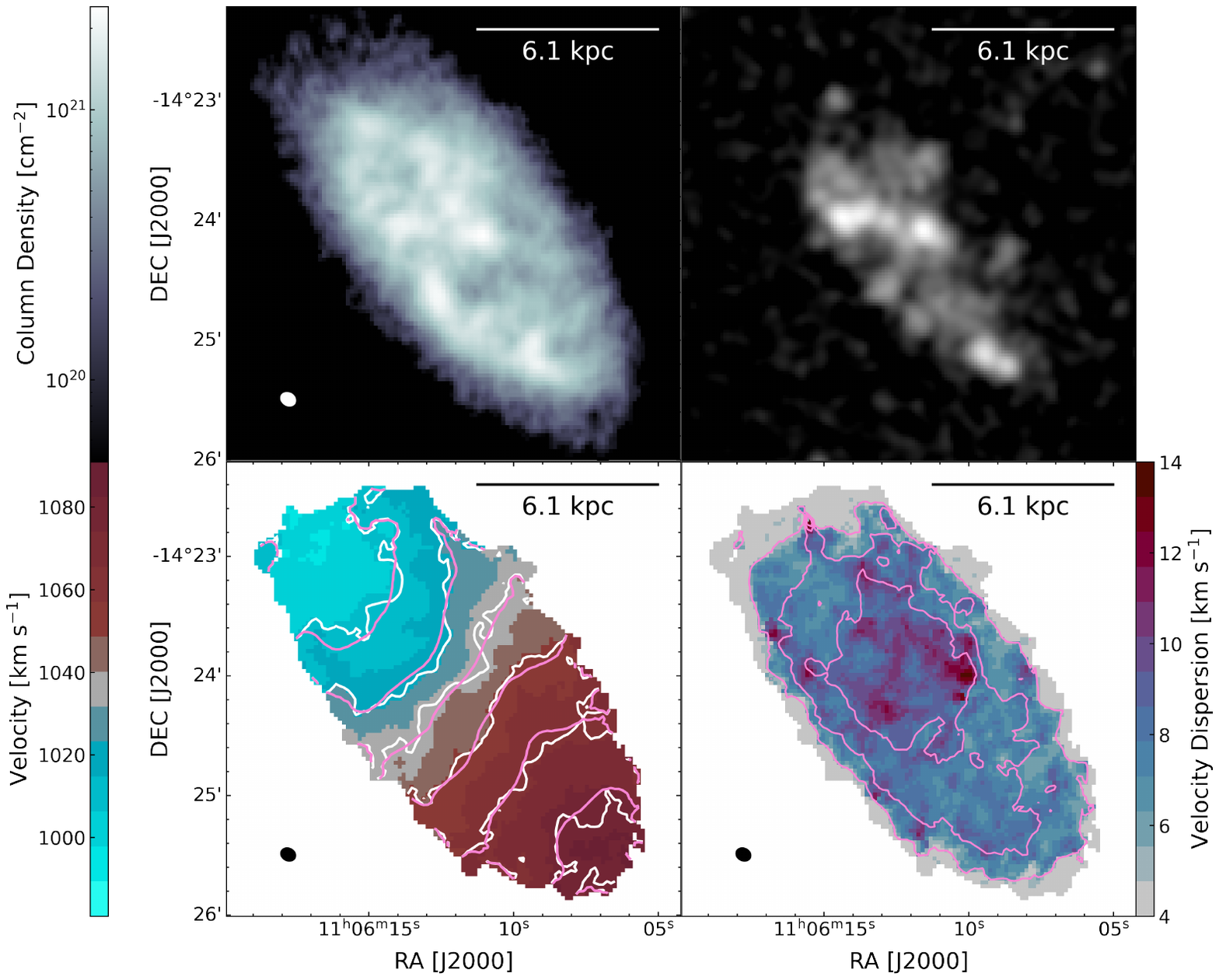}
\caption{High-resolution maps of [KKS2000]23. \textit{Top left:} \HI\ column density maps from the highest resolution cubes. \textit{Top right:} The GALEX FUV images. \textit{Bottom left:} Velocity field models (pink contours) and observed (white contours) overlaid on velocity field maps (observed). \textit{Bottom right:} Dispersion map models (pink contours) overlaid on the data. The velocity field and dispersion map models were derived from the kinematic analysis described in Section \ref{sec:kinematics}. Contours for the velocity field are set at 980.9, 994.7, 1008.4, 1022.2, 1035.9, 1049.7, 1063.4, 1077.2, and 1090.9 km\,s$^{-1}$, and contours for the velocity dispersion are set at 5, 7, 8, 10, and 12 km\,s$^{-1}$. The ellipses in the lower right corner of each map represent the beam sizes.} 
\label{fig:mom0high1}
\end{figure*}

Fig.~\ref{fig:mom0high1} presents the highest-resolution \HI\ column density map of [KKS2000]23. The map reveals a more fragmented distribution with localized high-density regions scattered across the disk. These regions align closely with the bright FUV emission in the GALEX maps \citep{2005ApJ...619L...1M} shown in Fig.~\ref{fig:mom0high}. The velocity and dispersion maps included in Fig.~\ref{fig:mom0high} will be analyzed in detail in Section~\ref{sec:kinematics}.

To characterize the \HI\ distribution, we calculated the effective \HI\ diameter from the highest-resolution maps by using the method described above. The \HI\ effective diameter was measured as 16.0 kpc (4$\farcm$0). Based on the method used to determine the stellar edge of ESO444–G084, we estimated a stellar diameter of 2.83 kpc for [KKS2000]23. This yields an H~I-to-optical size ratio of approximately 1.7.

\begin{figure*}
\centering
   \includegraphics[width=7.0cm]{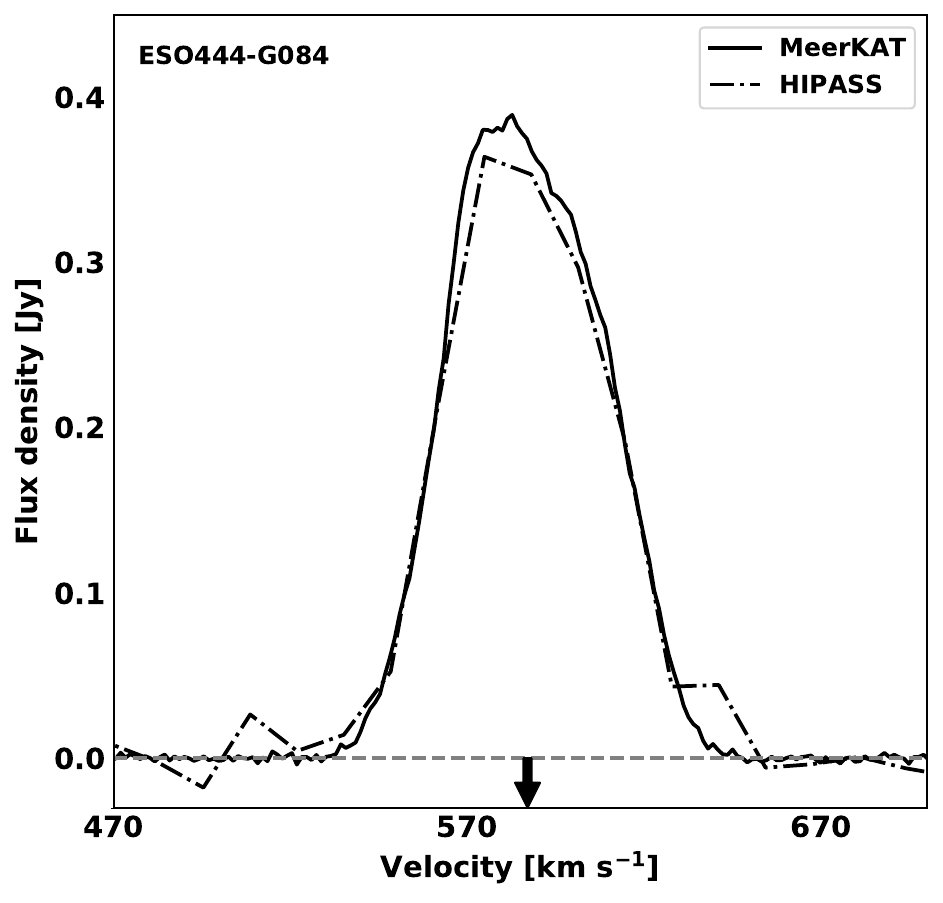}
   \includegraphics[width=7.5cm]{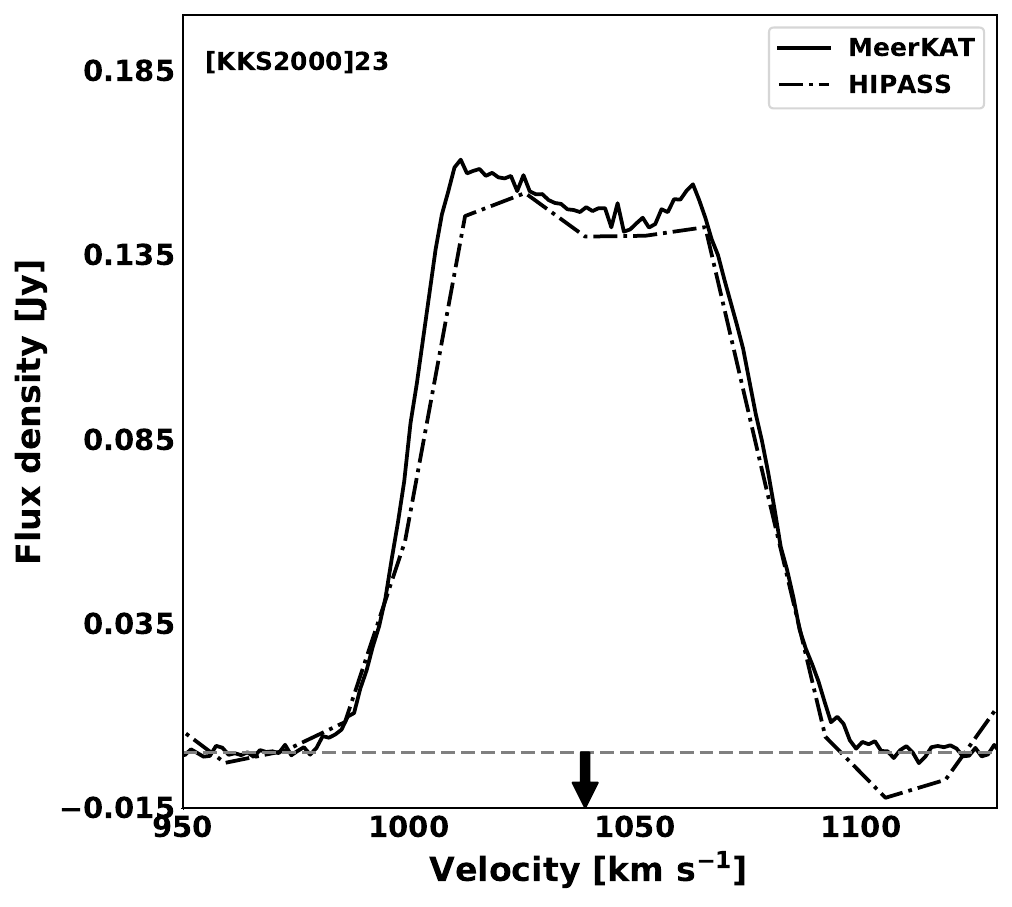}
\caption{Integrated \HI\ spectra of ESO444--G084 (left) and [KKS2000]23 (right) from MeerKAT (black solid line) and HIPASS \citep[black dashed line,][]{Koribalski2004}. The black arrow on each figure shows the kinematic systemic velocity of each galaxy derived in Section~\ref{sec:kinematics}. The horizontal dashed grey lines indicate the zero level. These profiles are derived from the high-resolution data.}
\label{fig:profile}
\end{figure*}

The primary beam corrected global \HI\ profile of [KKS2000]23 is shown in Fig.~\ref{fig:profile}. The profile displays a notable asymmetry, with a higher \HI\ mass on the approaching side. The integrated flux derived from the \HI\ profile is \(12.3 \pm 0.1\) Jy\kms, consistent with \citet{2024A&A...688A.109D} , but 13\% lower than the GBT value of \(14.2\) Jy\kms \citep{Sardone2021}. The discrepancy may result from the reported noise in the GBT spectra, which made it challenging to determine if the low column density gas was real emission or simply artefacts from the observations. At a distance of 13.9 Mpc, the measured flux corresponds to an \HI\ mass of \((6.1 \pm 0.3) \times 10^{8} \mathrm{M}_{\odot}\). The velocity widths at the 20\% and 50\% flux levels, calculated using the \textsc{GIPSY} task \texttt{profglob}, were measured as \(94.2 \pm 0.92\) \kms\ and \(79.7 \pm 1.4\) \kms, respectively. A summary of the MeerKAT \HI\ properties of [KKS2000]23 is provided in Table~\ref{tablehi}.

\begin{table*}
\centering
\caption{Basic \HI\ Properties of ESO444--G084 and [KKS2000]23 at different angular resolutions and column density sensitivities.}
\begin{tabular}{ccccc} 
\hline   
Galaxy & Resolution ($^{\prime \prime}$) & $N_{\textsc{HI}}$ (S/N = 3) (cm$^{-2}$) & \HI\ diameter (kpc) & Flux (Jy \kms) \\   
\hline
ESO444--G084 & 8.0 $\times$ 7.3 & 9.1 $\times$ 10$^{19}$ & 7.6 & 20.2\\
    & 13.3 $\times$ 10.1 & 3.1 $\times$ 10$^{19}$ & 7.8 & 21.2\\
    & 25.7 $\times$ 18.5 & 7.0 $\times$ 10$^{18}$ & 8.4 & 21.0\\
    & 33.9 $\times$ 26.1 & 3.8 $\times$ 10$^{18}$ & 9.5 & 21.2\\
    & 66.0 $\times$ 64.1 & 1.7 $\times$ 10$^{18}$ & 11.5 & 21.2\\
    & 95.2 $\times$ 91.3 & 7.6 $\times$ 10$^{17}$ & 12.2 & 21.5\\ 
\hline   
[KKS2000]~23 & 8.4 $\times$ 6.9 & 8.7 $\times$ 10$^{19}$ & 17.3 & 12.3\\
    & 14.4 $\times$ 9.3 & 2.9 $\times$ 10$^{19}$ & 18.5 & 12.3\\
    & 26.2 $\times$ 17.6 & 7.5 $\times$ 10$^{18}$ & 20.2 & 12.2\\
    & 33.8 $\times$ 24.5 & 4.3 $\times$ 10$^{18}$ & 21.8 & 12.3\\
    & 64.7 $\times$ 64.1 & 1.6 $\times$ 10$^{18}$ & 27.8 & 12.2\\
    & 94.1 $\times$ 91.1 & 1.2 $\times$ 10$^{18}$ & 30.7 & 12.3\\ 
\hline
\end{tabular}   
\label{table11}
\end{table*}


\vspace{1em} 

\begin{table}[ht]
\centering
\caption{MeerKAT \HI\ properties of ESO444--G084 and [KKS2000]23.}
\label{tablehi}
\footnotesize
\setlength{\tabcolsep}{3pt} 
\renewcommand{\arraystretch}{0.95} 
\begin{tabular}{p{4.2cm}@{\hskip 2pt}c@{\hskip 2pt}c} 
\hline \hline
Parameter & ESO444--G084 & [KKS2000]23 \\
\hline
Total \HI\ flux (Jy\,km\,s$^{-1}$) & 21.2 $\pm$ 0.2 & 12.3 $\pm$ 0.1 \\
Total \HI\ mass (M$_{\odot}$) & $(1.1 \pm 0.1) \times 10^{8}$ & $(6.1 \pm 0.3) \times 10^{8}$ \\
Profile width at 20\% (km\,s$^{-1}$) & $76.3 \pm 1.2$ & $94.2 \pm 0.9$ \\
Profile width at 50\% (km\,s$^{-1}$) & $55.8 \pm 2.1$ & $79.7 \pm 1.4$ \\
\hline
\end{tabular}
\end{table}
\section{\HI\ kinematics}\label{sec:kinematics}
To examine both the inner and outer kinematics of ESO444--G084 and [KKS2000]23, we derived the rotation curves by fitting tilted-ring models to the highest resolution ($\sim$ 8$^{\prime \prime}$) and mid-resolution ($\sim$ 23$^{\prime \prime}$) 3D data cubes using \textsc{PyFAT} \citep{2024ascl.soft07002K} and \textsc{TiRiFiC} \citep{2012ascl.soft08008J}.  

\textsc{PyFAT} is a Python wrapper for \textsc{TiRiFiC} that automates the fitting of tilted-ring models. The tilted-ring model is a widely used method for modeling the kinematics of disk galaxies by dividing the gas disk into concentric rings, each characterized by parameters such as rotation velocity, inclination, and position angle. \textsc{PyFAT} initially assumes a flat-disk model, meaning that the inclination and position angle remain constant across all rings, representing a uniformly rotating disk with no significant warps or distortions. This simplification works well for symmetric disks but may not fully capture more complex kinematic structures and may fail for galaxies with complicated kinematics. To run \textsc{PyFAT}, the only required input is the 3D data cube, from which the code automatically extracts initial kinematic parameters and iteratively fits a tilted-ring model across the disk. In the case of [KKS2000]23, \textsc{PyFAT} produced a reasonable kinematic model, successfully capturing the overall velocity structure of the galaxy. However, \textsc{PyFAT} generates a single axisymmetric rotation curve that averages the approaching and receding sides of the disk. In this study, our aim is to analyze the shapes of the rotation curves for the approaching and receding sides separately, leading to an additional modeling step using \textsc{TiRiFiC}.

For ESO444--G084, \textsc{PyFAT} did not converge to a satisfactory solution, likely due to its complex velocity field, which caused the initial flat-disk assumption to produce a poor fit, a result that was also reported in \citet{2015MNRAS.452.3139K}. To refine the model, we used \textsc{TiRiFiC} in standalone mode. We began with the simplest possible configuration, assuming constant inclination and position angle across all rings. After visually inspecting the model and comparing it with the observations, we performed multiple iterations, introducing variations in key parameters by either fixing or allowing radial changes in position angle and inclination, while also incorporating radial motions where needed. The best-fit model was obtained with a position angle and inclination that remained fixed in the central regions but varied at larger radii. At each stage, we evaluated the fit by comparing the model channel maps, moment maps, and position–velocity diagrams with the data, and it was accepted when it reproduced the large-scale \HI\ morphology and kinematics without introducing systematic mismatches. The gas disk was modeled using two distinct components fitted separately, with the exception of the velocity dispersion, which was treated as a single component to ensure a well-constrained model. Interpolation was applied to maintain smooth transitions between adjacent rings, minimizing artificial fluctuations in the rotation curves.
\subsection{Pressure gradient correction}
\begin{figure*}
\centering
   \advance\leftskip0cm
   \includegraphics[height=6cm]{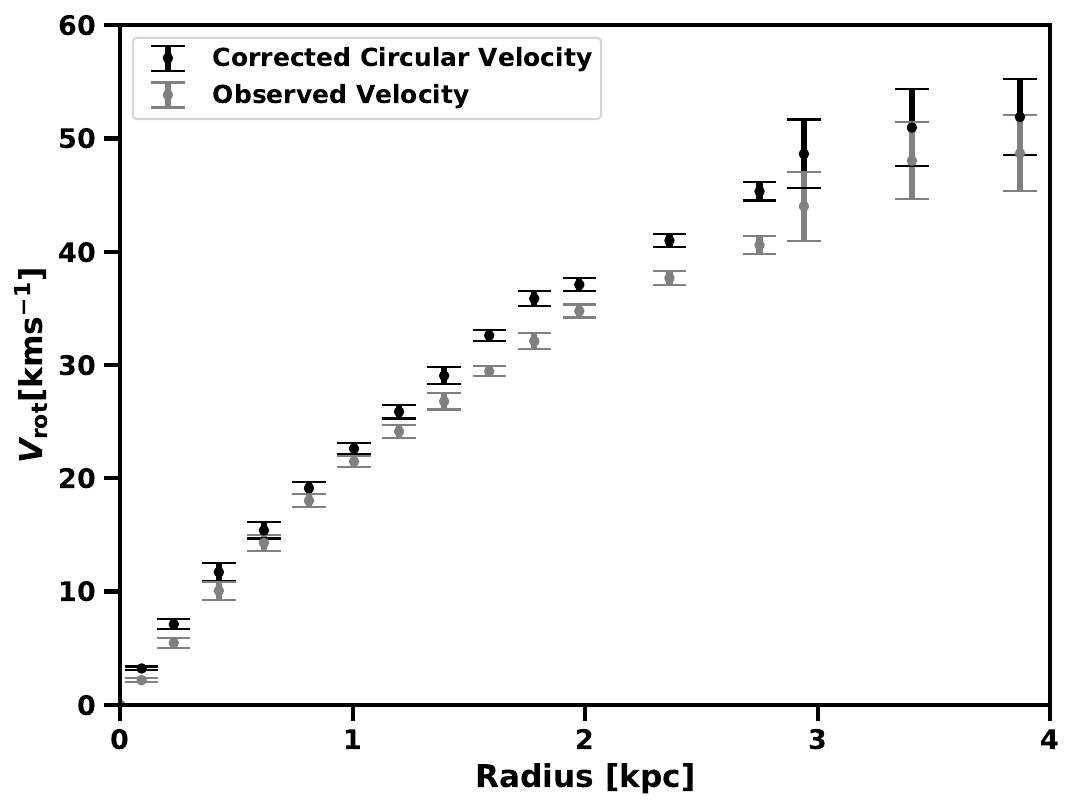}
    \includegraphics[height=6cm]{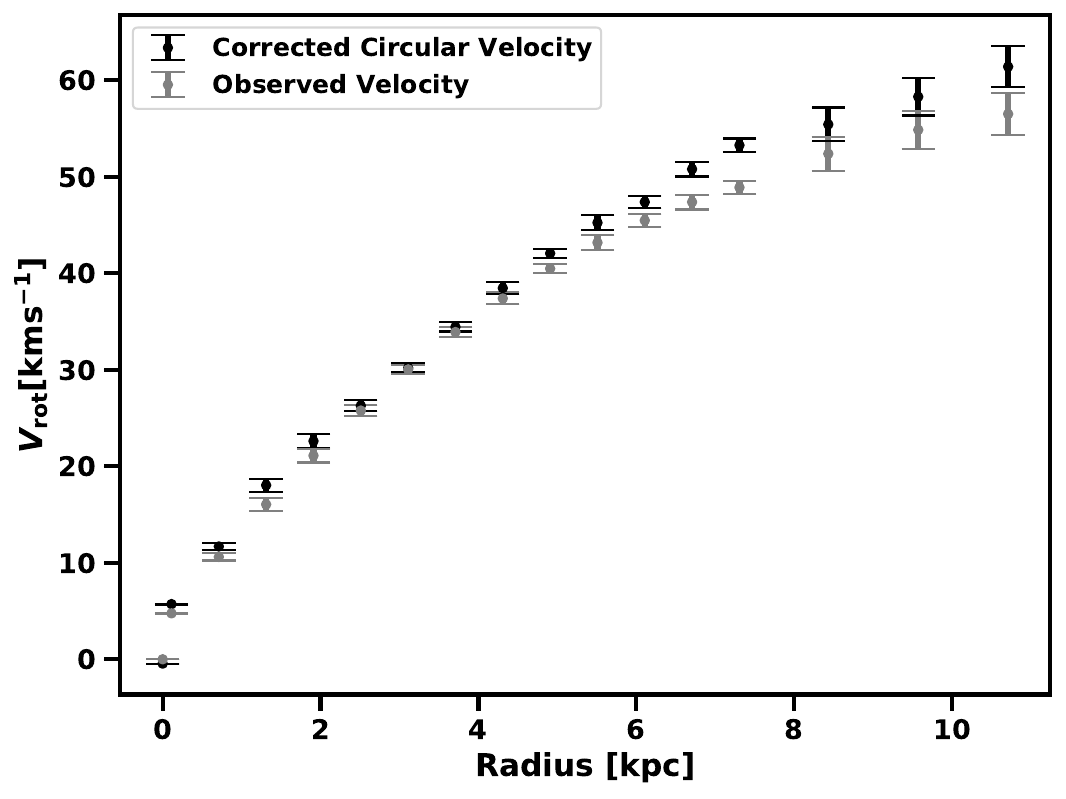}
\caption{Observed rotation curves (grey) and asymmetric drift-corrected rotation curves (black) for ESO444--G084 (left) and [KKS2000]23 (right).} 
\label{fig:asy}
\end{figure*}

Due to the gravitational interaction of gas on circular orbits, the stability of a galaxy against gravitational collapse is maintained by both rotation and pressure gradients. As a result, the observed rotation velocities are typically lower than the true circular velocities associated with the gravitational potential. To accurately trace the mass distribution of galaxies, we apply a pressure gradient correction\footnote{Often referred to as an asymmetric drift correction in analogy with the mathematically similar concept from collisionless dynamics.} to the derived rotation velocities following \citet{1996AJ....111.1551M}:
\begin{equation}
V^{2}_{\text{c}} = V^{2}_{\text{rot}} + \sigma^{2}_{D}
\end{equation}
where \( \sigma_{D} \) represents the asymmetric drift, given by:
\begin{equation}
\sigma^{2}_{D} = -R b^{2} \Bigg[\frac{\partial \ln \Sigma_{g}}{\partial R} + 2\frac{\partial \ln b}{\partial R} - \frac{\partial \ln h_{z}}{\partial R} \Bigg]
\end{equation}
where \( V_{\text{c}} \) is the circular velocity, \( V_{\text{rot}} \) is the rotational velocity, \( \Sigma_{g} \) is the \HI\ surface density including helium, \( R \) is the radius, and \( b \) is the velocity dispersion. The velocity dispersion values are derived from \textsc{PyFAT} and \textsc{TiRiFiC}. The asymmetric drift is computed under the assumption that the scale height gradient is negligible, i.e., \( \frac{\partial \ln h_{z}}{\partial R} = 0 \). Fig. \ref{fig:asy} shows the corrected circular velocities, and Tables \ref{tab:asy1} and \ref{tab:asy2} present the asymmetric drift-corrected rotation curves for ESO444--G084 and [KKS2000]23, respectively.

\begin{table}[h]
    \centering
    \setlength{\tabcolsep}{3pt} 
    \caption{Derived rotation velocities (V$_{\mathrm{rot}}$) of ESO444--G084 and asymmetric drift-corrected velocities (V$_{\mathrm{c}}$) with associated errors (V$_{\mathrm{errors}}$).}
    \label{tab:asy1}
    \begin{tabular}{cccc}
        \hline
        Radius (kpc) & V$_{\mathrm{rot}}$ (\kms) & V$_{\mathrm{c}}$ (\kms) & V$_{\mathrm{errors}}$ (\kms) \\
        \hline
        0.00 & 0.00  & 0.00  & 0.00 \\
        0.09 & 2.17  & 3.21  & 0.17 \\
        0.23 & 5.46  & 7.11  & 0.44 \\
        0.42 & 10.07 & 11.71 & 0.80 \\
        0.61 & 14.29 & 15.40 & 0.72 \\
        0.81 & 18.02 & 19.13 & 0.55 \\
        1.00 & 21.48 & 22.62 & 0.52 \\
        1.19 & 24.14 & 25.88 & 0.60 \\
        1.39 & 26.79 & 29.06 & 0.73 \\
        1.59 & 29.45 & 32.61 & 0.46 \\
        1.78 & 32.11 & 35.87 & 0.69 \\
        1.97 & 34.77 & 37.10 & 0.58 \\
        2.36 & 37.68 & 40.99 & 0.61 \\
        2.75 & 40.59 & 45.33 & 0.80 \\
        2.94 & 44.01 & 48.64 & 3.03 \\
        3.40 & 48.04 & 50.97 & 3.41 \\
        3.87 & 48.69 & 51.90 & 3.36 \\
        \hline
    \end{tabular}
\end{table}

\begin{table}[h]
    \centering
    \setlength{\tabcolsep}{3pt}
    \caption{Derived rotation velocities (V$_{\mathrm{rot}}$) of [KKS2000]2 and asymmetric drift-corrected velocities (V$_{\mathrm{c}}$) with associated errors (V$_{\mathrm{errors}}$).}
    \label{tab:asy2}
    \begin{tabular}{cccc}
        \hline
        Radius (kpc) & V$_{\mathrm{rot}}$ (\kms) & V$_{\mathrm{c}}$ (\kms) & V$_{\mathrm{errors}}$ (\kms) \\
        \hline
        0.00  & 0.00  & 0.00  & 0.00 \\
        0.11  & 4.76  & 5.72  & 0.01 \\
        0.71  & 10.60 & 11.69 & 0.38 \\
        1.31  & 16.05 & 18.04 & 0.70 \\
        1.91  & 21.11 & 22.62 & 0.70 \\
        2.51  & 25.77 & 26.29 & 0.54 \\
        3.11  & 30.04 & 30.19 & 0.39 \\
        3.71  & 33.91 & 34.45 & 0.51 \\
        4.31  & 37.40 & 38.47 & 0.72 \\
        4.91  & 40.48 & 42.07 & 0.48 \\
        5.51  & 43.18 & 45.25 & 0.74 \\
        6.11  & 45.48 & 47.39 & 0.73 \\
        6.71  & 47.38 & 50.80 & 0.85 \\
        7.31  & 48.90 & 53.28 & 0.88 \\
        8.43  & 52.87 & 55.43 & 3.00 \\
        9.57  & 54.87 & 58.30 & 2.85 \\
        10.71 & 56.51 & 61.42 & 2.24 \\
        \hline
    \end{tabular}
\end{table}

\subsection{Kinematic results}
The highest-resolution cubes provides a reliable determination of the \HI\ rotation curve out to $\sim$ 2.8 kpc for ESO444--G084, whereas the mid-resolution cubes allowed for an extended measurement out to $\sim$ 3.8 kpc, capturing additional kinematic information. For [KKS2000]23, the highest-resolution data constrained the rotation curve to $\sim$ 7.8 kpc, whereas the lowest-resolution cubes enabled its extension to $\sim$ 10.2 kpc, offering a more comprehensive view of the galaxy's outer dynamics. The findings in this section are presented in Figs.~\ref{fig:mom0high}, \ref{fig:mom0high1}, \ref{fig:kinematic1}, \ref{fig:kinematic11}, \ref{fig:channel1}, \ref{fig:channel2}, \ref{fig:pv13} and \ref{fig:pv11}.

\subsubsection{ESO444--G084}
The bottom-left panel of Fig.~\ref{fig:mom0high} presents the observed and model velocity fields of ESO444--G084. The central region does not exhibit strictly parallel and symmetric isovelocity contours. These contours also lack the significant S-shaped or X-shaped distortions that are typically associated with bar-like structures (e.g., \citealt{1999ApJ...522..699A}). To explore the possibility of a bar, we incorporated a bar component into the kinematic modeling using the method prescribed in\footnote{\url{https://github.com/PeterKamphuis/pk_common_functions}}; however, the results did not provide strong evidence for a significant bar. This suggests that ESO444--G084 is unlikely to host a prominent bar, and the observed central kinematics may instead be influenced by mild radial motions or weak non-circular gas flows. In the outer disk, a change in the position angle of the velocity gradient is apparent, suggesting the presence of a kinematic warp.  

\begin{figure*}
\centering
   \advance\leftskip0cm
   \includegraphics[height=18cm]{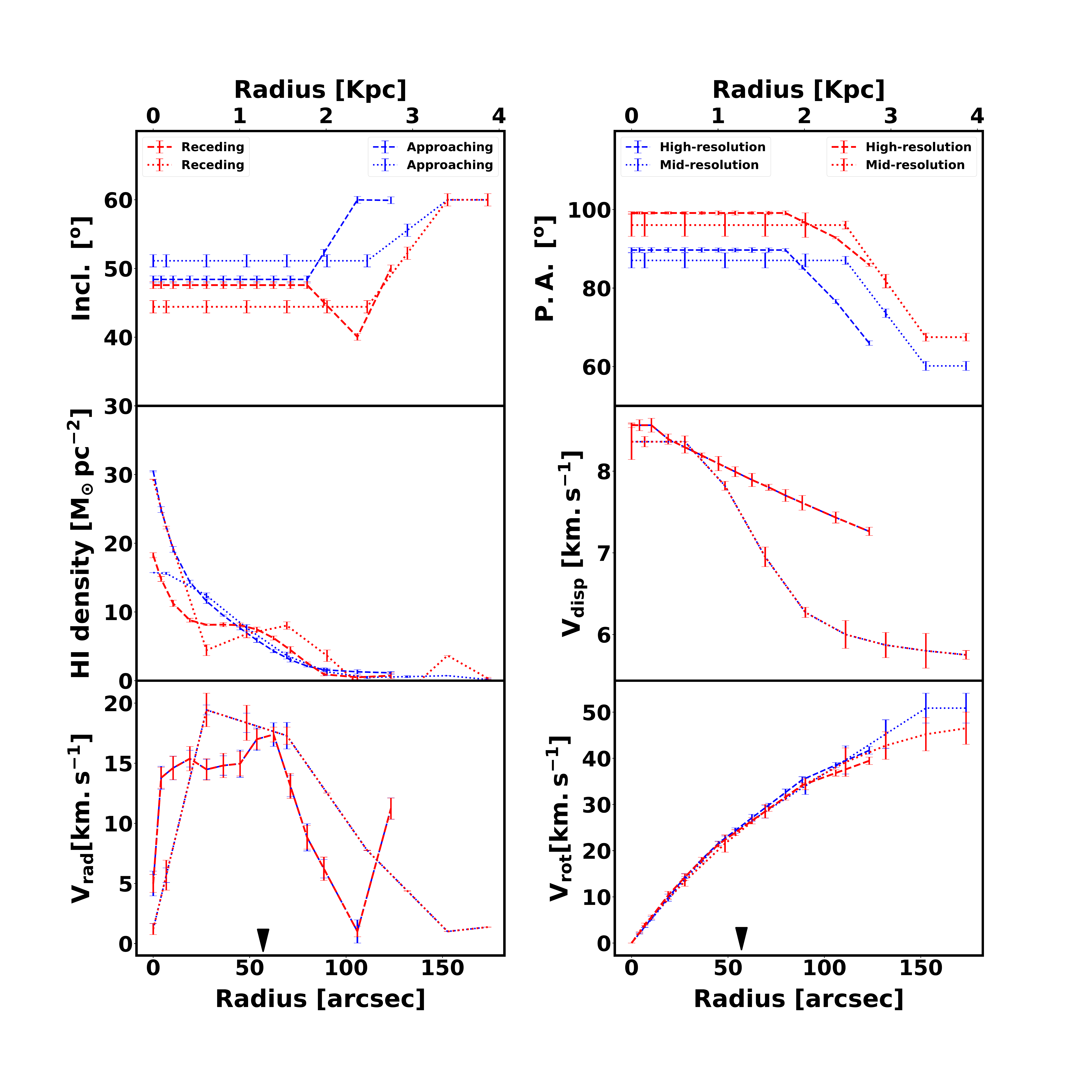}
\caption{Best-fit kinematic parameters of ESO444--G084 derived from the \HI\ data cube using \textsc{TiRiFiC}. The panels display inclination angle, position angle, \HI\ surface density, radial velocity, and rotation velocity. The blue and red lines correspond to the approaching and receding sides, respectively. The dashed line represents high resolution, while the dotted line represents mid-resolution. The black arrows represent the 90th percentile stellar radius obtained from the WISE 3.4$\mu$m band surface brightness profile.}
\label{fig:kinematic1}
\end{figure*}

From the highest-resolution model, the dynamical center was determined at \( \alpha,\delta \) (J2000) = \( 13^{\mathrm{h}}37^{\mathrm{m}}19.8^{\mathrm{s}}, -28^{\circ}02^{\mathrm{m}}46.4^{\mathrm{s}} \), consistent with the optical center (Table~\ref{table1}). The derived systemic velocity of \( 586.0 \pm 2.1 \) \kms closely matches the \( 586.7 \) \kms reported by \citet{2024A&A...688A.109D}. Using the profiles in Fig.~\ref{fig:kinematic1}, we adopted an average inclination of \( 49.0 \pm 1.2^{\circ} \) and a position angle of \( 90.0 \pm 2.4^{\circ} \).  

The rotation curve exhibits a gradual rise in the inner region (\( R < 2 \) kpc) rather than the steep, abrupt increase expected for an NFW-like halo. This shape is characteristic of a system dominated by rotational motion. The radial velocity profile in Fig.~\ref{fig:kinematic1} reveals significant non-circular motions in the inner regions, reaching up to \( \sim20 \) \kms. These deviations are possibly linked to weak inflows or minor warps. Beyond \( \sim3 \) kpc, these radial motions gradually diminish, marking a transition to a more rotation-dominated outer disk.

\begin{figure*}
\centering
   \advance\leftskip0cm
   \includegraphics[height=10cm]{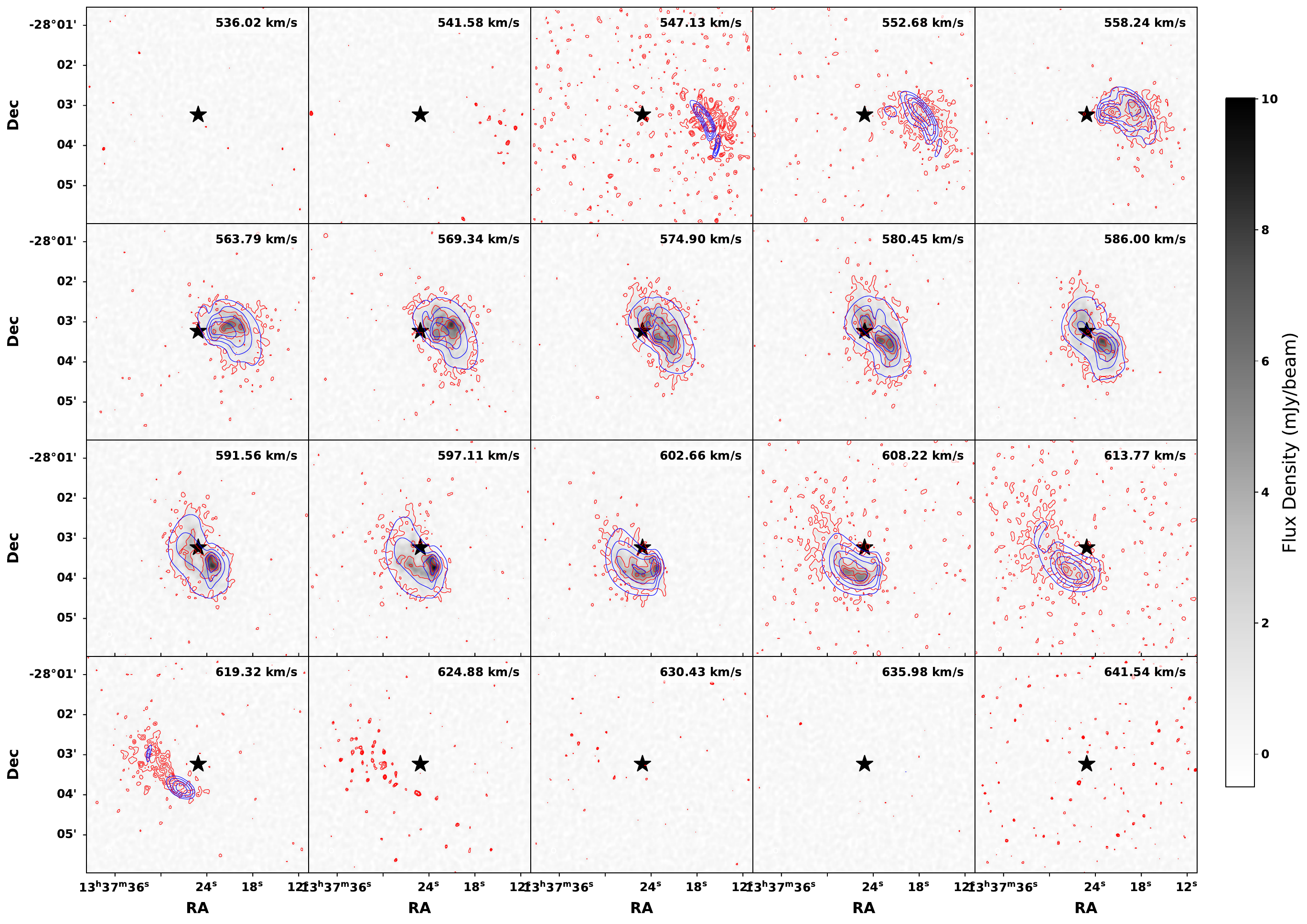} 
\caption{Individual channel maps of ESO444-G084 from the MeerKAT high-resolution cube. Blue contours represent the channel maps from the 3D modeling, while the red contours show the observed data. Contour levels denote values of (3, 6, 9, 12) $\times \sigma$. Black crosses represent the kinematic center of the observed data.}
\label{fig:channel1}
\end{figure*}

\begin{figure*}
\centering
   \advance\leftskip0cm
    \includegraphics[height=6cm]{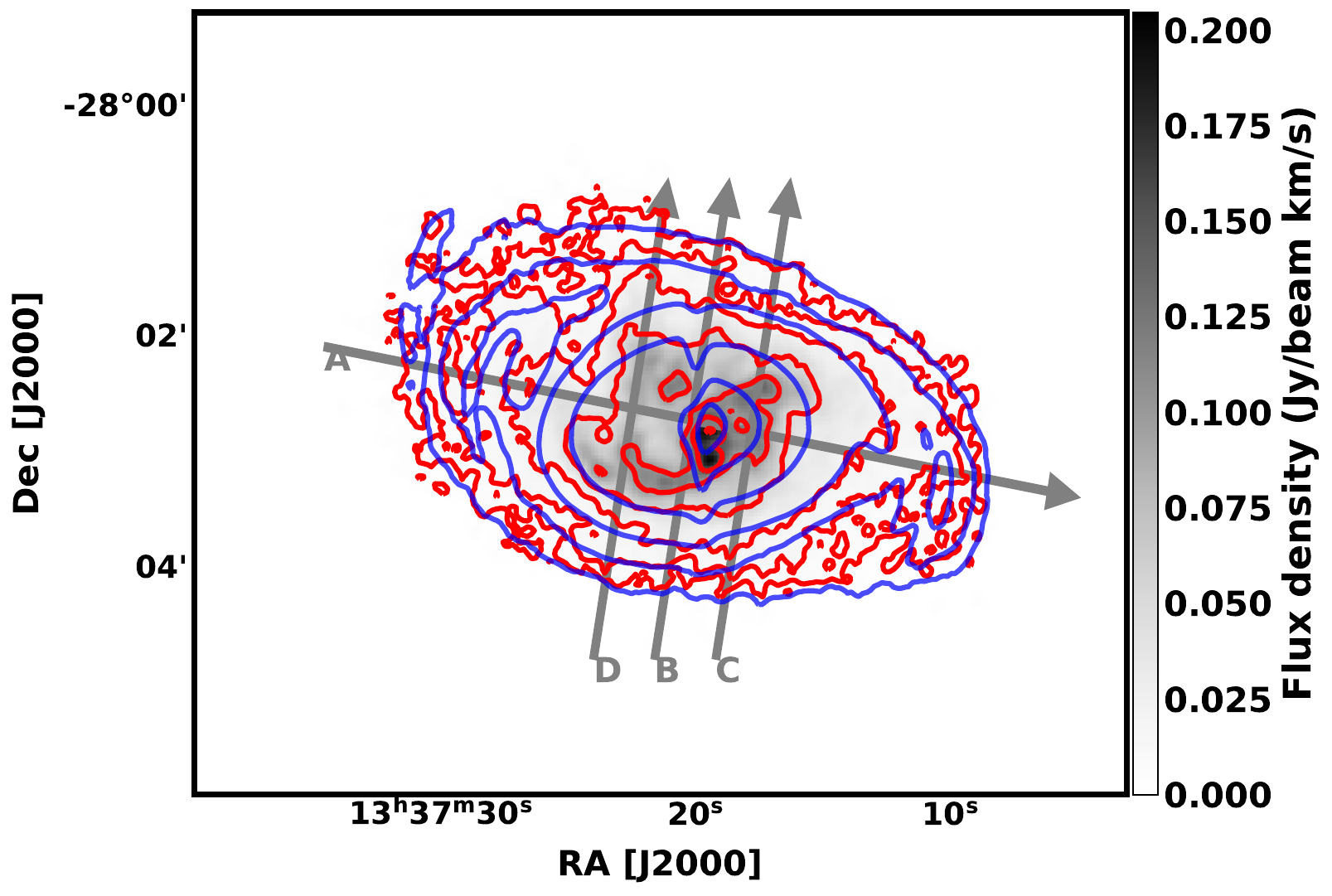}
   \includegraphics[height=13cm]{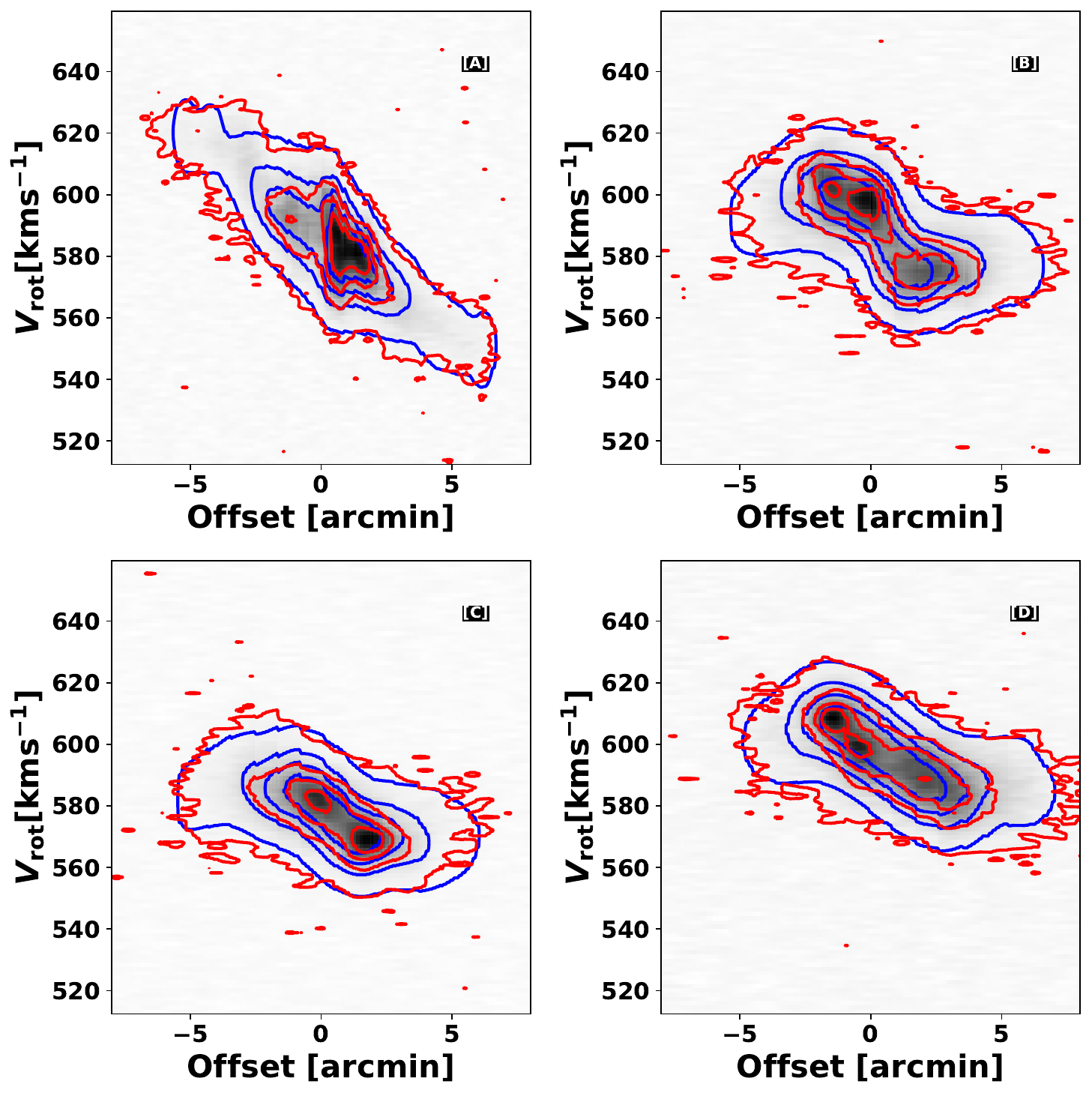}
\caption{Top: The moment-0 map of ESO444-G084 from the MeerKAT high-resolution cube, displaying grey arrows that indicate the positions of slices along which the position-velocity diagrams (bottom) were extracted. The arrows indicate the locations of the PV slices, with one aligned along the kinematic major axis (A) to capture the bulk of rotational motion and another along the minor axis (B) to examine deviations from pure rotation. Additional slices (C and D) were chosen to explore variations in the \HI\ kinematics and overall disk structure. The red contours represent the data, while the blue contours represent the model. The contour levels are set at (3, 6, 9, 15, 18) $\times$ $\sigma$.}
\label{fig:pv13}
\end{figure*}

The channel maps in Fig.~\ref{fig:channel1} and the PV diagrams in Fig.~\ref{fig:pv13} show good agreement between the observed data and the kinematic model across the disk. This agreement suggests that the model effectively captures the large-scale kinematics of the galaxy, with only minor deviations in localized regions. The summary of the kinematic results is shown in Table~\ref{table:kinematics}.
\subsubsection{[KKS2000]23}
The bottom-left panel of Fig.~\ref{fig:mom0high1} shows the velocity field of [KKS2000]23, which exhibits smooth, well-ordered isovelocity contours, suggesting a coherent rotation pattern. The transition from blue-shifted to red-shifted velocities follows a clear gradient, indicative of a regularly rotating disk. There are no obvious kinematic twists, strong distortions, or significant asymmetries that would suggest major disturbances from external interactions.

\begin{figure*}
\centering
   \advance\leftskip0cm
   \includegraphics[height=18cm]{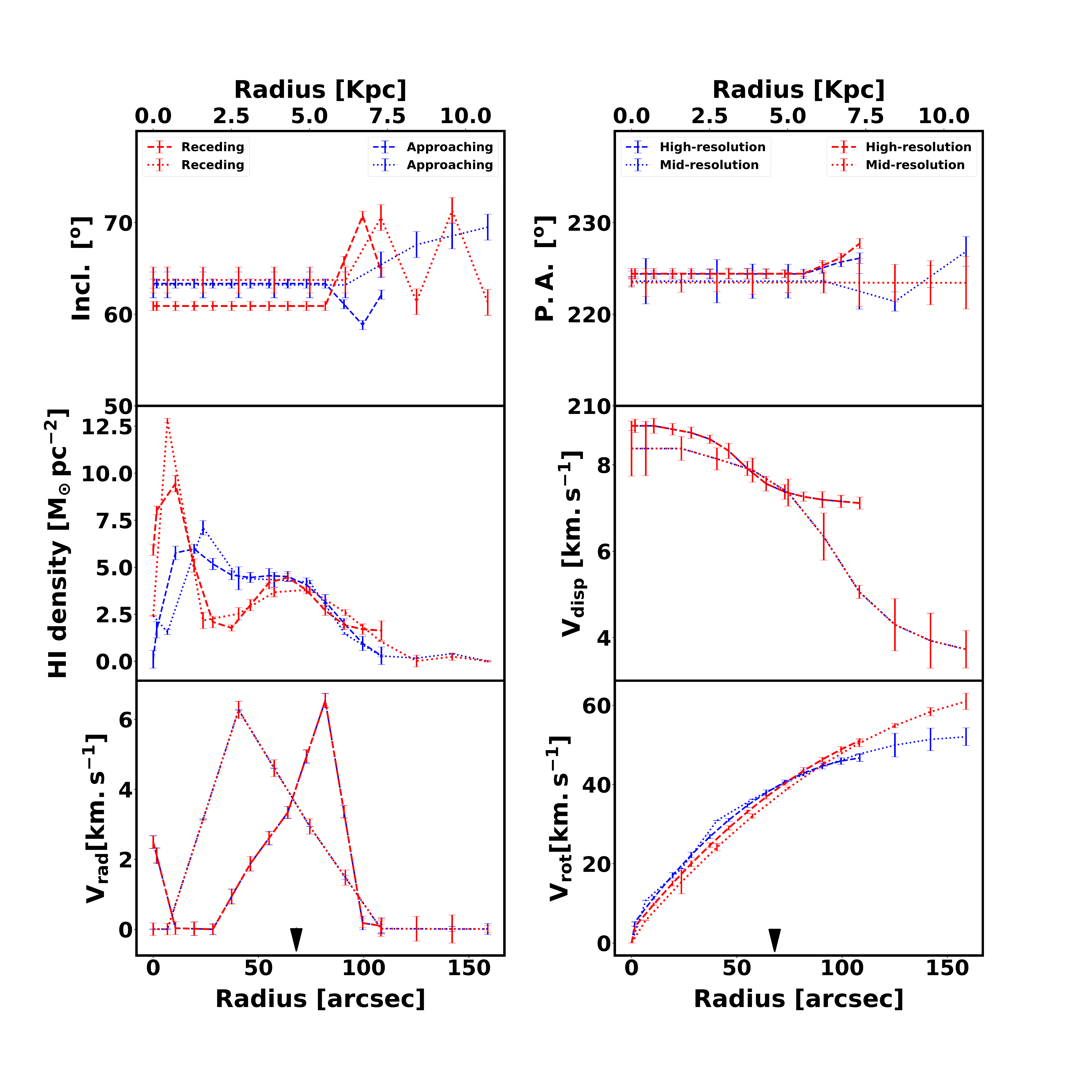}
\caption{Best-fit kinematic parameters of [KKS2000]23 derived from the \HI\ data cube using  \textsc{PyFAT} and \textsc{TiRiFiC}. The panels display inclination angle, position angle, \HI\ surface density, radial velocity, and rotation velocity. The blue and red lines correspond to the approaching and receding sides, respectively. The dashed line represents high resolution, while the dotted line represents mid-resolution. The black arrows represent the 90th percentile stellar radius obtained from the WISE 3.4$\mu$m band surface brightness profile.}
\label{fig:kinematic11}
\end{figure*}

Using the highest-resolution model, we derived a systemic velocity of \( 1038.0 \pm 0.3 \) km s\(^{-1}\), close to the value of \( 1035.9 \) km s\(^{-1}\) reported by \citet{2024A&A...688A.109D}. The dynamical center, determined at \( \alpha,\delta \) (J2000) = \( 11^{\mathrm{h}}06^{\mathrm{m}}11.8^{\mathrm{s}}, -14^{\circ}24^{\mathrm{m}}10.5^{\mathrm{s}} \). We find a slight offset between the optical and kinematic centers of KKS2000-23. As shown in Fig.~\ref{fig:kinematic11}, the position angle and inclination remain relatively stable across the disk, with an average inclination of \( 62.0 \pm 1.0^{\circ} \) and a position angle of \( 224.0 \pm 2.0^{\circ} \).

\begin{figure*}
\centering
   \advance\leftskip0cm
    \includegraphics[height=10cm]{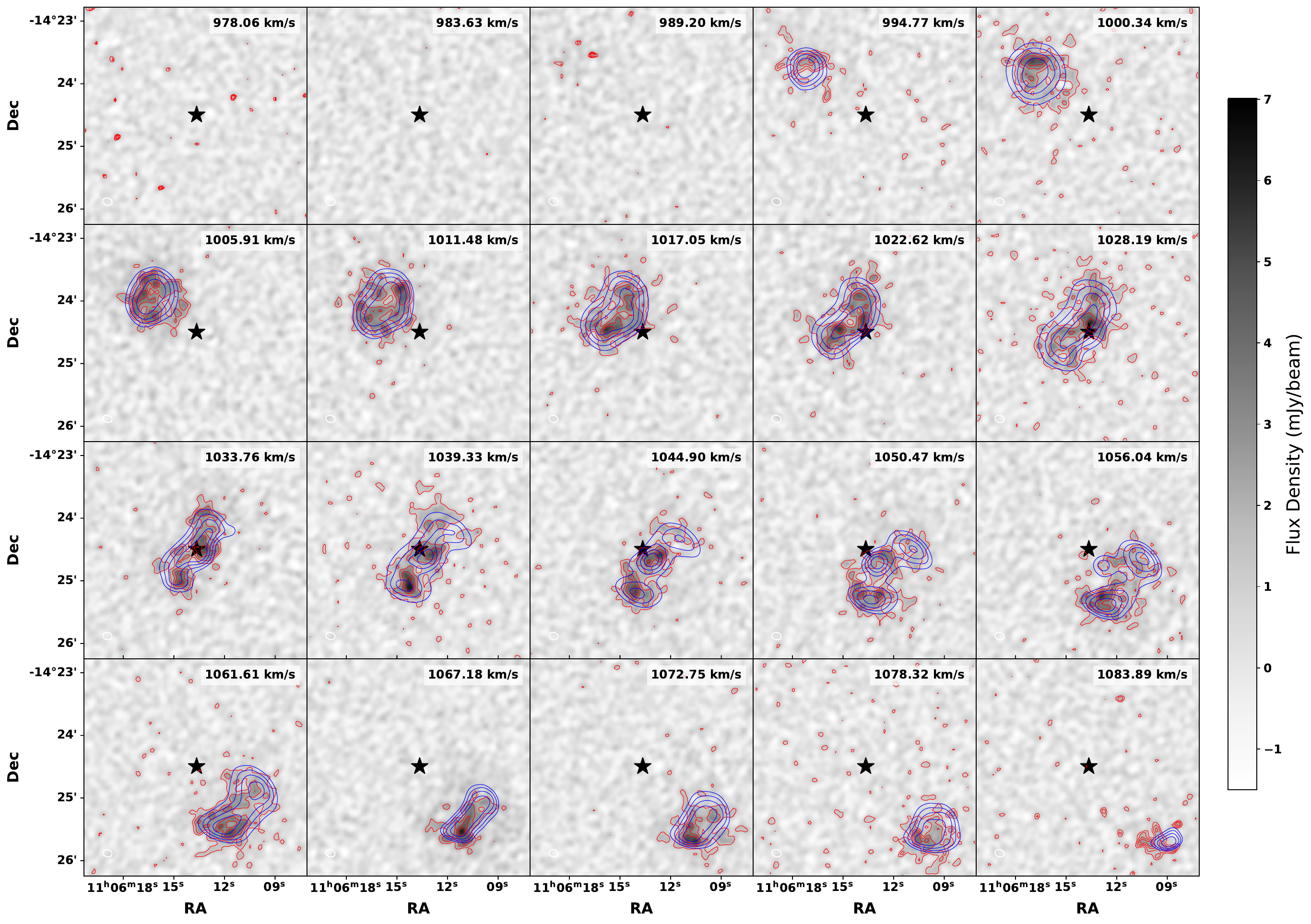}
\caption{Individual channel maps of [KKS2000]23 from the MeerKAT high-resolution cube. Blue contours represent the channel maps from the 3D modeling, while the red contours show the observed data. Contour levels denote values of (3, 6, 9, 12) $\times \sigma$. Black crosses represent the kinematic center of the observed data.}
\label{fig:channel2}
\end{figure*}

\begin{figure*}
\centering
   \advance\leftskip0cm
    \includegraphics[height=6cm]{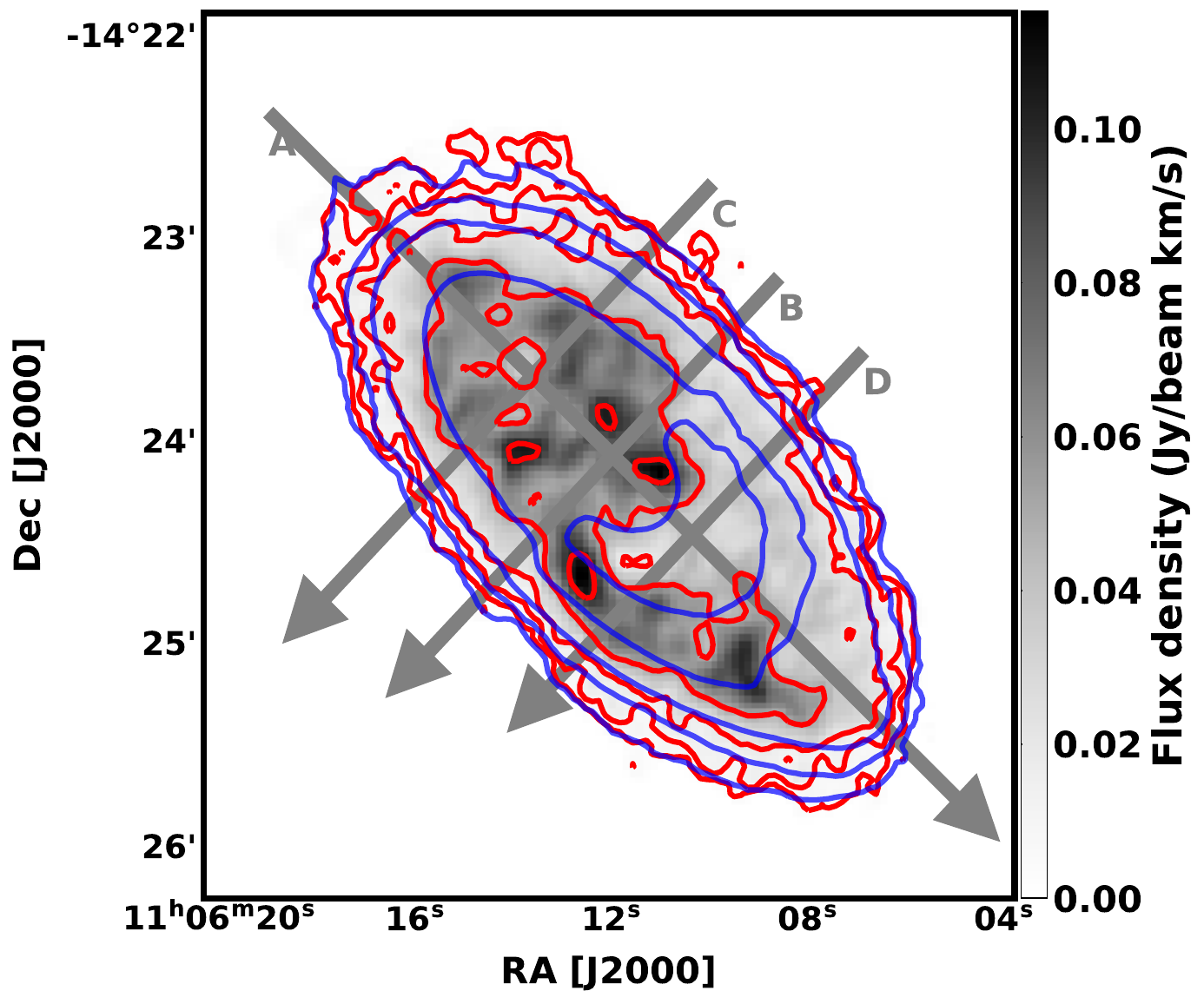}
   \includegraphics[height=13cm]{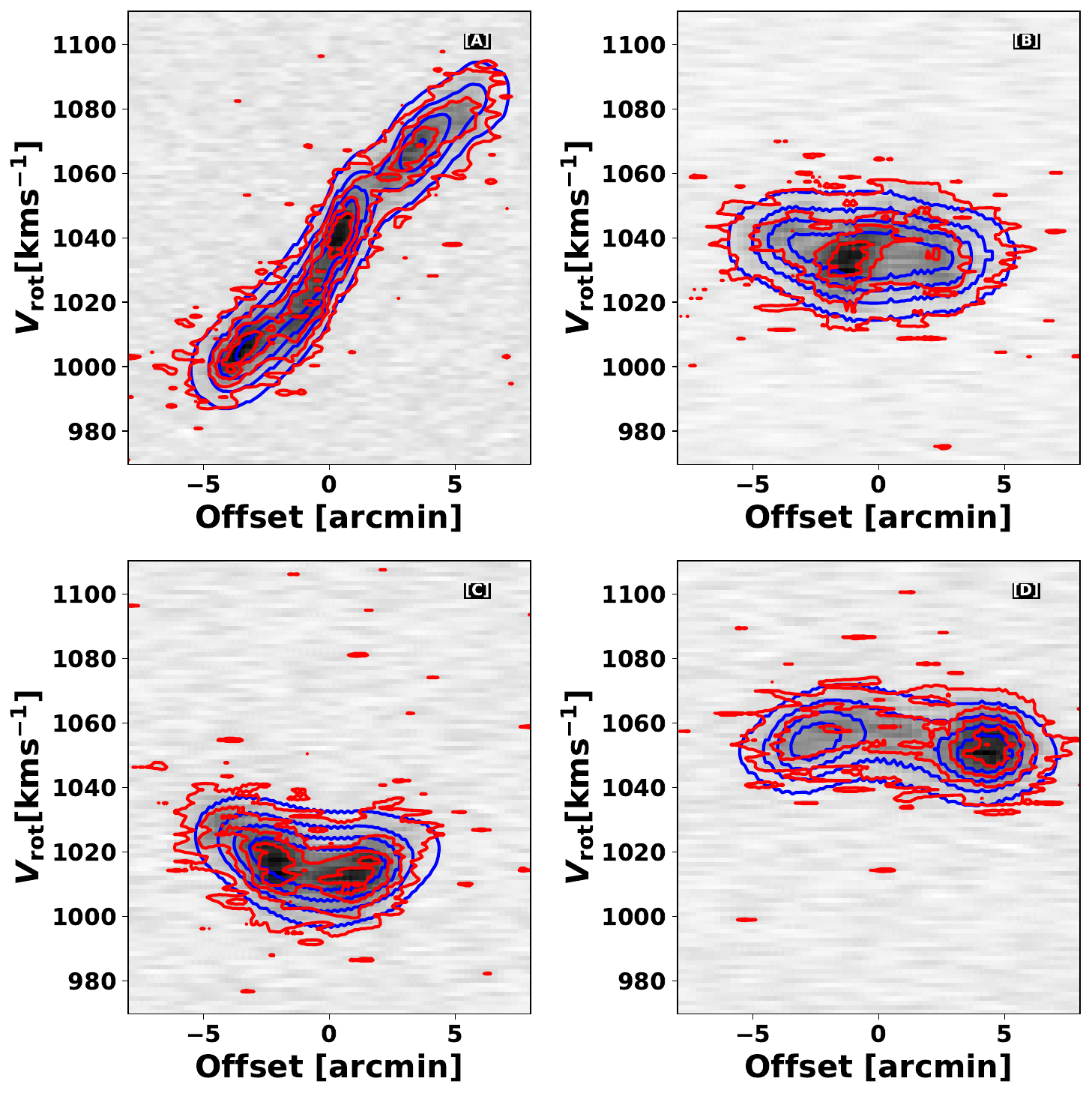}
 \vspace{1pt}
\caption{Top: The moment-0 map of [KKS2000]23 from the MeerKAT high-resolution cube, displaying grey arrows that indicate the positions of slices along which the position-velocity diagrams (bottom) were extracted. The arrows indicate the locations of the PV slices, with one aligned along the kinematic major axis (A) to capture the bulk of rotational motion and another along the minor axis (B) to examine deviations from pure rotation. Additional slices (C and D) were chosen to explore variations in the \HI\ kinematics and overall disk structure. The red contours represent the data, while the blue contours represent the model. The contour levels are set at (3, 6, 9, 15, 18) $\times$ $\sigma$.}
\label{fig:pv11}
\end{figure*}

[KKS2000]23 exhibits a more gradual rise in its rotation curve, continuing to increase in the outer regions, suggesting a more extended and diffuse dark matter distribution. 
The radial velocity residuals for [KKS2000]23 remain below approximately 6 \kms across most of the disk. The channel maps in Fig.~\ref{fig:channel2} and PV diagrams in Fig.~\ref{fig:pv11} show good agreement between the model and observed data. The summary of the kinematic results is shown in Table~\ref{table:kinematics}.

\begin{table*}
    \centering
    \caption{Kinematic properties of ESO444--G084 and [KKS2000]23.}
    \label{table:kinematics}
    \begin{tabular}{lcc}
        \hline\hline
        Parameter & ESO444--G084 & [KKS2000]23 \\
        \hline
        Dynamical Center (J2000) & $13^{\mathrm{h}}37^{\mathrm{m}}19.8^{\mathrm{s}}, -28^{\circ}02^{\mathrm{m}}46.4^{\mathrm{s}}$ & $11^{\mathrm{h}}06^{\mathrm{m}}11.8^{\mathrm{s}}, -14^{\circ}24^{\mathrm{m}}10.5^{\mathrm{s}}$ \\
        Systemic Velocity (km s$^{-1}$) & $586.0 \pm 2.1$ & $1038.0 \pm 0.3$ \\
        Mean Inclination ($^\circ$) & $49.0 \pm 1.2$ & $62.0 \pm 1.0$ \\
        Mean Position Angle ($^\circ$) & $90.0 \pm 2.4$ & $224.0 \pm 2.0$ \\
        Kinematic Warp & Present & None detected \\
        \hline
    \end{tabular}
\end{table*}
\section{Mass modeling}\label{sec:massmodel}
Under the assumption of axisymmetry and dynamical equilibrium, the circular velocity reflects the total gravitational potential of a galaxy, which includes contributions from gas, stars, and dark matter. To determine the dark matter distribution in ESO444--G084 and [KKS2000]23, we decompose the observed circular velocities into separate dynamical components using:
\begin{equation}
V_{\text{c}} = \sqrt{V_{\text{gas}}^{2} + V_{\text{disk}}^{2} + V_{\text{halo}}^{2}},
\end{equation}
where \( V_{\text{c}} \) is the circular velocity, \( V_{\text{gas}} \) represents the velocity contribution from the gas mass distribution, \( V_{\text{disk}} \) corresponds to the stellar disk, and \( V_{\text{halo}} \) represents the velocity associated with the dark matter halo density distribution. The contributions from gas and stars are computed using the methods described below.

\subsection{Stellar component}
We use the Wide-field Infrared Survey Explorer (WISE\footnote{\url{https://wise2.ipac.caltech.edu/docs/release/allsky/}}, \citealt{2010AJ....140.1868W})
 3.4-micron surface brightness profiles to calculate the stellar mass distribution. At this wavelength, the light is less affected by dust, allowing us to probe the older stellar disk population and obtain a more accurate estimate of the stellar mass. These profiles were provided by Prof. Tom Jarrett (private communication). Following the method outlined in \citet{2015AJ....149..180O}, the stellar surface brightness profiles in mag \ arcsec$^{-2}$ were converted to luminosity densities in \( \mathrm{L}_{\odot}\ \text{pc}^{-2} \) and then to mass densities using the equation:
\begin{equation}
\Sigma \, [M_{\odot} \, \text{pc}^{-2}] = (M/L)^{*}_{3.4\,\mu\text{m}} \times 10^{-0.4 (\mu_{3.4\,\mu\text{m}} - C_{3.4\,\mu\text{m}})},
\end{equation}
where \( (M/L)_{3.4\,\mu\text{m}}^{*} \) is the stellar mass-to-light ratio in the 3.4-micron band, \( \mu_{3.4\,\mu\text{m}} \) is the surface brightness in mag\ arcsec$^{-2}$, and \( C_{3.4\,\mu\text{m}} \) is a conversion constant that relates mag\ arcsec$^{-2}$ to \( \mathrm{L}_{\odot}\ \text{pc}^{-2} \). The conversion constant is calculated as:
\begin{equation}
C_{3.4} = M_{\odot,3.4\,\mu\text{m}} + 21.56,
\end{equation}
where \( M_{\odot,3.4\,\mu\text{m}} = 3.24 \) represents the absolute magnitude of the Sun in the 3.4-micron band. The \textsc{GIPSY} task \texttt{ROTMOD} was then used to convert the stellar surface density profile into a rotation curve. \texttt{ROTMOD} assumes that the stellar component forms an infinitely thin, axisymmetric disk and numerically solves Poisson's equation to calculate the gravitational potential generated by the stellar mass distribution. The task then derives the circular velocity at each radius by computing the radial derivative of this potential, thereby translating the observed stellar density into the corresponding circular velocity needed to support the disk. To estimate the mass-to-light ratio, we employed two approaches:
1) WISE-based estimation: the mass-to-light ratio using the WISE surface brightness and the empirical relation from \citet{2014ApJ...782...90C}, given by the following equation.
\begin{equation}
\log (M/L)^{*}_{3.4\,\mu\text{m}} = -1.93 (W_{3.4\,\mu\text{m}} - W_{4.6\,\mu\text{m}}) - 0.04,
\end{equation}
where \( W_{3.4\,\mu\text{m}} - W_{4.6\,\mu\text{m}} = 0.36 \) for [KKS2000]23 and \( 0.32 \) for ESO444--G084. 
2) Adopted literature value: As an alternative approach, we adopted a fixed mass-to-light ratio of \( (M/L)^{*}_{3.6\,\mu\text{m}} = 0.5 \), a commonly used value reported in the literature \citep{2016AJ....152..157L}.

\subsection{Gas component}
The contribution of the gaseous disk to the rotation curve was determined using the \HI\ surface brightness profiles shown in Fig.~\ref{fig:kinematic1} and Fig.~\ref{fig:kinematic11}. The \HI\ surface densities were derived from the 3D kinematic modeling described in Section \ref{sec:kinematics}. To account for the presence of helium, the surface densities were scaled by a factor of 1.4. This correction factor is widely adopted in galaxy mass modeling (e.g., \citealt{2008AJ....136.2782L, 2008AJ....136.2563W,2008AJ....136.2648D,2008MNRAS.386.1667B,2015AJ....149..180O}). The corresponding gas rotation velocities were then computed using the \texttt{ROTMOD} task in \textsc{GIPSY}. Here, we ignore molecular gas and include only \HI\ and helium in the gas component. This approach is justified given that dwarf galaxies such as ESO444–G084 and [KKS2000]23 typically have very low molecular gas content, with CO detections often weak or absent (e.g., \citealt{2012AJ....143..138S}). Several previous mass modeling studies of dwarf and low-mass galaxies also exclude molecular gas for this reason (e.g., \citealt{2008AJ....136.2648D,2015AJ....149..180O}).
\subsection{Dark matter halo}
Over the past three decades, numerous dark matter halo models have been developed, reflecting advances in galaxy dynamics and cosmological simulations. High-resolution simulations incorporating baryonic processes such as supernova feedback, gas outflows, and star formation suggest that these mechanisms can reshape central dark matter distributions, potentially transforming cuspy halos into core-like structures \citep{2005MNRAS.356..107R, 2012MNRAS.421.3464P}.

Alternative dark matter models have also gained attention in recent years. Self-interacting dark matter (SIDM) suggests non-negligible interactions between dark matter particles, naturally leading to core-like density profiles \citep{2000PhRvL..84.3760S, 2013MNRAS.430...81R}. Fuzzy dark matter (FDM), composed of ultra-light axions, predicts wave-like effects that suppress small-scale structure formation and produce cored profiles \citep{2000PhRvL..85.1158H, 2020PrPNP.11303787N}. These models provide testable predictions that extend beyond the standard $\Lambda$CDM framework. 

Given the increasing evidence for core-like dark matter distributions, particularly in low-mass galaxies, the pseudo-isothermal (ISO) model remains a useful benchmark for describing observed rotation curves. The ISO model provides a simple representation of galaxies with constant-density cores \citep{1991MNRAS.249..523B}. This model has been widely used in mass modeling studies due to its ability to match the observed kinematics of dwarf galaxies and low surface brightness (LSB) galaxies, where cuspy profiles often fail \citep{1994ApJ...427L...1F,2011AJ....141..193O}. Moreover, the ISO profile serves as a limiting case when testing alternative halo models, allowing for direct comparisons between core-forming mechanisms driven by baryonic feedback or self-interacting dark matter. In this study, we adopt the pseudo-isothermal model as an observationally-motivated profile to describe the dark matter distribution in ESO444--G084 and [KKS2000]23.

The density profile of the isothermal halo model with a constant-density core is given by
\begin{equation}
\rho_{\text{iso}} (r) = \frac{\rho_{0}}{1 + (r/r_{c})^{2}}
\end{equation}
where \( \rho_{0} \) is the core density and \( r_{c} \) is the core radius of the halo. The corresponding circular velocity is given by:
\begin{equation}
V_{\text{iso}} (r) =  \sqrt{4\pi \rho_{0} r_{c}^{2} \left[1 - \frac{r_{c}}{r} \tan^{-1} \left(\frac{r}{r_{c}}\right)\right]}
\end{equation}
The total dark matter mass enclosed within a given radius $r$ can be computed by integrating the density profile. For the pseudo-isothermal halo, the enclosed mass is given by:
\begin{equation}
M_{\text{halo}}(r) = 4\pi \rho_{0} r_{c}^{3} \left[\frac{r}{r_{c}} - \tan^{-1} \left( \frac{r}{r_{c}} \right)\right]
\end{equation}
The fitted parameters $\rho_0$ and $r_c$ are obtained from the rotation curve decomposition using the best-fit ISO model, and $r$ represents the maximum radius reached by the observed rotation curve.
\begin{figure*}
\centering
   \advance\leftskip0cm
   \includegraphics[height=7cm]{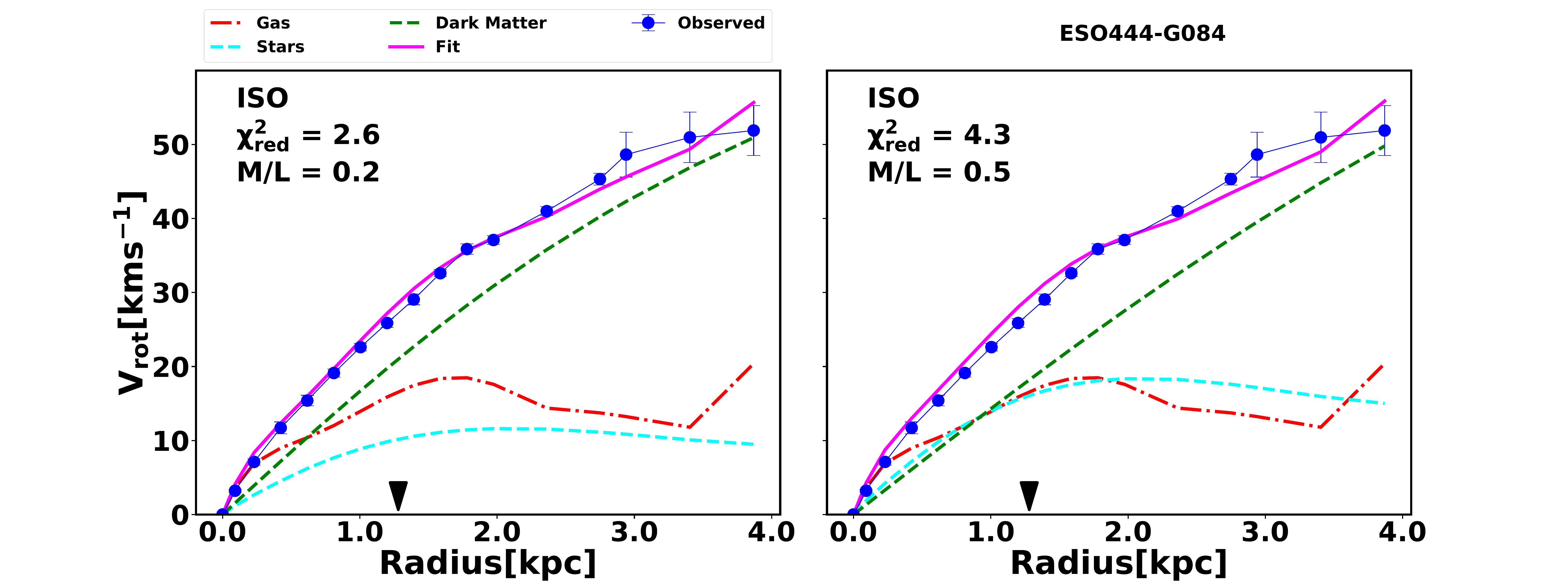}
\caption{Isothermal (ISO) mass modeling results for ESO444--G084. The decomposition for both galaxies was performed under two different assumptions of the mass-to-light ratio, as indicated in each figure. The blue circles represent the derived rotation curve, while the magenta lines show the fitted rotation curve. The green dashed lines indicate the dark matter contribution to the rotation velocities, while the red dot-dashed and cyan dashed lines correspond to the rotational velocities of the gas and stellar components, respectively. The black arrows represent the 90th percentile stellar radius obtained from the WISE 3.4$\mu$m band surface brightness profile.}
\label{fig:massmodel}
\end{figure*}

\begin{figure*}
\centering
   \advance\leftskip0cm
   \includegraphics[height=7cm]{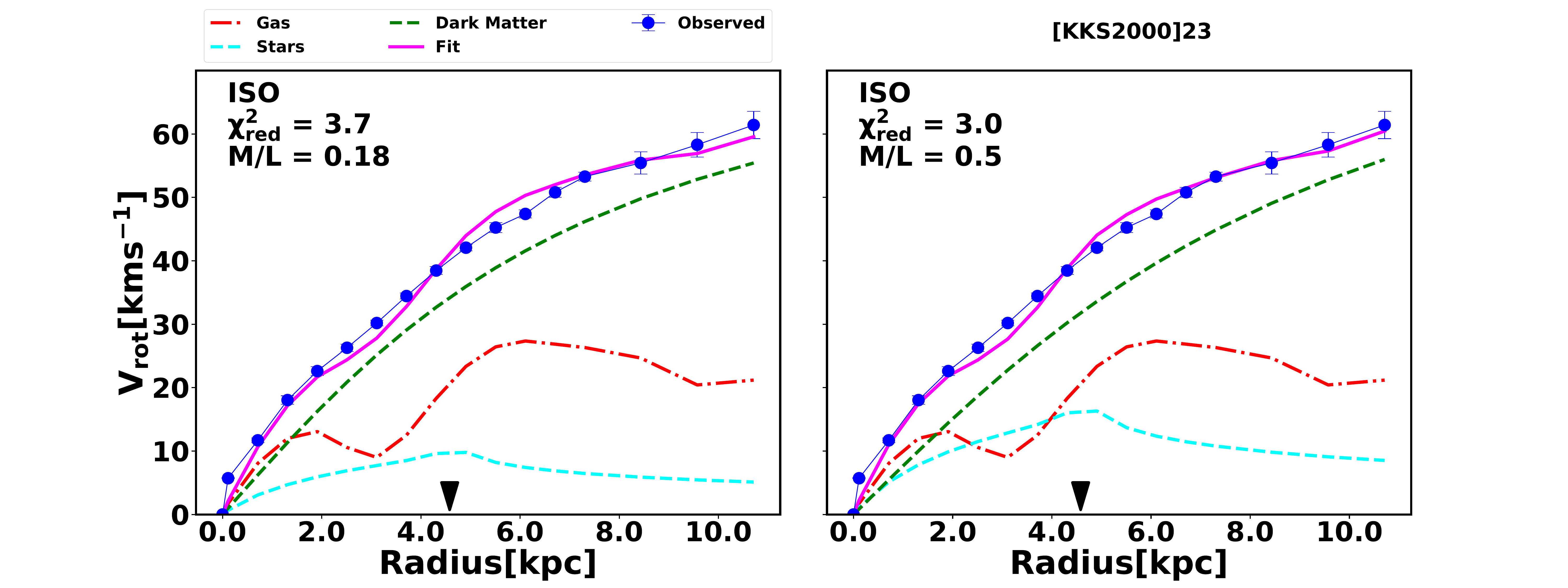}
\caption{Isothermal (ISO) mass modeling results for [KKS2000]23. The decomposition for both galaxies was performed under two different assumptions of the mass-to-light ratio, as indicated in each figure. The blue circles represent the derived rotation curve, while the magenta lines show the fitted rotation curve. The green dashed lines indicate the dark matter contribution to the rotation velocities, while the red dot-dashed and cyan dashed lines correspond to the rotational velocities of the gas and stellar components, respectively. The black arrows represent the 90th percentile stellar radius obtained from the WISE 3.4$\mu$m band surface brightness profile.}
\label{fig:massmodel1}
\end{figure*}
\subsection{Constructing the mass models and results}
The dark matter halo was modeled within \texttt{ROTMAS} using the ISO halo profile. This profile is characterized by two free parameters: the core radius (\(r_C\)) and the central density (\(\rho_0\)). The fitting procedure involved keeping the baryonic contributions fixed according to the assumed stellar mass-to-light ratio $(M/L)^{*}$ while adjusting the dark matter distribution to match the observed rotation curve. Uncertainties in the observed rotation curve were propagated through the fitting process, and the quality of the fits was assessed using the reduced chi-square statistic (\(\chi_{\rm red}^2\)). The results in this section are shown in Figs.~\ref{fig:massmodel} and \ref{fig:massmodel1}.
\subsubsection{ESO444--G084}
The mass modeling results for ESO444--G084 are summarized in Fig.~\ref{fig:massmodel} and Table~\ref{tab:halo_parameters}. The best-fit dark matter halo parameters were derived for two different \( (M/L)^{*} \) assumptions. For \( (M/L)_{3.4\,\mu\text{m}}^{*} = 0.2 \), the halo is characterized by a core radius of \( 3.48 \pm 0.57 \) kpc and a central density of \( (16.05 \pm 1.45) \times 10^{-3} \, \mathrm{M}_{\odot} \, \text{pc}^{-3} \). The enclosed dark matter mass within the last measured point of the rotation curve (\(3.8\) kpc) is \( (2.23 \pm 0.17) \times 10^9 \, M_{\odot} \). Increasing the mass-to-light ratio to \( (M/L)_{3.6\,\mu\text{m}}^{*} = 0.5 \) results in a larger core radius of \( 5.84 \pm 2.64 \) kpc and a lower central density of \( (11.47 \pm 1.46) \times 10^{-3} \, M_{\odot} \, \text{pc}^{-3} \). While both models yield core radii within the expected range of \( \sim 1 \)--\( 6 \) kpc \citep{2001ApJ...552L..23D, 2015AJ....149..180O}, the central densities are higher than the typical \( \sim (2 - 6) \times 10^{-3} \, \mathrm{M}_{\odot} \, \text{pc}^{-3} \) found in dwarf galaxies \citep{2015AJ....149..180O}. It is worth noting that in the \( (M/L)_{3.6\,\mu\text{m}}^{*} \) model the derived core radius is similar to or even larger than the outermost data point, indicating that the available observations do not strongly constrain the inner mass profile. Consequently, the core size estimate should be regarded as a lower limit rather than a definitive measurement. The \( (M/L)_{3.4\,\mu\text{m}}^{*} \) model provides a better fit to the observed rotation curve (\(\chi_{\rm red}^2 = 2.6\)) compared to the \( (M/L)_{3.6\,\mu\text{m}}^{*} \) model (\(\chi_{\rm red}^2 = 4.3\)).

\subsubsection{[KKS2000]23}
The derived dark matter halo parameters for [KKS2000]23 are shown in Fig.~\ref{fig:massmodel1} and Table~\ref{tab:halo_parameters}. For the WISE-derived \( (M/L)_{3.4\,\mu\text{m}}^{*} = 0.18 \), the best-fit model yields a core radius of \( 5.49 \pm 0.60 \) kpc, and a central density of \( (4.29 \pm 0.43) \times 10^{-3} \, M_{\odot} \, \text{pc}^{-3} \). The enclosed dark matter mass within the last measured point of the rotation curve (10.8 kpc) is \( (7.73 \pm 0.74) \times 10^9 \, M_{\odot} \). These values align well with previous studies reporting core radii of \( \sim 1 \)--\( 6 \) kpc and central densities of \( \sim (2 - 6) \times 10^{-3} \, M_{\odot} \, \text{pc}^{-3} \) \citep{2001ApJ...552L..23D, 2015AJ....149..180O}. When adopting the higher \( (M/L)_{3.6\,\mu\text{m}}^{*} = 0.5 \), the model produces a slightly larger core radius \( (7.11 \pm 0.86) \) kpc and a lower central density \( (3.30 \pm 0.30) \times 10^{-3} \, M_{\odot} \, \text{pc}^{-3} \).
\begin{table*}
    \centering
    \caption{Best-fit dark matter halo parameters for ESO444--G084 and [KKS2000]23. The core radius (\(r_c\)) and central density (\(\rho_0\)) are derived for different \((M/L)^*\) assumptions. The enclosed halo mass \(M_{\text{halo}}(<r)\) is computed within the last measured point of the rotation curve.}
    \label{tab:halo_parameters}
    \begin{tabular}{lccccc}
        \hline
        Galaxy & $(M/L)^{*}$ & \(r_c\) (kpc) & \(\rho_0\) (\(10^{-3} M_{\odot} \, \text{pc}^{-3}\)) & \(\chi_{\rm red}^2\) & \(M_{\text{halo}}(<r)\) \((10^9\,M_{\odot})\) \\
        \hline
        \multirow{2}{*}{ESO444--G084} 
        & 0.2  & \(3.48 \pm 0.57\) & \(16.05 \pm 1.45\) & 2.6 & \(2.23 \pm 0.17\) \\
        & 0.5  & \(5.84 \pm 2.64\) & \(11.47 \pm 1.46\) & 4.3 & \(2.12 \pm 0.45\) \\
        \hline
        \multirow{2}{*}{[KKS2000]23} 
        & 0.18 & \(5.49 \pm 0.60\) & \(4.29 \pm 0.43\) & 3.4 & \(7.73 \pm 0.74\) \\
        & 0.5  & \(7.11 \pm 0.86\) & \(3.30 \pm 0.30\) & 3.06 & \(7.91 \pm 1.20\) \\
        \hline
    \end{tabular}
\end{table*}

\section{Gravitational instabilities and star formation}\label{sec:starformation}
Dwarf galaxies, characterized by their low densities, provide an ideal environment for studying both local and global disk instabilities \citep{2008AJ....136.2782L}. The extended \HI\ reservoirs in ESO444--G084 and [KKS2000]23 allow us to examine whether their gaseous disks are stable against gravitational collapse and how the regions of high \HI\ column density relate to active star formation. While other tracers, such as molecular gas and stellar kinematics play a role in assessing the disk stability, \HI\ provides critical insight into the large-scale stability of the gas disk, particularly in environments where molecular gas is not detected. By analyzing the \HI\ distribution and kinematics, we can assess whether these galaxies host large-scale gravitational instabilities and how efficiently their gas-rich regions are forming stars. The stability of a rotating gas disk is defined using the Toomre-\(Q\) parameter, which indicates whether the disk is susceptible to gravitational collapse \citep{1964ApJ...139.1217T}. It is given by:
\begin{equation}
    Q_{\text{gas}} = \frac{\kappa \sigma_{\rm HI}}{\pi G (1.4 \Sigma_{\rm \text{gas}})}
\end{equation}  
where \(\sigma_{\rm HI}\) is the gas velocity dispersion, \(G\) is the gravitational constant, and \(\Sigma_{\rm gas}\) represents the gas surface density. The \(\sigma_{\rm HI}\) and \(\Sigma_{\rm HI}\) are derived using both \textsc{PyFAT} and \textsc{TiRiFiC}. The factor 1.4 accounts for helium, and \(\kappa\), the epicyclic frequency, describes the differential rotation of the disk, defined as:  
\begin{equation}
\kappa = \sqrt{2 \Biggr[\frac{V^{2}}{R^{2}} + \frac{V}{R} \frac{dV}{dR} \Biggr]}
\end{equation}  
When \(Q_{\text{gas}} < 1\), the gas is gravitationally unstable and susceptible to collapse, leading to star formation. However, in galaxies with thicker disks, additional vertical support means that gravitational collapse may require a slightly higher threshold, with instability occurring at approximately \(1.5 Q_{\text{gas}}\) \citep{2013MNRAS.433.1389R}.

The efficiency of star formation is strongly linked to the gas density, with previous studies showing that galaxies with low gas surface densities tend to exhibit reduced star formation rates \citep{1989ApJ...344..685K,2008AJ....136.2782L,2021AJ....161...71H}. One way to assess whether a galaxy's disk is able to sustain star formation is by comparing the gas surface density to the critical threshold, \(\Sigma_{\text{crit}}\), given by:
\begin{equation}
    \Sigma_{\text{crit}} = \frac{1.5 \kappa \sigma_{\rm HI}}{\pi G}
\end{equation}  
where \(\kappa\) is the epicyclic frequency,  \(\sigma_{\rm HI}\) is the velocity dispersion, and \(G\) is the gravitational constant. When the ratio \(\Sigma_{\text{gas}}/\Sigma_{\text{crit}}\) exceeds unity, the gas is expected to be unstable to gravitational collapse, potentially triggering star formation. However, star formation is not only determined by this condition. Additional factors such as the presence of molecular gas, turbulence, and feedback processes influence how efficiently gas converts into stars. To explore the relationship between gas stability and star formation, we examine the spatial distribution of \HI\ column densities alongside star formation tracers such as H\(\alpha\) and far-ultraviolet (FUV) emission. H\(\alpha\) emission traces ionized gas from recent star formation (1–3 Myr), while FUV emission captures longer timescales (up to 100 Myr), providing insight into the sustained star formation activity. 

To derive the stability of the gas disk, we determined the pixel-to-pixel Toomre-\(Q\) parameter and the \(\Sigma_{\text{gas}}/\Sigma_{\text{crit}}\) ratio across both galaxies. These quantities depend on the underlying rotation curve, which we parametrized using a functional form that describes how the rotational velocity varies with radius \citep{2008AJ....136.2782L}. We adopted the rotation curves derived from the best-fit mass model using the pseudo-isothermal (ISO) halo profile. The parameterization of the rotation curves is given by:
\begin{equation}  
V_{\text{rot}} = V_{\text{flat}} \left[1 - \exp \left(-\frac{r}{l_{\text{flat}}}\right)\right]
\label{eq:star}
\end{equation}
 where \( V_{\text{rot}} \) represents the rotation velocity at radius \( r \). \( V_{\text{flat}} \) and \( l_{\text{flat}} \) are free parameters describing the velocity at which the rotation curve flattens, and the scale length at which this velocity is reached. The derived velocity maps were used to calculate the epicyclic frequency \( \kappa \), which, when combined with the gas surface density maps and the mean velocity dispersion derived from Section~\ref{sec:kinematics}, allowed for the construction of spatially resolved \( Q_{\text{gas}} \) and \( \Sigma_{\text{gas}} / \Sigma_{\text{crit}} \) maps.
  \begin{figure*}
\centering
   \advance\leftskip0cm
   \includegraphics[height=6cm]{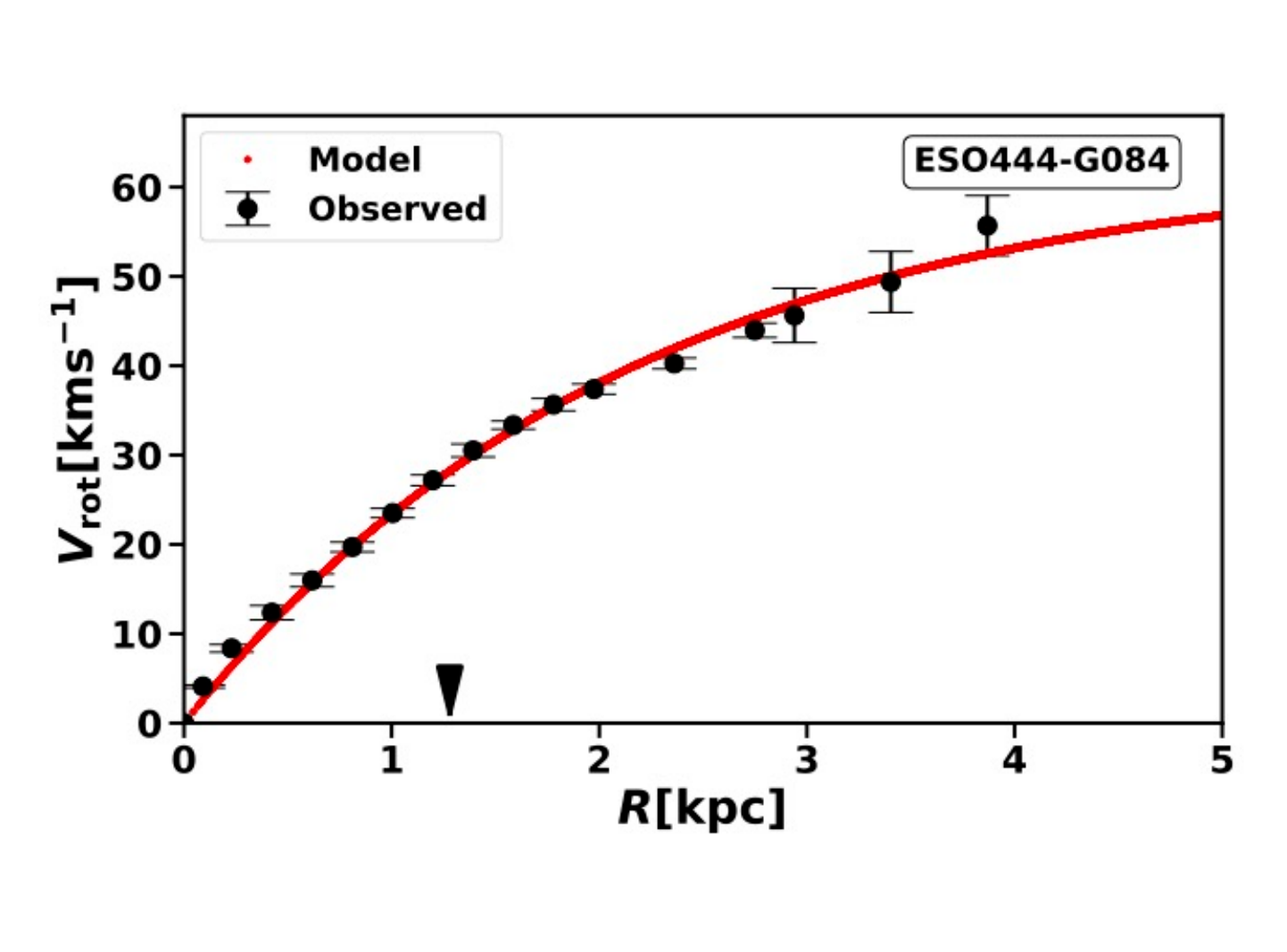}
   \includegraphics[height=6cm]{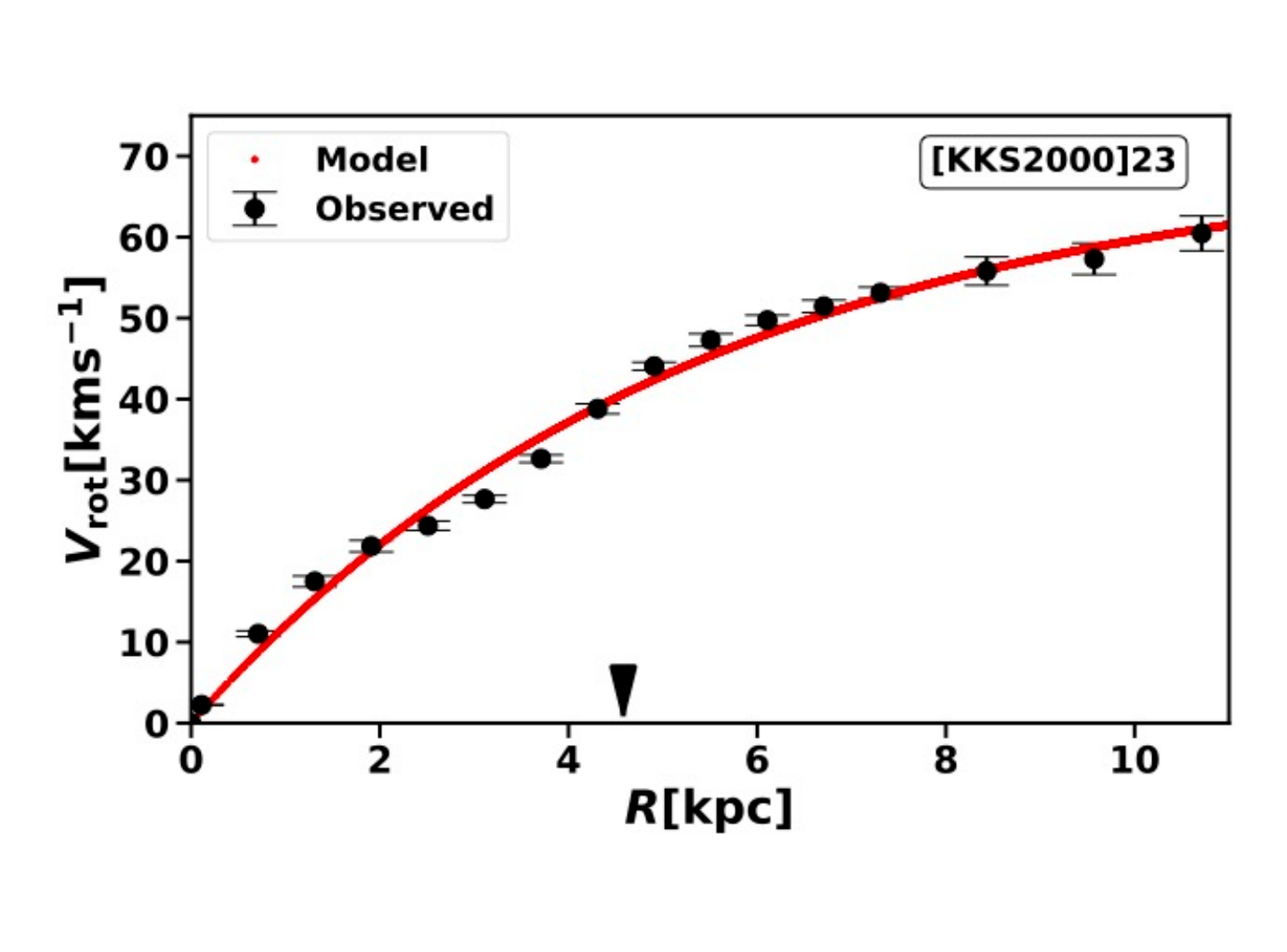}
\caption{The black points represent the best-fit rotation velocities from the ISO mass model, while the red points correspond to the rotation curve fit obtained using Equation~\ref{eq:star}. The black arrows represent the 90th percentile stellar radius obtained from the WISE 3.4$\mu$m band surface brightness profile.}
\label{fig:fit}
\end{figure*}

\begin{figure*}
\centering
   \advance\leftskip0cm
   \includegraphics[height=7cm]{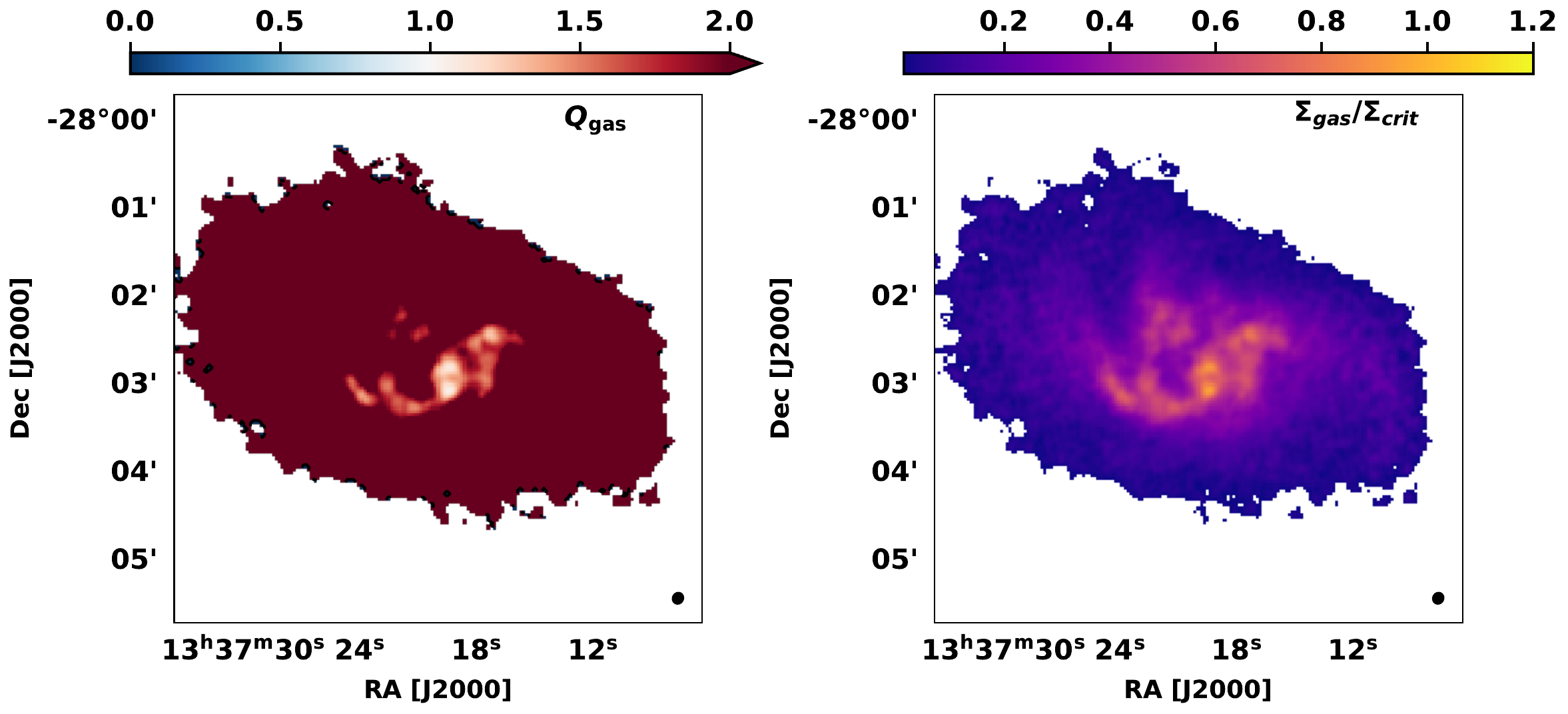} 
   \includegraphics[height=7cm]{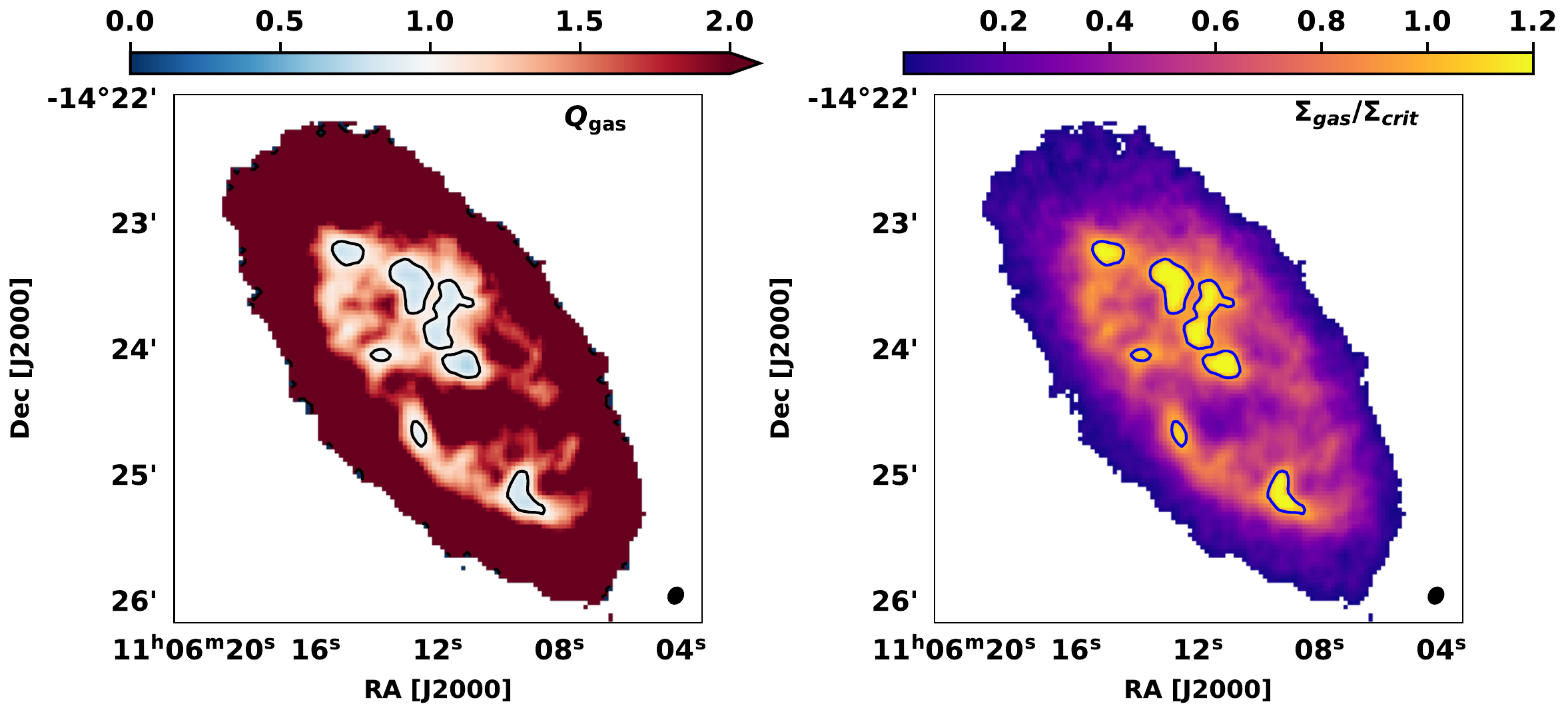}
\caption{MeerKAT high resolution maps of the Q$_{\text{gas}}$ parameter and $\Sigma_{\text{gas}}$/$\Sigma_{ \rm crit}$ for ESO444-G084 (top) and [KKS2000]23 (bottom). The black contours present values of $Q<1$ while the blue contours show  $\Sigma_{ \rm gas}$/$\Sigma_{ \rm crit}$ >1. Note: The color scale for Q is limited to a maximum of Q = 2 for visual clarity; values above this appear saturated in red. The arrow on the colorbar indicates this saturation. The apparent flattening in the outer regions is a visual effect and does not imply that Q is physically constant.}
\label{fig:crit}
\end{figure*}
We also calculated the associated uncertainty maps to evaluate the reliability of these quantities. The uncertainty in $Q_{\mathrm{gas}}$ is propagated as:
\begin{equation}
\delta Q_{\text{gas}} = \left| \frac{\partial Q_{\text{gas}}}{\partial \kappa} \delta \kappa + \frac{\partial Q_{\text{gas}}}{\partial \Sigma_{\mathrm{gas}}} \delta \Sigma_{\mathrm{gas}} + \frac{\partial Q}{\partial \sigma_{\mathrm{HI}}} \delta \sigma_{\mathrm{HI}} \right|
\end{equation}
Similarly, the uncertainty in the stability ratio is calculated by:
\begin{equation}
\delta \left( \frac{\Sigma_{\mathrm{gas}}}{\Sigma_{\mathrm{crit}}} \right) = \left| \frac{1}{\Sigma_{\mathrm{crit}}} \, \delta \Sigma_{\mathrm{gas}} + \frac{\Sigma_{\mathrm{gas}}}{\Sigma_{\mathrm{crit}}^2} \, \delta \Sigma_{\mathrm{crit}} \right|
\end{equation}
with:
\begin{equation}
\delta \Sigma_{\mathrm{crit}} = \left| \frac{\partial \Sigma_{\mathrm{crit}}}{\partial \kappa} \, \delta \kappa + \frac{\partial \Sigma_{\mathrm{crit}}}{\partial \sigma_{\mathrm{HI}}} \, \delta \sigma_{\mathrm{HI}} \right|
\end{equation}
These uncertainties incorporate contributions from the rotation curve, the velocity dispersion $\sigma_{\mathrm{HI}}$, and the gas surface density $\Sigma_{\mathrm{gas}}$.
\begin{figure*}
\centering
   \advance\leftskip0cm
   \includegraphics[height=7cm]{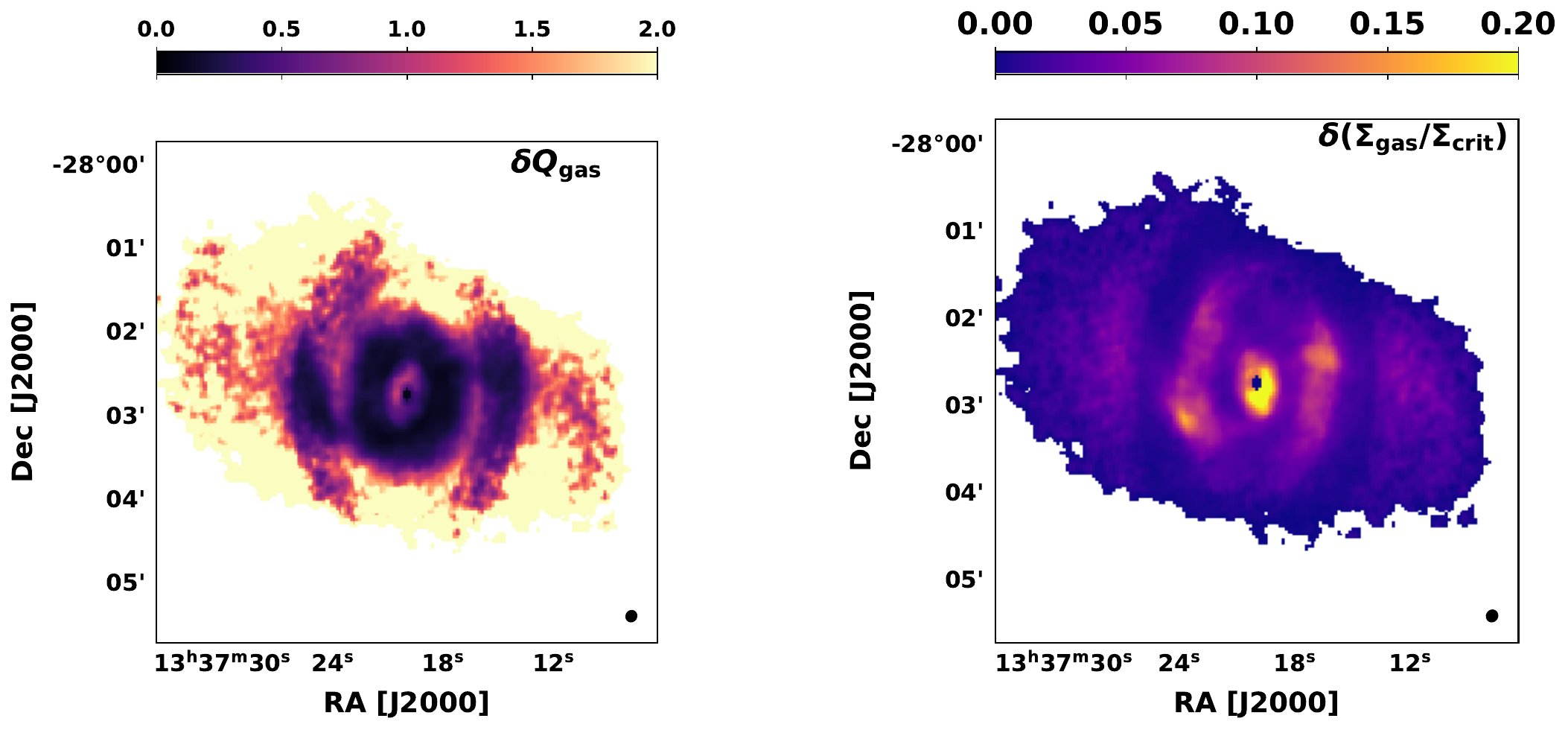} 
   \includegraphics[height=7cm]{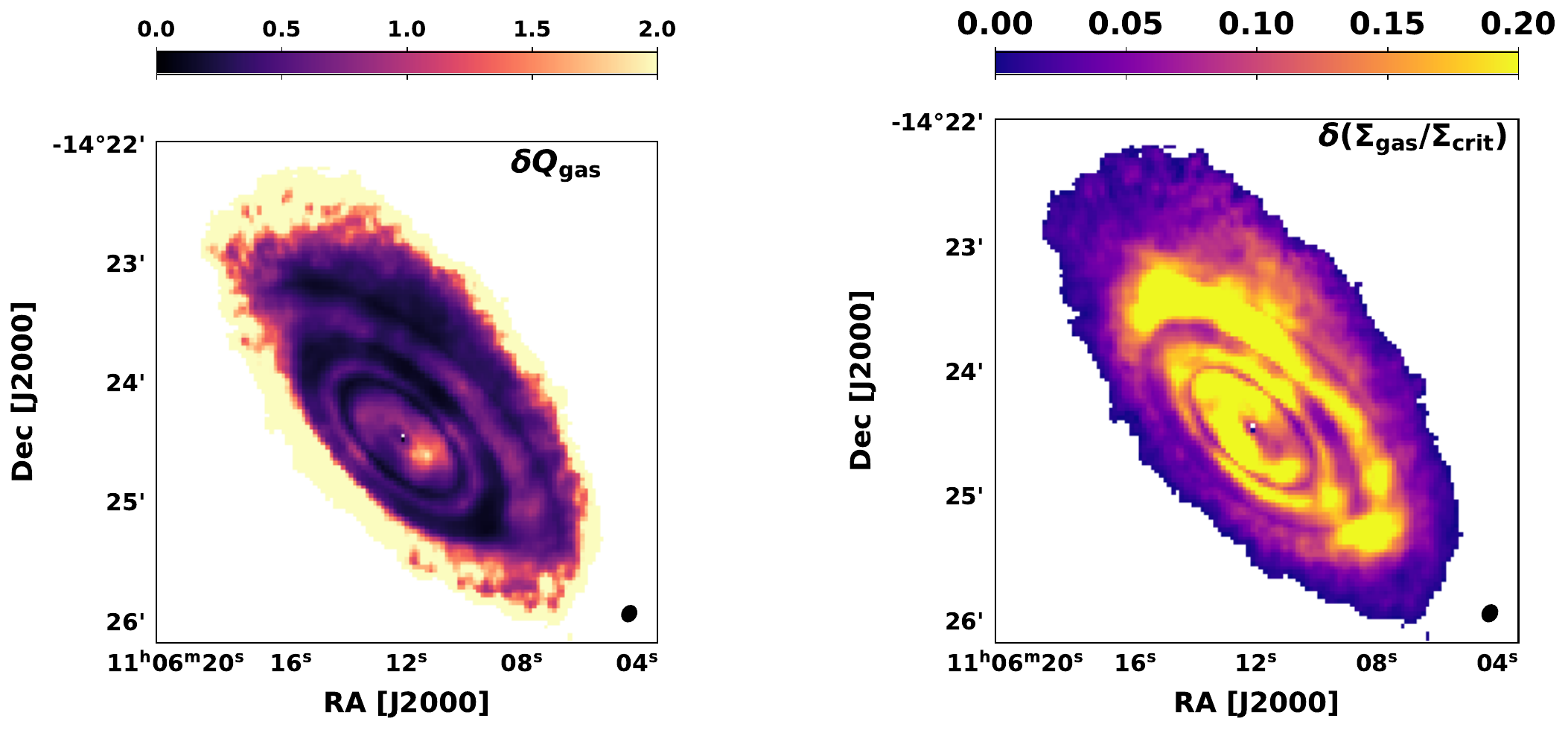}
\caption{Spatial distribution of uncertainties associated with the Toomre $Q_{\mathrm{gas}}$ parameter (left panels) and the $\Sigma_{\mathrm{gas}} / \Sigma_{\mathrm{crit}}$ ratio (right panels) for ESO444--G084 (top) and [KKS2000]23 (bottom).The beam size is shown in the bottom-right of each panel.}
\label{fig:qer}
\end{figure*}
\subsection{Results}
The results in this section are highlighted in Fig.~\ref{fig:crit}, Fig.~\ref{fig:qer}, and Fig.~\ref{fig:cont}.
\subsubsection{ESO444--G084}
Fig.~\ref{fig:crit} shows that the \HI\ disk of ESO444--G084 is stable against global gravitational instability, suggesting that large-scale gravitational processes are not the primary drivers of disk dynamics. The ratio \( \Sigma_{\text{gas}} / \Sigma_{\text{crit}} \) approaches unity in some central regions but remains below 1 across most of the disk. Fig.~\ref{fig:qer} presents the associated uncertainty maps for both $Q_{\mathrm{gas}}$ and $\Sigma_{\mathrm{gas}}/\Sigma_{\mathrm{crit}}$. Despite this stability, star formation is observed near the center of ESO444--G084, as traced by H\( \alpha \) and FUV emission (Fig.~\ref{fig:cont}). These tracers indicate ongoing and recent star formation and coincide with regions of high \HI\ column density. This suggests that localized gas compression, rather than large-scale gravitational instabilities, is responsible for triggering star formation in this galaxy, indicating that while localized conditions may support star formation, large-scale gravitational effects are not the dominant mechanism governing disk dynamics. According to \citet{2012MNRAS.419.1051L}, the SFR derived from FUV emission is \( 0.0093 \pm 0.0011 \) \(M_{\odot}\) yr\(^{-1}\), while the H\(\alpha\)-based SFR is lower at \( 0.0022 \pm 0.0002 \) \(M_{\odot}\) yr\(^{-1}\). This discrepancy suggests that recent star formation has declined relative to longer-term activity, consistent with a scenario where localized gas compression, rather than sustained large-scale gravitational instabilities, is responsible for triggering star formation in ESO444-G084.
\subsubsection{[KKS2000]23}
The analysis reveals that the \HI\ disk of [KKS2000]23 shows clear signatures of gravitational instability across several regions. The spatial correlation between low values of \( Q_{\text{gas}} < 1 \) and high \(\Sigma_{\text{gas}} / \Sigma_{\text{crit}} > 1\) in Fig.~\ref{fig:crit} suggests that these regions are gravitationally unstable and likely to undergo star formation. Fig.~\ref{fig:qer} presents the associated uncertainty maps for both $Q_{\mathrm{gas}}$ and $\Sigma_{\mathrm{gas}}/\Sigma_{\mathrm{crit}}$.

\begin{figure*}
\centering
   \advance\leftskip0cm
   \includegraphics[width=8.5cm]{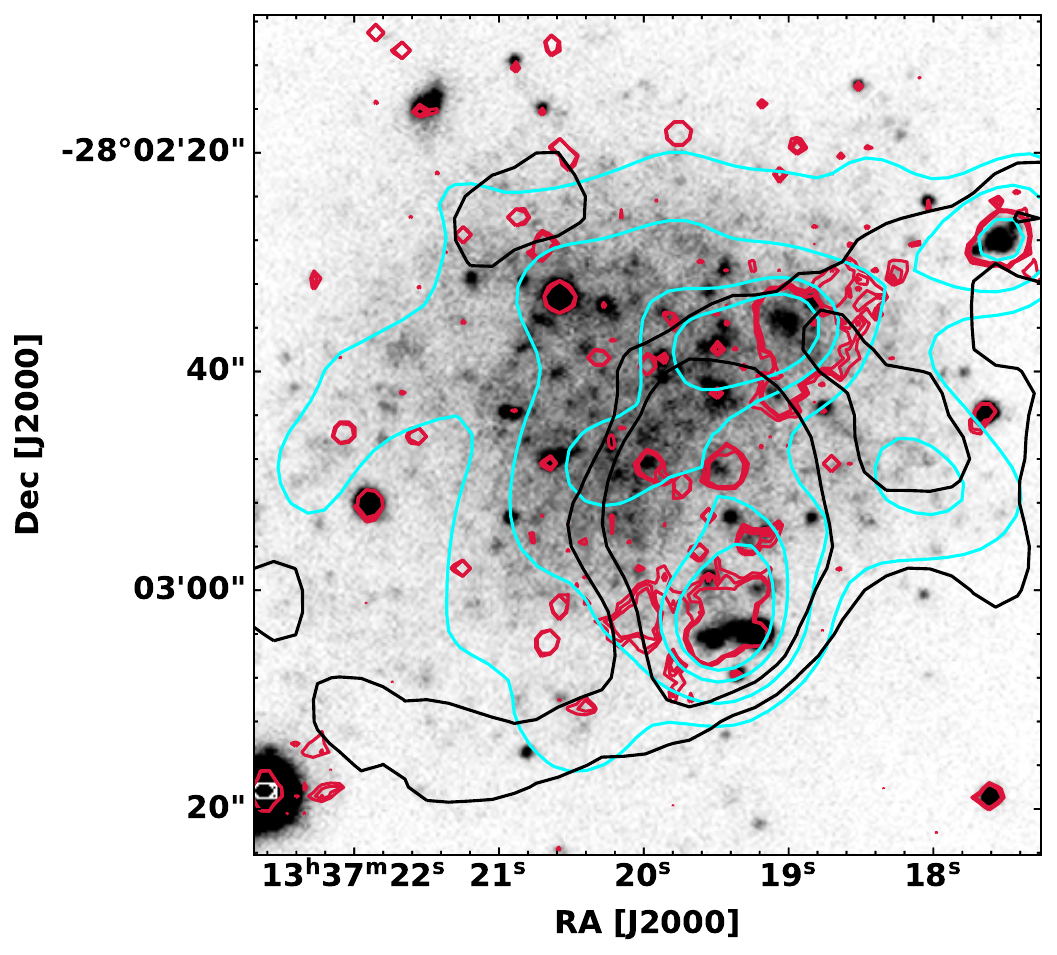}
     \includegraphics[width=8.5cm]{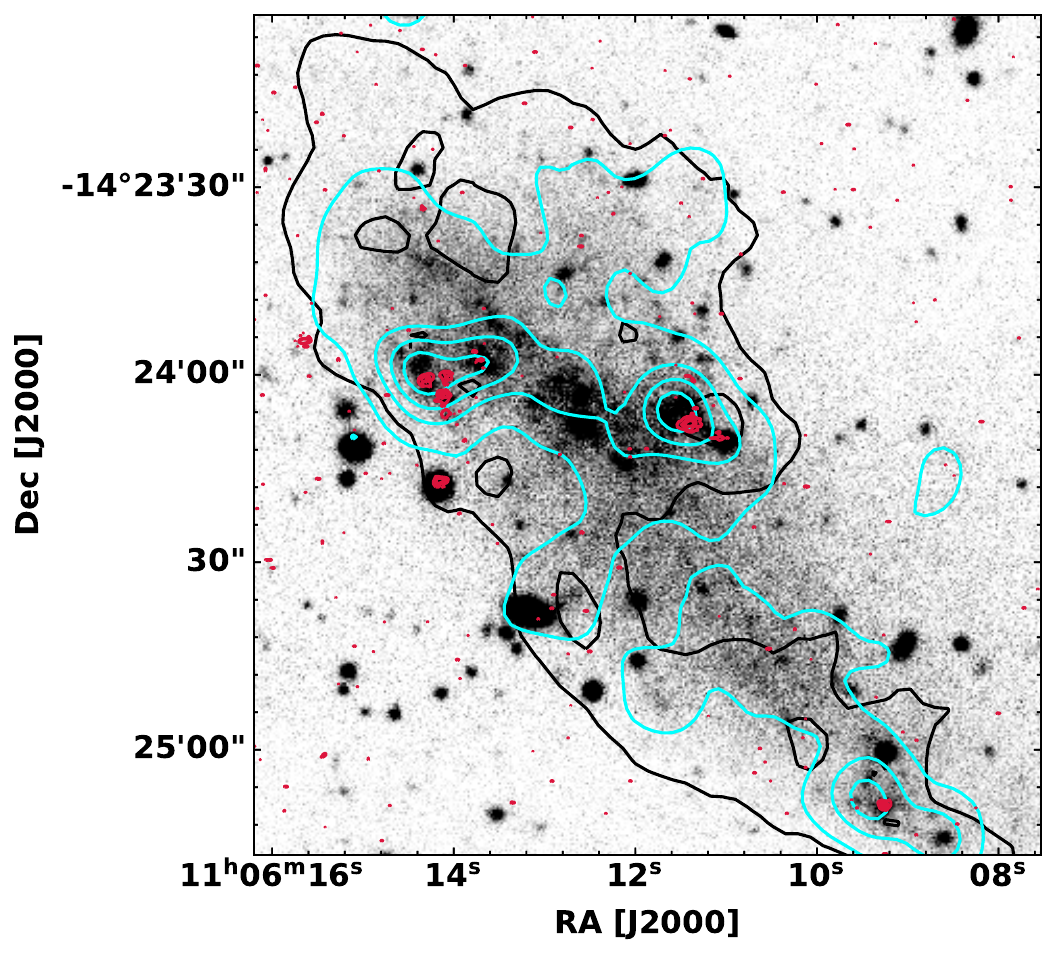}
\caption{High-resolution MeerKAT \HI\ column density maps (black contours), GALEX FUV (cyan contours), and H$\alpha$ (crimson contours) of ESO444--G084 (left) and [KKS2000]23 (right) overlaid on DECaLS optical maps. The \HI\ column density contours are set at (2.0, 2.5, 4.0) $\times 10^{21}$ cm$^{-2}$ for ESO444--G084 and (1.0, 2.0, 2.5) $\times 10^{21}$ cm$^{-2}$ for [KKS2000]23. H$\alpha$ contours are drawn at intensity levels of (70, 90, 110) for ESO444--G084 and (0.02, 0.04, 0.06) for [KKS2000]23. The GALEX FUV contours are plotted at (0.003, 0.005, 0.008, 0.01) for ESO444--G084 and (0.02, 0.04, 0.06) for [KKS2000]23. H$\alpha$ and FUV contours are arbitrary units.}
\label{fig:cont}
\end{figure*}

Fig.~\ref{fig:cont} shows a strong spatial correlation between high \HI\ column densities and regions of recent star formation, as traced by FUV and H$\alpha$ emission.  Although H$\alpha$ emission is weaker in this system, indicative of fewer ionized gas regions associated with ongoing star formation, the FUV emission aligns well with regions identified as gravitationally unstable. This suggests that ongoing star formation in [KKS2000]23 is more extended and likely sustained over longer timescales, as FUV traces star formation over the past ~100 Myr. A notable feature is the presence of star formation in both the central disk and the outskirts. Its occurrence in the outer regions, despite lower average \HI\ gas densities, suggests that localized enhancements in gas surface density can trigger gravitational instabilities, leading to star formation.

\section{Discussion} \label{sec:discussion}
\subsection{\HI\  Morphology and Distribution}
The MHONGOOSE survey provides an unprecedented view of dwarf galaxies, probing low column density \HI\ down to \( < 10^{19} \) cm\(^{-2}\). Both ESO444--G084 and [KKS2000]23 exhibit extended \HI\ distributions that stretch well beyond their optical radii.  

ESO444--G084 exhibits a symmetric \HI\ distribution in its global profile with the highest \HI\ emission at its center, indicating that the system is dynamically settled,  whereas [KKS2000]23 shows significant asymmetries and clumpiness, with high-density gas appearing in irregular locations. The absence of nearby companions suggests that these irregularities originate from internal processes rather than external interactions, although the possibility of past tidal interactions or minor mergers contributing to its kinematic distortions cannot be ruled out \citep{2018MNRAS.478.1611K}. These results align with the idea where most isolated dwarfs exhibit well-ordered HI distribution unless significantly shaped by internal feedback \citep{2020A&A...643A.147D}, which show that most isolated dwarfs exhibit well-ordered \HI\ distributions unless significantly shaped by internal feedback. Similar trends have been observed in previous studies \citep{2008MNRAS.386.1667B, 2011AJ....142..121H}, where \HI\ asymmetries in dwarf galaxies have been linked to internal turbulence, localized instabilities, and feedback from star formation.  

A comparison with single-dish flux measurements is crucial for determining whether interferometric observations recover the total \HI\ content. As highlighted in \citet{2024A&A...688A.109D}, the MeerKAT fluxes for ESO444--G084 and [KKS2000]23 are in good agreement with the GBT and HIPASS single-dish measurements, within the expected calibration uncertainties \citep{Koribalski2004, Barnes2001}. The overall agreement between MeerKAT and single-dish measurements supports the reliability of MeerKAT interferometric observations in recovering total \HI\ flux for these two systems which are small in angular size.

Interestingly, the \HI\ fluxes measured at column densities of \( 10^{19} \) cm\(^{-2}\) remain consistent with those at \( 10^{18} \) cm\(^{-2}\), indicating that the observed increase in the \HI\ diameter at lower column densities is primarily due to the increase in beam size rather than the detection of significant additional gas. The contrasting \HI\ distributions of ESO444--G084 and [KKS2000]23 further emphasize the diversity of gas-rich dwarf galaxies.

\subsection{\HI\ Kinematics}
The kinematics of ESO444--G084 and [KKS2000]23 shed light on the dynamical evolution of dwarf galaxies and the interplay between internal and external processes. The kinematic warp in ESO444--G084 indicates ongoing dynamical activity, which may involve internal gas flows or past accretion events. Warps and asymmetries in \HI\ disks are common in low-mass galaxies and often arise from interactions, internal feedback, or misaligned gas accretion \citep{Swaters2009}.  

[KKS2000]23 exhibits regular rotation with minimal perturbations, pointing to a settled kinematic structure. This suggests a history of steady gas accretion and a sufficiently deep potential well that helps mitigate major dynamical disturbances \citep{ 2017ARA&A..55..343B}. Observational studies find that galaxies with deeper potential wells tend to show smoother rotation curves and fewer kinematic irregularities \citep{Keres2005}. 

ESO444--G084’s rotation curve exhibits a relatively fast rise in the inner region, indicating a centrally concentrated mass distribution and suggesting a dynamically evolved system with a significant mass concentration. However, the rise is not as steep as expected for an NFW-like halo. [KKS2000]23’s more gradual rise in its rotation curve indicates a more diffuse mass profile, which may reflect differences in its evolutionary history.  

Velocity fields confirm disk-like rotation in both galaxies but highlight notable variations. ESO444--G084 exhibits a position angle shift beyond \(\sim1.8\) kpc, consistent with a kinematic warp. Subtle deviations in its isovelocity contours suggest weak non-circular motions or small-scale perturbations \citep{2006MNRAS.365..555S, 2021A&A...654A...7K}. Meanwhile, [KKS2000]23’s isovelocity contours remain largely undisturbed, reinforcing its rotation-dominated nature.  

Radial velocity profiles further illustrate these differences. ESO444--G084’s moderate radial motions (\(\sim15\)-\(20\) km s\(^{-1}\)) indicate weak internal perturbations, while [KKS2000]23’s weaker radial inflows (\(\sim6\) km s\(^{-1}\)) align with a more quiescent gas distribution \citep{2008AJ....136.2563W, 2012AJ....144..134H}.  

The absence of a strong bar in ESO444--G084 suggests that bar-driven streaming motions are not a dominant feature of its kinematics. However, mild perturbations in its isovelocity contours could be linked to weak oval distortions or residual radial flows. Some galaxies exhibit bar-like distortions in gas kinematics without a prominent stellar bar, highlighting the need for further investigation \citep{1984MNRAS.210..547P, 1999ApJ...522..699A, 2017A&A...606A..47F,2018MNRAS.476.2168M,2023MNRAS.522.3318D}.   
\subsection{Dark matter distribution}  
The mass modeling results provide valuable insights into the cusp-core problem in dwarf galaxies, a long-standing issue in galaxy formation theory \citep{2010AdAst2010E...5D, 2015MNRAS.452.3650O, 2017ARA&A..55..343B,2022NatAs...6..897S}. ESO444--G084’s high central density and relatively compact halo structure suggest a centrally concentrated dark matter distribution. However, its rotation curve does not rise steeply enough to be fully consistent with a cuspy profile, instead aligning more closely with a cored distribution. This suggests that, while some central mass concentration is present, feedback processes or other mechanisms may have partially redistributed the inner dark matter. The moderate core size indicates that if supernova-driven outflows have influenced the central mass distribution, their effect has been limited. Our analysis does not show strong evidence for extensive core expansion, though the precise impact of feedback remains uncertain \citep{1996MNRAS.283L..72N,2005MNRAS.356..107R, 2016MNRAS.456.3542T}. Some studies suggest that dwarf galaxies with centrally concentrated but non-cuspy profiles may have undergone moderate levels of gas outflows, which could contribute to the formation of a core rather than a steep cusp \citep{2002MNRAS.333..299G}. [KKS2000]23's lower central density and larger core radius suggest a cored dark matter profile, consistent with models in which repeated bursts of star formation and feedback driven gas outflows displace dark matter from the central regions \citep{2012MNRAS.422.1231G,2012MNRAS.421.3464P}. Although our analysis based on the ISO model offers only preliminary support for this interpretation, future work testing models such as DC14 \citep{2014MNRAS.441.2986D} or coreNFW \citep{2016MNRAS.459.2573R} will provide a more robust assessment.

While ESO444--G084 exhibits a more centrally concentrated dark matter halo, [KKS2000]23's shallower density profile suggests that its dark matter distribution may have been altered over time, potentially through cumulative effects of stellar feedback \citep{2019MNRAS.484.1401R,2020MNRAS.497.2393L}. The observed differences between these two galaxies further contribute to the ongoing discussion on whether CDM adequately explains dwarf galaxy mass profiles or if an alternative model, such as self-interacting dark matter (SIDM), could provide additional insight into the observed variations \citep{1992ApJ...398...43C,2018PhR...730....1T}.  

Another key factor in mass modeling is the assumed stellar mass-to-light ratio \((M/L)^{*}\). Lower \((M/L)^{*}\) values yield more physically motivated halo parameters, particularly for ESO444--G084, where best-fit models suggest a higher central density than expected for typical dwarfs \citep{2013A&A...557A.131M, 2020A&A...634A.135G}. Meanwhile, for [KKS2000]23, the results align with expectations from dwarf galaxy studies, reinforcing the idea of a more diffuse dark matter halo shaped by feedback processes \citep{2019MNRAS.482.5606D}.  

Although the maximum rotation velocities of ESO444–G084 and [KKS2000]23 are similar, the marked differences in their core radii and inner mass distributions emphasize the diverse impact of baryonic processes on dwarf galaxy halos. Studies such as \citet{2015MNRAS.452.3650O} and \citet{2020MNRAS.495...58S} further suggest that variations in baryonic surface density, star formation efficiency, and gas outflow dynamics are key to understanding this diversity.

\subsection{Gravitational instabilities and star formation}
The contrasting properties of ESO444--G084 and [KKS2000]23 illustrate that gravitational stability alone does not fully determine the star formation history of a galaxy. Whilst gravitational instability is a key mechanism for triggering star formation, additional factors such as turbulence, feedback, and gas depletion must be considered \citep{2012MNRAS.421.3488H}. ESO444--G084 appears globally stable, yet localized perturbations continue to sustain residual star formation, indicating that gravitational stability does not prevent all star formation, but regulates its extent and efficiency. [KKS2000]23 remains gravitationally unstable, yet ongoing star formation is weak, suggesting that turbulent gas motions or past feedback processes may be suppressing star formation despite instability. These findings reinforce well-established results that dwarf galaxy evolution is shaped by episodic bursts of star formation and quiescence, regulated by gravitational collapse, gas depletion, and feedback-driven turbulence \citep{2002ApJ...577..206E, 2012MNRAS.421.3488H, 2023A&A...677A..44R}.

The presence of H$\alpha$ and FUV emission in ESO444--G084 suggests sustained star formation over the past $\sim$100 Myr. The alignment of these tracers with high \HI\ column density supports the idea that localized gas compression continues to regulate star formation. Evidently, large-scale gravitational collapse is not fueling widespread star formation, however, minor dynamical perturbations still allow localized cloud collapse \citep{2008AJ....136.2782L, 2012ARA&A..50..531K}. Meanwhile, [KKS2000]23 exhibits gravitational instability but lacks strong H$\alpha$ emission, with only FUV predominantly detected across the disk. This suggests that recent star formation has significantly declined, possibly due to past gas depletion or turbulence disrupting dense cloud formation, aligning with post-burst dwarf galaxies where a previous episode consumed or dispersed available cold gas, leaving behind an unstable but currently quiescent disk \citep{1999ApJ...513..142M, 2012ARA&A..50..531K}.  

These findings highlight the importance of interpreting star formation across different timescales. H$\alpha$ emission traces massive stars formed in the past 1–3 Myr, while FUV emission tracks star formation up to 100 Myr. The detection of FUV with little H$\alpha$ in [KKS2000]23 suggests that recent star formation has significantly declined, leaving a residual UV signature \citep{2002ApJ...577..206E, 2012MNRAS.421.3488H}. In contrast, the continued presence of both tracers in ESO444--G084 suggests ongoing star formation. These variations indicate that ESO444--G084 maintains ongoing star formation, while [KKS2000]23 is experiencing a phase of declining star formation, where previous activity left the disk unstable but with little recent star formation. This aligns with studies emphasizing the cyclical nature of star formation in dwarfs, where gravitational instabilities trigger star formation, followed by gas depletion and stabilization \citep{2008AJ....136.2846B}.  

\section{Conclusions} \label{sec:conclude}

In this study, we utilized high-resolution MeerKAT \HI\ observations to investigate the \HI\ distribution, kinematics, mass modeling, and disk stability of the dwarf galaxies ESO444--G084 and [KKS2000]23. Our analysis offers new insights into the interplay between dark matter, baryonic processes, and star formation in these systems. Below, we summarize our key findings:

\begin{itemize}
    \item[] {\HI\ morphology and distribution:}
    \begin{itemize}
        \item ESO444--G084 exhibits a smooth and well-ordered \HI\ distribution, with high-column-density gas concentrated centrally and distinct spiral arm structures not visible in optical and UV images.
        \item {[KKS2000]23} displays a clumpier and more irregular \HI\ distribution, with localized high-column-density regions scattered throughout the disk, indicating a fragmented gas structure.
        \item The \HI\ fluxes measured at column densities of \(10^{19}\) cm\(^{-2}\) are consistent with those at \(10^{18}\) cm\(^{-2}\), suggesting that the observed increase in \HI\ diameter is primarily due to beam size rather than additional gas detection.
        \item The total \HI\ masses are (1.1 $\pm$ 0.1) $\times$ 10$^{8}$ M$_{\odot}$ for ESO444--G084 and (6.1 $\pm$ 0.3) $\times$ 10$^{8}$ M$_{\odot}$ for [KKS2000]23, indicating a significantly higher gas content in the latter.
    \end{itemize}
    \item[] {\HI\ kinematics:}
    \begin{itemize}
        \item Deviations in the velocity field of ESO444--G084 within the inner $\sim$1.8 kpc suggest a weak kinematic warp, possibly influenced by internal gas flows or past accretion events.
        \item The isovelocity contours in ESO444--G084's central region deviate slightly from parallel, indicating weak non-circular motions. The absence of characteristic S-shaped or X-shaped distortions suggests these motions are not bar-driven but may result from mild internal perturbations or gas flows.
        \item {[KKS2000]23} exhibits regular rotation with minimal perturbations, indicating a stable kinematic structure.
        \item The derived rotation curves show a relatively fast rise in ESO444--G084’s rotation velocity within the inner 2 kpc, reaching $\sim$50 km s$^{-1}$. [KKS2000]23’s rotation curve rises more gradually, peaking at $\sim$60 km s$^{-1}$ beyond 6 kpc.
    \end{itemize}
    \item[] {Dark matter halo:}
    \begin{itemize}
        \item The central dark matter density (\(\rho_0\)) is significantly higher for ESO444--G084 (\(\rho_0 = 16.05 \pm 1.45 \times 10^{-3} M_{\odot} \text{pc}^{-3}\)) at \((M/L)^{*}_{3.4 \mu m} = 0.2\), compared to [KKS2000]23 (\(\rho_0 = 4.29 \pm 0.43 \times 10^{-3} M_{\odot} \text{pc}^{-3}\)), reflecting a more concentrated dark matter halo in ESO444--G084.
        \item The core radius (\(r_C\)) is smaller for ESO444--G084 (\(r_C = 3.48 \pm 0.57\) kpc), indicating a more compact dark matter distribution. [KKS2000]23 has a larger core radius (\(r_C = 5.49 \pm 0.60\) kpc), suggesting a less concentrated mass distribution and the likely presence of a constant density core.
    \end{itemize}
    \item[] {Gravitational instabilities and star formation:}
    \begin{itemize}
        \item {[KKS2000]23} shows \(Q < 1\) regions, indicating active gravitational instabilities that may be triggering ongoing star formation, whereas ESO444--G084 remains stable (\(Q > 1\)), suggesting a different star formation mechanism such as turbulence or stellar feedback.
        \item ESO444--G084 exhibits both H$\alpha$ and FUV emission, confirming ongoing star formation, while [KKS2000]23 is dominated by FUV emission, implying that its last major star formation episode occurred over $\sim$10 Myr ago.
        \item Despite MeerKAT’s high sensitivity, no significant gas inflows were detected in either galaxy, suggesting that internal processes (e.g., turbulence, gravitational instabilities) rather than external accretion regulate their star formation.
    \end{itemize}
\end{itemize}
In summary, our comprehensive analysis of ESO444--G084 and [KKS2000]23 reveals distinct differences in their \HI\ morphology, kinematics, and dark matter distributions. These differences underscore the varied evolutionary pathways of dwarf galaxies and highlight the complex interplay between dark matter, gas dynamics, and star formation processes.
\begin{acknowledgements}
The MeerKAT telescope is operated by the South African Radio Astronomy Observatory, which is a facility of the National Research Foundation, an agency of the Department of Science and Innovation. BN acknowledges the grant PTA2023-023268-I funded by MICIU/AEI10.13039/501100011033 and byESF+. BN, AM,RI and LVM acknowledge financial support from the grant PID2021-123930OB-C21 funded by MICIU/AEI/ 10.13039/501100011033 and by ERDF/EU and grant CEX2021-001131-S funded by MICIU/AEI.  BN, RD, and XN acknowledge the South African Research Chairs Initiative, through the South African Radio Astronomy Observatory (SARAO, grant ID 77948), which is a facility of the National Research Foundation (NRF), an agency of the Department of Science, Technology, and Innovation (DSTI) of South Africa. This work has received funding from the European Research Council (ERC) under the European Union’s Horizon 2020 research and innovation programme (grant agreement No. 882793 "MeerGas"). P.K. is partially supported by the BMBF project 05A23PC1 for D-MeerKAT. JH acknowledges support from the UK SKA Regional Centre (UKSRC). The UKSRC is a collaboration between the University of Cambridge, University of Edinburgh, Durham University, University of Hertfordshire, University of Manchester, University College London, and the UKRI Science and Technology Facilities Council (STFC) Scientific Computing at RAL. The UKSRC is supported by funding from the UKRI STFC. KS acknowledges support from the Natural Science and Engineering Research Council of Canada (NSERC). KAO acknowledges support by the Royal Society through Dorothy Hodgkin Fellowship DHF/R1/231105. EA and AB gratefully acknowledge support from the Centre National d’Etudes Spatiales (CNES), France. L.C. acknowledges financial support from the Chilean Agencia Nacional de Investigación y Desarrollo (ANID) through Fondo Nacional de Desarrollo Científico y Tecnologico (FONDECYT) Regular Project 1210992. SPS acknowledges funding from NRF/SARAO. 
\end{acknowledgements}

\vspace{2em}
\noindent
\rule{9.5cm}{0.4pt} 
\vspace{1em}
{\scriptsize
\begin{tabular}{@{}r p{9cm}@{}}
\textsuperscript{1}  & Instituto de Astrofísica de Andalucía (CSIC), Glorieta de la Astronomía s/n, 18008 Granada, Spain\\
\textsuperscript{2}  & Wits Centre for Astrophysics, School of Physics, University of the Witwatersrand, 1 Jan Smuts Avenue, 2000, South Africa\\
\textsuperscript{3}  & Australia Telescope National Facility, CSIRO, Space and Astronomy, P.O. Box 76, NSW 1710, Epping, Australia\\
\textsuperscript{4}  & Western Sydney University, Locked Bag 1797, Penrith, NSW 2751, Australia\\
\textsuperscript{5}  & Aix Marseille Univ, CNRS, CNES, LAM, Marseille, France\\
\textsuperscript{6}  & Department of Astronomy, University of Cape Town, Private Bag X3, Rondebosch 7701, South Africa\\
\textsuperscript{7}  & Département de physique, Université de Montréal, Complexe des sciences MIL, 1375 Avenue Thérèse-Lavoie-Roux, \& Montréal, QC, Canada H2V 0B3\\
\textsuperscript{8}  & Observatoire d'Astrophysique de l'Université Ouaga I Pr Joseph Ki-Zerbo (ODAUO), BP 7021, Ouaga 03, Burkina Faso\\
\textsuperscript{9}  & Max-Planck Institut für Radioastronomie, Auf dem Hügel 69, 53121 Bonn, Germany\\
\textsuperscript{10} & Department of Physics and Electronics, Rhodes University, PO Box 94, Makhanda 6140, South Africa\\
\textsuperscript{11} & Ruhr University Bochum, Faculty of Physics and Astronomy, Astronomical Institute (AIRUB), 44780 Bochum, Germany\\
\textsuperscript{12} & Department of Physics, University of Pretoria, Hatfield, Pretoria, 0028, South Africa\\
\textsuperscript{13} & Astrophysics Research Centre, University of KwaZulu-Natal, Durban, 3696, South Africa\\
\textsuperscript{14} & School of Mathematics, Statistics \& Computer Science, University of KwaZulu-Natal, Westville Campus, Durban 4041, South Africa\\
\textsuperscript{15} & Centre for Astrophysics Research, University of Hertfordshire, College Lane, Hatfield, AL10 9AB, UK\\
\textsuperscript{16} & Université de Strasbourg, CNRS, Observatoire astronomique de Strasbourg, UMR 7550, 67000 Strasbourg, France\\
\textsuperscript{17} & Instituto de Astrofísica, Departamento de Ciencias Físicas, Universidad Andrés Bello, Fernández Concha 700,\\ & Las Condes, Santiago, Chile\\
\textsuperscript{18} & Observatoire de Paris, Collège de France, Université PSL, Sorbonne Université, CNRS, LERMA, Paris, France\\
\textsuperscript{19} & Netherlands Institute for Radio Astronomy (ASTRON), Oude Hoogeveensedijk 4, 7991 PD Dwingeloo, the Netherlands\\
\textsuperscript{20} & Kapteyn Astronomical Institute, University of Groningen, PO Box 800, 9700 AV Groningen, The Netherlands\\
\textsuperscript{21} & Department of Physics, Engineering Physics and Astronomy, Queen’s University, Kingston, ON K7L 3N6, Canada\\
\textsuperscript{22} & Department of Physics and Astronomy, University of Manitoba, Winnipeg, Manitoba, Canada, R3T 2N2\\
\textsuperscript{23} & Jodrell Bank Centre for Astrophysics, School of Physics and Astronomy, University of Manchester, Oxford Road, Manchester M13 9PL, UK\\
\textsuperscript{24} & United Kingdom SKA Regional Centre (UKSRC), UK\\
\textsuperscript{25} & INAF – Padova Astronomical Observatory, Vicolo dell’Osservatorio 5, I-35122 Padova, Italy\\
\textsuperscript{26} & Department of Astronomy, Case Western Reserve University, 10900 Euclid Avenue, Cleveland, OH 44106, USA\\
\textsuperscript{27} & Institute for Computational Cosmology, Department of Physics, Durham University, South Road, Durham DH1 3LE, UK\\
\textsuperscript{28} & Centre for Extragalactic Astronomy, Department of Physics, Durham University, South Road, Durham DH1 3LE, UK\\
\textsuperscript{29} & Department of Physics, Engineering Physics and Astronomy, Queen’s University, Kingston, ON K7L 3N6, Canada\\
\textsuperscript{30} & CSIRO, Space \& Astronomy, PO Box 1130, Bentley WA 6102, Australia\\
\textsuperscript{31} & International Centre for Radio Astronomy Research, The University of Western Australia, 35 Stirling Highway, Crawley, WA 6009, Australia\\
\end{tabular}
}

\begin{thebibliography}{106}
\expandafter\ifx\csname natexlab\endcsname\relax\def\natexlab#1{#1}\fi

\bibitem[{{Athanassoula} \& {Bureau}(1999)}]{1999ApJ...522..699A}
{Athanassoula}, E. \& {Bureau}, M. 1999, \apj, 522, 699

\bibitem[{{Barnes} {et~al.}(2001){Barnes}, {Staveley-Smith}, {de Blok}, {Oosterloo}, {Stewart}, {Wright}, {Banks}, {Bhathal}, {Boyce}, {Calabretta}, {Disney}, {Drinkwater}, {Ekers}, {Freeman}, {Gibson}, {Green}, {Haynes}, {te Lintel Hekkert}, {Henning}, {Jerjen}, {Juraszek}, {Kesteven}, {Kilborn}, {Knezek}, {Koribalski}, {Kraan-Korteweg}, {Malin}, {Marquarding}, {Minchin}, {Mould}, {Price}, {Putman}, {Ryder}, {Sadler}, {Schr{\"o}der}, {Stootman}, {Webster}, {Wilson}, \& {Ye}}]{Barnes2001}
{Barnes}, D.~G., {Staveley-Smith}, L., {de Blok}, W.~J.~G., {et~al.} 2001, \mnras, 322, 486

\bibitem[{{Begeman}(1989)}]{1989A&A...223...47B}
{Begeman}, K.~G. 1989, \aap, 223, 47

\bibitem[{{Begeman} {et~al.}(1991){Begeman}, {Broeils}, \& {Sanders}}]{1991MNRAS.249..523B}
{Begeman}, K.~G., {Broeils}, A.~H., \& {Sanders}, R.~H. 1991, \mnras, 249, 523

\bibitem[{{Begum} {et~al.}(2008){Begum}, {Chengalur}, {Karachentsev}, {Sharina}, \& {Kaisin}}]{2008MNRAS.386.1667B}
{Begum}, A., {Chengalur}, J.~N., {Karachentsev}, I.~D., {Sharina}, M.~E., \& {Kaisin}, S.~S. 2008, \mnras, 386, 1667

\bibitem[{{Bigiel} {et~al.}(2008){Bigiel}, {Leroy}, {Walter}, {Brinks}, {de Blok}, {Madore}, \& {Thornley}}]{2008AJ....136.2846B}
{Bigiel}, F., {Leroy}, A., {Walter}, F., {et~al.} 2008, \aj, 136, 2846

\bibitem[{{Blais-Ouellette} {et~al.}(2004){Blais-Ouellette}, {Amram}, {Carignan}, \& {Swaters}}]{2004A&A...420..147B}
{Blais-Ouellette}, S., {Amram}, P., {Carignan}, C., \& {Swaters}, R. 2004, \aap, 420, 147

\bibitem[{{Bosma}(1978)}]{1978PhDT.......195B}
{Bosma}, A. 1978, PhD thesis, University of Groningen, Netherlands

\bibitem[{{Bullock} \& {Boylan-Kolchin}(2017)}]{2017ARA&A..55..343B}
{Bullock}, J.~S. \& {Boylan-Kolchin}, M. 2017, \araa, 55, 343

\bibitem[{{Camilo} {et~al.}(2018){Camilo}, {Scholz}, {Serylak}, {Buchner}, {Merryfield}, {Kaspi}, {Archibald}, {Bailes}, {Jameson}, {van Straten}, {Sarkissian}, {Reynolds}, {Johnston}, {Hobbs}, {Abbott}, {Adam}, {Adams}, {Alberts}, {Andreas}, {Asad}, {Baker}, {Baloyi}, {Bauermeister}, {Baxana}, {Bennett}, {Bernardi}, {Booisen}, {Booth}, {Botha}, {Boyana}, {Brederode}, {Burger}, {Cheetham}, {Conradie}, {Conradie}, {Davidson}, {De Bruin}, {de Swardt}, {de Villiers}, {de Villiers}, {de Villiers}, {de Villiers}, {De Waal}, {Dikgale}, {du Toit}, {du Toit}, {Esterhuyse}, {Fanaroff}, {Fataar}, {Foley}, {Foster}, {Fourie}, {Gamatham}, {Gatsi}, {Geschke}, {Goedhart}, {Grobler}, {Gumede}, {Hlakola}, {Hokwana}, {Hoorn}, {Horn}, {Horrell}, {Hugo}, {Isaacson}, {Jacobs}, {Jansen van Rensburg}, {Jonas}, {Jordaan}, {Joubert}, {Joubert}, {J{\'o}zsa}, {Julie}, {Julius}, {Kapp}, {Karastergiou}, {Karels}, {Kariseb}, {Karuppusamy}, {Kasper}, {Knox-Davies}, {Koch}, {Kotz{\'e}}, {Krebs}, {Kriek}, {Kriel}, {Kusel}, {Lamoor},
  {Lehmensiek}, {Liebenberg}, {Liebenberg}, {Lord}, {Lunsky}, {Mabombo}, {Macdonald}, {Macfarlane}, {Madisa}, {Mafhungo}, {Magnus}, {Magozore}, {Mahgoub}, {Main}, {Makhathini}, {Malan}, {Malgas}, {Manley}, {Manzini}, {Marais}, {Marais}, {Marais}, {Maree}, {Martens}, {Matshawule}, {Matthysen}, {Mauch}, {McNally}, {Merry}, {Millenaar}, {Mjikelo}, {Mkhabela}, {Mnyand u}, {Moeng}, {Mokone}, {Monama}, {Montshiwa}, {Moss}, {Mphego}, {New}, {Ngcebetsha}, {Ngoasheng}, {Niehaus}, {Ntuli}, {Nzama}, {Obies}, {Obrocka}, {Ockards}, {Olyn}, {Oozeer}, {Otto}, {Padayachee}, {Passmoor}, {Patel}, {Paula}, {Peens-Hough}, {Pholoholo}, {Prozesky}, {Rakoma}, {Ramaila}, {Rammala}, {Ramudzuli}, {Rasivhaga}, {Ratcliffe}, {Reader}, {Renil}, {Richter}, {Robyntjies}, {Rosekrans}, {Rust}, {Salie}, {Sambu}, {Schollar}, {Schwardt}, {Seranyane}, {Sethosa}, {Sharpe}, {Siebrits}, {Sirothia}, {Slabber}, {Smirnov}, {Smith}, {Sofeya}, {Songqumase}, {Spann}, {Stappers}, {Steyn}, {Steyn}, {Strong}, {Struthers}, {Stuart}, {Sunnylall}, {Swart},
  {Taljaard}, {Tasse}, {Taylor}, {Theron}, {Thondikulam}, {Thorat}, {Tiplady}, {Toruvanda}, {van Aardt}, {van Balla}, {van den Heever}, {van der Byl}, {van der Merwe}, {van der Merwe}, {van Niekerk}, {van Rooyen}, {van Staden}, {van Tonder}, {van Wyk}, {Wait}, {Walker}, {Wallace}, {Welz}, {Williams}, {Xaia}, {Young}, \& {Zitha}}]{2018ApJ...856..180C}
{Camilo}, F., {Scholz}, P., {Serylak}, M., {et~al.} 2018, \apj, 856, 180

\bibitem[{{Carignan} \& {Freeman}(1985)}]{1985ApJ...294..494C}
{Carignan}, C. \& {Freeman}, K.~C. 1985, \apj, 294, 494

\bibitem[{{Carignan} \& {Freeman}(1988)}]{1988ApJ...332L..33C}
{Carignan}, C. \& {Freeman}, K.~C. 1988, \apjl, 332, L33

\bibitem[{{Carignan} \& {Purton}(1998)}]{1998ApJ...506..125C}
{Carignan}, C. \& {Purton}, C. 1998, \apj, 506, 125

\bibitem[{{Carlson} {et~al.}(1992){Carlson}, {Machacek}, \& {Hall}}]{1992ApJ...398...43C}
{Carlson}, E.~D., {Machacek}, M.~E., \& {Hall}, L.~J. 1992, \apj, 398, 43

\bibitem[{{Chamba} {et~al.}(2024){Chamba}, {Marcum}, {Saintonge}, {Borlaff}, {Hayes}, {Le Gouellec}, {Chojnowski}, \& {Fanelli}}]{2024ApJ...974..247C}
{Chamba}, N., {Marcum}, P.~M., {Saintonge}, A., {et~al.} 2024, \apj, 974, 247

\bibitem[{{Cluver} {et~al.}(2014){Cluver}, {Jarrett}, {Hopkins}, {Driver}, {Liske}, {Gunawardhana}, {Taylor}, {Robotham}, {Alpaslan}, {Baldry}, {Brown}, {Peacock}, {Popescu}, {Tuffs}, {Bauer}, {Bland-Hawthorn}, {Colless}, {Holwerda}, {Lara-L{\'o}pez}, {Leschinski}, {L{\'o}pez-S{\'a}nchez}, {Norberg}, {Owers}, {Wang}, \& {Wilkins}}]{2014ApJ...782...90C}
{Cluver}, M.~E., {Jarrett}, T.~H., {Hopkins}, A.~M., {et~al.} 2014, \apj, 782, 90

\bibitem[{{C{\^o}t{\'e}} {et~al.}(2000){C{\^o}t{\'e}}, {Carignan}, \& {Freeman}}]{2000AJ....120.3027C}
{C{\^o}t{\'e}}, S., {Carignan}, C., \& {Freeman}, K.~C. 2000, \aj, 120, 3027

\bibitem[{{Dale} {et~al.}(2009){Dale}, {Cohen}, {Johnson}, {Schuster}, {Calzetti}, {Engelbracht}, {Gil de Paz}, {Kennicutt}, {Lee}, {Begum}, {Block}, {Dalcanton}, {Funes}, {Gordon}, {Johnson}, {Marble}, {Sakai}, {Skillman}, {van Zee}, {Walter}, {Weisz}, {Williams}, {Wu}, \& {Wu}}]{2009ApJ...703..517D}
{Dale}, D.~A., {Cohen}, S.~A., {Johnson}, L.~C., {et~al.} 2009, \apj, 703, 517

\bibitem[{{de Blok}(2010)}]{2010AdAst2010E...5D}
{de Blok}, W.~J.~G. 2010, Advances in Astronomy, 2010, 789293

\bibitem[{{de Blok} {et~al.}(2020){de Blok}, {Athanassoula}, {Bosma}, {Combes}, {English}, {Heald}, {Kamphuis}, {Koribalski}, {Meurer}, {Rom{\'a}n}, {Sardone}, {Verdes-Montenegro}, {Bigiel}, {Brinks}, {Chemin}, {Fraternali}, {Jarrett}, {Kleiner}, {Maccagni}, {Pisano}, {Serra}, {Spekkens}, {Amram}, {Carignan}, {Dettmar}, {Gibson}, {Holwerda}, {J{\'o}zsa}, {Lucero}, {Oosterloo}, {Ramaila}, {Ramatsoku}, {Sheth}, {Walter}, {Wong}, {Zijlstra}, {Bloemen}, {Groot}, {Le Poole}, {Klein-Wolt}, {K{\"o}rding}, {McBride}, {Paterson}, {Pieterse}, {Vreeswijk}, \& {Woudt}}]{2020A&A...643A.147D}
{de Blok}, W.~J.~G., {Athanassoula}, E., {Bosma}, A., {et~al.} 2020, \aap, 643, A147

\bibitem[{{de Blok} {et~al.}(2024){de Blok}, {Healy}, {Maccagni}, {Pisano}, {Bosma}, {English}, {Jarrett}, {Marasco}, {Meurer}, {Veronese}, {Bigiel}, {Chemin}, {Fraternali}, {Holwerda}, {Kamphuis}, {Kl{\"o}ckner}, {Kleiner}, {Leroy}, {Mogotsi}, {Oman}, {Schinnerer}, {Verdes-Montenegro}, {Westmeier}, {Wong}, {Zabel}, {Amram}, {Carignan}, {Combes}, {Brinks}, {Dettmar}, {Gibson}, {Jozsa}, {Koribalski}, {McGaugh}, {Oosterloo}, {Spekkens}, {Schr{\"o}der}, {Adams}, {Athanassoula}, {Bershady}, {Beswick}, {Blyth}, {Elson}, {Frank}, {Heald}, {Henning}, {Kurapati}, {Loubser}, {Lucero}, {Meyer}, {Namumba}, {Oh}, {Sardone}, {Sheth}, {Smith}, {Sorgho}, {Walter}, {Williams}, {Woudt}, \& {Zijlstra}}]{2024A&A...688A.109D}
{de Blok}, W.~J.~G., {Healy}, J., {Maccagni}, F.~M., {et~al.} 2024, \aap, 688, A109

\bibitem[{{de Blok} {et~al.}(2001{\natexlab{a}}){de Blok}, {McGaugh}, {Bosma}, \& {Rubin}}]{2001ApJ...552L..23D}
{de Blok}, W.~J.~G., {McGaugh}, S.~S., {Bosma}, A., \& {Rubin}, V.~C. 2001{\natexlab{a}}, \apjl, 552, L23

\bibitem[{{de Blok} {et~al.}(2001{\natexlab{b}}){de Blok}, {McGaugh}, \& {Rubin}}]{2001AJ....122.2396D}
{de Blok}, W.~J.~G., {McGaugh}, S.~S., \& {Rubin}, V.~C. 2001{\natexlab{b}}, \aj, 122, 2396

\bibitem[{{de Blok} {et~al.}(2008){de Blok}, {Walter}, {Brinks}, {Trachternach}, {Oh}, \& {Kennicutt}}]{2008AJ....136.2648D}
{de Blok}, W.~J.~G., {Walter}, F., {Brinks}, E., {et~al.} 2008, \aj, 136, 2648

\bibitem[{{de Vaucouleurs} {et~al.}(1991){de Vaucouleurs}, {de Vaucouleurs}, {Corwin}, {Buta}, {Paturel}, \& {Fouque}}]{1991rc3..book.....D}
{de Vaucouleurs}, G., {de Vaucouleurs}, A., {Corwin}, Herold~G., J., {et~al.} 1991, {Third Reference Catalogue of Bright Galaxies}

\bibitem[{{Dey} {et~al.}(2019){Dey}, {Schlegel}, {Lang}, {Blum}, {Burleigh}, {Fan}, {Findlay}, {Finkbeiner}, {Herrera}, {Juneau}, {Landriau}, {Levi}, {McGreer}, {Meisner}, {Myers}, {Moustakas}, {Nugent}, {Patej}, {Schlafly}, {Walker}, {Valdes}, {Weaver}, {Y{\`e}che}, {Zou}, {Zhou}, {Abareshi}, {Abbott}, {Abolfathi}, {Aguilera}, {Alam}, {Allen}, {Alvarez}, {Annis}, {Ansarinejad}, {Aubert}, {Beechert}, {Bell}, {BenZvi}, {Beutler}, {Bielby}, {Bolton}, {Brice{\~n}o}, {Buckley-Geer}, {Butler}, {Calamida}, {Carlberg}, {Carter}, {Casas}, {Castander}, {Choi}, {Comparat}, {Cukanovaite}, {Delubac}, {DeVries}, {Dey}, {Dhungana}, {Dickinson}, {Ding}, {Donaldson}, {Duan}, {Duckworth}, {Eftekharzadeh}, {Eisenstein}, {Etourneau}, {Fagrelius}, {Farihi}, {Fitzpatrick}, {Font-Ribera}, {Fulmer}, {G{\"a}nsicke}, {Gaztanaga}, {George}, {Gerdes}, {Gontcho}, {Gorgoni}, {Green}, {Guy}, {Harmer}, {Hernandez}, {Honscheid}, {Huang}, {James}, {Jannuzi}, {Jiang}, {Joyce}, {Karcher}, {Karkar}, {Kehoe}, {Kneib}, {Kueter-Young}, {Lan},
  {Lauer}, {Le Guillou}, {Le Van Suu}, {Lee}, {Lesser}, {Perreault Levasseur}, {Li}, {Mann}, {Marshall}, {Mart{\'\i}nez-V{\'a}zquez}, {Martini}, {du Mas des Bourboux}, {McManus}, {Meier}, {M{\'e}nard}, {Metcalfe}, {Mu{\~n}oz-Guti{\'e}rrez}, {Najita}, {Napier}, {Narayan}, {Newman}, {Nie}, {Nord}, {Norman}, {Olsen}, {Paat}, {Palanque-Delabrouille}, {Peng}, {Poppett}, {Poremba}, {Prakash}, {Rabinowitz}, {Raichoor}, {Rezaie}, {Robertson}, {Roe}, {Ross}, {Ross}, {Rudnick}, {Safonova}, {Saha}, {S{\'a}nchez}, {Savary}, {Schweiker}, {Scott}, {Seo}, {Shan}, {Silva}, {Slepian}, {Soto}, {Sprayberry}, {Staten}, {Stillman}, {Stupak}, {Summers}, {Sien Tie}, {Tirado}, {Vargas-Maga{\~n}a}, {Vivas}, {Wechsler}, {Williams}, {Yang}, {Yang}, {Yapici}, {Zaritsky}, {Zenteno}, {Zhang}, {Zhang}, {Zhou}, \& {Zhou}}]{2019AJ....157..168D}
{Dey}, A., {Schlegel}, D.~J., {Lang}, D., {et~al.} 2019, \aj, 157, 168

\bibitem[{{Di Cintio} {et~al.}(2014){Di Cintio}, {Brook}, {Dutton}, {Macci{\`o}}, {Stinson}, \& {Knebe}}]{2014MNRAS.441.2986D}
{Di Cintio}, A., {Brook}, C.~B., {Dutton}, A.~A., {et~al.} 2014, \mnras, 441, 2986

\bibitem[{{Di Teodoro} \& {Fraternali}(2015)}]{2015MNRAS.451.3021D}
{Di Teodoro}, E.~M. \& {Fraternali}, F. 2015, \mnras, 451, 3021

\bibitem[{{Downing} \& {Oman}(2023)}]{2023MNRAS.522.3318D}
{Downing}, E.~R. \& {Oman}, K.~A. 2023, \mnras, 522, 3318

\bibitem[{{Dutton} {et~al.}(2019){Dutton}, {Obreja}, \& {Macci{\`o}}}]{2019MNRAS.482.5606D}
{Dutton}, A.~A., {Obreja}, A., \& {Macci{\`o}}, A.~V. 2019, \mnras, 482, 5606

\bibitem[{{Elmegreen}(2002)}]{2002ApJ...577..206E}
{Elmegreen}, B.~G. 2002, \apj, 577, 206

\bibitem[{{Elmegreen} \& {Hunter}(2015)}]{2015ApJ...805..145E}
{Elmegreen}, B.~G. \& {Hunter}, D.~A. 2015, \apj, 805, 145

\bibitem[{{Elson} {et~al.}(2012){Elson}, {de Blok}, \& {Kraan-Korteweg}}]{2012AJ....143....1E}
{Elson}, E.~C., {de Blok}, W.~J.~G., \& {Kraan-Korteweg}, R.~C. 2012, \aj, 143, 1

\bibitem[{{Flores} \& {Primack}(1994)}]{1994ApJ...427L...1F}
{Flores}, R.~A. \& {Primack}, J.~R. 1994, \apjl, 427, L1

\bibitem[{{Fragkoudi} {et~al.}(2017){Fragkoudi}, {Di Matteo}, {Haywood}, {G{\'o}mez}, {Combes}, {Katz}, \& {Semelin}}]{2017A&A...606A..47F}
{Fragkoudi}, F., {Di Matteo}, P., {Haywood}, M., {et~al.} 2017, \aap, 606, A47

\bibitem[{{Ghosh} \& {Jog}(2018)}]{2018NewA...63...38G}
{Ghosh}, S. \& {Jog}, C.~J. 2018, \na, 63, 38

\bibitem[{{Girelli} {et~al.}(2020){Girelli}, {Pozzetti}, {Bolzonella}, {Giocoli}, {Marulli}, \& {Baldi}}]{2020A&A...634A.135G}
{Girelli}, G., {Pozzetti}, L., {Bolzonella}, M., {et~al.} 2020, \aap, 634, A135

\bibitem[{{Gnedin} \& {Zhao}(2002)}]{2002MNRAS.333..299G}
{Gnedin}, O.~Y. \& {Zhao}, H. 2002, \mnras, 333, 299

\bibitem[{{Governato} {et~al.}(2012){Governato}, {Zolotov}, {Pontzen}, {Christensen}, {Oh}, {Brooks}, {Quinn}, {Shen}, \& {Wadsley}}]{2012MNRAS.422.1231G}
{Governato}, F., {Zolotov}, A., {Pontzen}, A., {et~al.} 2012, \mnras, 422, 1231

\bibitem[{{Healy} {et~al.}(2024){Healy}, {de Blok}, {Maccagni}, {Amram}, {Chemin}, {Combes}, {Holwerda}, {Kamphuis}, {Pisano}, {Schinnerer}, {Spekkens}, {Verdes-Montenegro}, {Walter}, {Adams}, {Gibson}, {Kleiner}, {Veronese}, {Zabel}, {English}, \& {Carignan}}]{2024A&A...687A.254H}
{Healy}, J., {de Blok}, W.~J.~G., {Maccagni}, F.~M., {et~al.} 2024, \aap, 687, A254

\bibitem[{{Hopkins} {et~al.}(2012){Hopkins}, {Quataert}, \& {Murray}}]{2012MNRAS.421.3488H}
{Hopkins}, P.~F., {Quataert}, E., \& {Murray}, N. 2012, \mnras, 421, 3488

\bibitem[{{Hu} {et~al.}(2000){Hu}, {Barkana}, \& {Gruzinov}}]{2000PhRvL..85.1158H}
{Hu}, W., {Barkana}, R., \& {Gruzinov}, A. 2000, \prl, 85, 1158

\bibitem[{{Huchtmeier} {et~al.}(2001{\natexlab{a}}){Huchtmeier}, {Karachentsev}, \& {Karachentseva}}]{Huchtmeier2001}
{Huchtmeier}, W.~K., {Karachentsev}, I.~D., \& {Karachentseva}, V.~E. 2001{\natexlab{a}}, \aap, 377, 801

\bibitem[{{Huchtmeier} {et~al.}(2001{\natexlab{b}}){Huchtmeier}, {Karachentsev}, \& {Karachentseva}}]{2001A&A...377..801H}
{Huchtmeier}, W.~K., {Karachentsev}, I.~D., \& {Karachentseva}, V.~E. 2001{\natexlab{b}}, \aap, 377, 801

\bibitem[{{Hunter} {et~al.}(2021){Hunter}, {Elmegreen}, {Goldberger}, {Taylor}, {Ermakov}, {Herrmann}, {Oh}, {Malko}, {Barandi}, \& {Jundt}}]{2021AJ....161...71H}
{Hunter}, D.~A., {Elmegreen}, B.~G., {Goldberger}, E., {et~al.} 2021, \aj, 161, 71

\bibitem[{{Hunter} {et~al.}(2011){Hunter}, {Elmegreen}, {Oh}, {Anderson}, {Nordgren}, {Massey}, {Wilsey}, \& {Riabokin}}]{2011AJ....142..121H}
{Hunter}, D.~A., {Elmegreen}, B.~G., {Oh}, S.-H., {et~al.} 2011, \aj, 142, 121

\bibitem[{{Hunter} {et~al.}(2012){Hunter}, {Ficut-Vicas}, {Ashley}, {Brinks}, {Cigan}, {Elmegreen}, {Heesen}, {Herrmann}, {Johnson}, {Oh}, {Rupen}, {Schruba}, {Simpson}, {Walter}, {Westpfahl}, {Young}, \& {Zhang}}]{2012AJ....144..134H}
{Hunter}, D.~A., {Ficut-Vicas}, D., {Ashley}, T., {et~al.} 2012, \aj, 144, 134

\bibitem[{{Jonas} \& {MeerKAT Team}(2016)}]{2016mks..confE...1J}
{Jonas}, J. \& {MeerKAT Team}. 2016, in MeerKAT Science: On the Pathway to the SKA, 1

\bibitem[{{J{\'o}zsa} {et~al.}(2007){J{\'o}zsa}, {Kenn}, {Klein}, \& {Oosterloo}}]{2007A&A...468..731J}
{J{\'o}zsa}, G.~I.~G., {Kenn}, F., {Klein}, U., \& {Oosterloo}, T.~A. 2007, \aap, 468, 731

\bibitem[{{J{\'o}zsa} {et~al.}(2012){J{\'o}zsa}, {Kenn}, {Oosterloo}, \& {Klein}}]{2012ascl.soft08008J}
{J{\'o}zsa}, G. I.~G., {Kenn}, F., {Oosterloo}, T.~A., \& {Klein}, U. 2012, {TiRiFiC: Tilted Ring Fitting Code}, Astrophysics Source Code Library, record ascl:1208.008

\bibitem[{{Kaldare} {et~al.}(2003){Kaldare}, {Colless}, {Raychaudhury}, \& {Peterson}}]{2003MNRAS.339..652K}
{Kaldare}, R., {Colless}, M., {Raychaudhury}, S., \& {Peterson}, B.~A. 2003, \mnras, 339, 652

\bibitem[{{Kamphuis}(2024)}]{2024ascl.soft07002K}
{Kamphuis}, P. 2024, {pyFAT: Python Fully Automated TiRiFiC}, Astrophysics Source Code Library, record ascl:2407.002

\bibitem[{{Kamphuis} {et~al.}(2015){Kamphuis}, {J{\'o}zsa}, {Oh}, {Spekkens}, {Urbancic}, {Serra}, {Koribalski}, \& {Dettmar}}]{2015MNRAS.452.3139K}
{Kamphuis}, P., {J{\'o}zsa}, G.~I.~G., {Oh}, S. .~H., {et~al.} 2015, \mnras, 452, 3139

\bibitem[{{Karachentsev} {et~al.}(2000){Karachentsev}, {Karachentseva}, {Suchkov}, \& {Grebel}}]{KKS2000}
{Karachentsev}, I.~D., {Karachentseva}, V.~E., {Suchkov}, A.~A., \& {Grebel}, E.~K. 2000, \aaps, 145, 415

\bibitem[{{Karachentsev} {et~al.}(2002){Karachentsev}, {Sharina}, {Dolphin}, {Grebel}, {Geisler}, {Guhathakurta}, {Hodge}, {Karachentseva}, {Sarajedini}, \& {Seitzer}}]{2002A&A...385...21K}
{Karachentsev}, I.~D., {Sharina}, M.~E., {Dolphin}, A.~E., {et~al.} 2002, \aap, 385, 21

\bibitem[{{Kennicutt}(1989)}]{1989ApJ...344..685K}
{Kennicutt}, Robert~C., J. 1989, \apj, 344, 685

\bibitem[{{Kennicutt} \& {Evans}(2012)}]{2012ARA&A..50..531K}
{Kennicutt}, R.~C. \& {Evans}, N.~J. 2012, \araa, 50, 531

\bibitem[{Kereš {et~al.}(2005)Kereš, Katz, Weinberg, \& Davé}]{Keres2005}
Kereš, D., Katz, N., Weinberg, D.~H., \& Davé, R. 2005, Monthly Notices of the Royal Astronomical Society, 363, 2, provided by SAO/NASA Astrophysics Data System

\bibitem[{{Khademi} {et~al.}(2021){Khademi}, {Yang}, {Hammer}, \& {Nasiri}}]{2021A&A...654A...7K}
{Khademi}, M., {Yang}, Y., {Hammer}, F., \& {Nasiri}, S. 2021, \aap, 654, A7

\bibitem[{{Koribalski} {et~al.}(2004){Koribalski}, {Staveley-Smith}, {Kilborn}, {Ryder}, {Kraan-Korteweg}, {Ryan-Weber}, {Ekers}, {Jerjen}, {Henning}, {Putman}, {Zwaan}, {de Blok}, {Calabretta}, {Disney}, {Minchin}, {Bhathal}, {Boyce}, {Drinkwater}, {Freeman}, {Gibson}, {Green}, {Haynes}, {Juraszek}, {Kesteven}, {Knezek}, {Mader}, {Marquarding}, {Meyer}, {Mould}, {Oosterloo}, {O'Brien}, {Price}, {Sadler}, {Schr{\"o}der}, {Stewart}, {Stootman}, {Waugh}, {Warren}, {Webster}, \& {Wright}}]{Koribalski2004}
{Koribalski}, B.~S., {Staveley-Smith}, L., {Kilborn}, V.~A., {et~al.} 2004, \aj, 128, 16

\bibitem[{{Koribalski} {et~al.}(2020){Koribalski}, {Staveley-Smith}, {Westmeier}, {Serra}, {Spekkens}, {Wong}, {Lee-Waddell}, {Lagos}, {Obreschkow}, {Ryan-Weber}, {Zwaan}, {Kilborn}, {Bekiaris}, {Bekki}, {Bigiel}, {Boselli}, {Bosma}, {Catinella}, {Chauhan}, {Cluver}, {Colless}, {Courtois}, {Crain}, {de Blok}, {D{\'e}nes}, {Duffy}, {Elagali}, {Fluke}, {For}, {Heald}, {Henning}, {Hess}, {Holwerda}, {Howlett}, {Jarrett}, {Jones}, {Jones}, {J{\'o}zsa}, {Jurek}, {J{\"u}tte}, {Kamphuis}, {Karachentsev}, {Kerp}, {Kleiner}, {Kraan-Korteweg}, {L{\'o}pez-S{\'a}nchez}, {Madrid}, {Meyer}, {Mould}, {Murugeshan}, {Norris}, {Oh}, {Oosterloo}, {Popping}, {Putman}, {Reynolds}, {Rhee}, {Robotham}, {Ryder}, {Schr{\"o}der}, {Shao}, {Stevens}, {Taylor}, {van{\^A} der Hulst}, {Verdes-Montenegro}, {Wakker}, {Wang}, {Whiting}, {Winkel}, \& {Wolf}}]{Koribalski2020}
{Koribalski}, B.~S., {Staveley-Smith}, L., {Westmeier}, T., {et~al.} 2020, \apss, 365, 118

\bibitem[{{Koribalski} {et~al.}(2018){Koribalski}, {Wang}, {Kamphuis}, {Westmeier}, {Staveley-Smith}, {Oh}, {L{\'o}pez-S{\'a}nchez}, {Wong}, {Ott}, {de Blok}, \& {Shao}}]{2018MNRAS.478.1611K}
{Koribalski}, B.~S., {Wang}, J., {Kamphuis}, P., {et~al.} 2018, \mnras, 478, 1611

\bibitem[{{Lazar} {et~al.}(2020){Lazar}, {Bullock}, {Boylan-Kolchin}, {Chan}, {Hopkins}, {Graus}, {Wetzel}, {El-Badry}, {Wheeler}, {Straight}, {Kere{\v{s}}}, {Faucher-Gigu{\`e}re}, {Fitts}, \& {Garrison-Kimmel}}]{2020MNRAS.497.2393L}
{Lazar}, A., {Bullock}, J.~S., {Boylan-Kolchin}, M., {et~al.} 2020, \mnras, 497, 2393

\bibitem[{{Lelli} {et~al.}(2016){Lelli}, {McGaugh}, \& {Schombert}}]{2016AJ....152..157L}
{Lelli}, F., {McGaugh}, S.~S., \& {Schombert}, J.~M. 2016, \aj, 152, 157

\bibitem[{{Leroy} {et~al.}(2008){Leroy}, {Walter}, {Brinks}, {Bigiel}, {de Blok}, {Madore}, \& {Thornley}}]{2008AJ....136.2782L}
{Leroy}, A.~K., {Walter}, F., {Brinks}, E., {et~al.} 2008, \aj, 136, 2782

\bibitem[{{L{\'o}pez-S{\'a}nchez} {et~al.}(2012){L{\'o}pez-S{\'a}nchez}, {Koribalski}, {van Eymeren}, {Esteban}, {Kirby}, {Jerjen}, \& {Lonsdale}}]{2012MNRAS.419.1051L}
{L{\'o}pez-S{\'a}nchez}, {\'A}.~R., {Koribalski}, B.~S., {van Eymeren}, J., {et~al.} 2012, \mnras, 419, 1051

\bibitem[{{Mac Low} \& {Ferrara}(1999)}]{1999ApJ...513..142M}
{Mac Low}, M.-M. \& {Ferrara}, A. 1999, \apj, 513, 142

\bibitem[{{Marasco} {et~al.}(2018){Marasco}, {Oman}, {Navarro}, {Frenk}, \& {Oosterloo}}]{2018MNRAS.476.2168M}
{Marasco}, A., {Oman}, K.~A., {Navarro}, J.~F., {Frenk}, C.~S., \& {Oosterloo}, T. 2018, \mnras, 476, 2168

\bibitem[{{Martin} {et~al.}(2005){Martin}, {Fanson}, {Schiminovich}, {Morrissey}, {Friedman}, {Barlow}, {Conrow}, {Grange}, {Jelinsky}, {Milliard}, {Siegmund}, {Bianchi}, {Byun}, {Donas}, {Forster}, {Heckman}, {Lee}, {Madore}, {Malina}, {Neff}, {Rich}, {Small}, {Surber}, {Szalay}, {Welsh}, \& {Wyder}}]{2005ApJ...619L...1M}
{Martin}, D.~C., {Fanson}, J., {Schiminovich}, D., {et~al.} 2005, \apjl, 619, L1


{Martinsson}, T. P.~K., {Verheijen}, M. A.~W., {Westfall}, K.~B., {et~al.} 2013, \aap, 557, A131

\bibitem[{{Meurer} {et~al.}(1996){Meurer}, {Carignan}, {Beaulieu}, \& {Freeman}}]{1996AJ....111.1551M}
{Meurer}, G.~R., {Carignan}, C., {Beaulieu}, S.~F., \& {Freeman}, K.~C. 1996, \aj, 111, 1551

\bibitem[{{Meurer} {et~al.}(2006){Meurer}, {Hanish}, {Ferguson}, {Knezek}, {Kilborn}, {Putman}, {Smith}, {Koribalski}, {Meyer}, {Oey}, {Ryan-Weber}, {Zwaan}, {Heckman}, {Kennicutt}, {Lee}, {Webster}, {Bland-Hawthorn}, {Dopita}, {Freeman}, {Doyle}, {Drinkwater}, {Staveley-Smith}, \& {Werk}}]{2006ApJS..165..307M}
{Meurer}, G.~R., {Hanish}, D.~J., {Ferguson}, H.~C., {et~al.} 2006, \apjs, 165, 307

\bibitem[{{Navarro} {et~al.}(1996){Navarro}, {Eke}, \& {Frenk}}]{1996MNRAS.283L..72N}
{Navarro}, J.~F., {Eke}, V.~R., \& {Frenk}, C.~S. 1996, \mnras, 283, L72

\bibitem[{{Niemeyer}(2020)}]{2020PrPNP.11303787N}
{Niemeyer}, J.~C. 2020, Progress in Particle and Nuclear Physics, 113, 103787

\bibitem[{{Oh} {et~al.}(2011){Oh}, {de Blok}, {Brinks}, {Walter}, \& {Kennicutt}}]{2011AJ....141..193O}
{Oh}, S.-H., {de Blok}, W.~J.~G., {Brinks}, E., {Walter}, F., \& {Kennicutt}, Jr., R.~C. 2011, \aj, 141, 193

\bibitem[{{Oh} {et~al.}(2008){Oh}, {de Blok}, {Walter}, {Brinks}, \& {Kennicutt}}]{2008AJ....136.2761O}
{Oh}, S.-H., {de Blok}, W.~J.~G., {Walter}, F., {Brinks}, E., \& {Kennicutt}, Robert~C., J. 2008, \aj, 136, 2761

\bibitem[{{Oh} {et~al.}(2015){Oh}, {Hunter}, {Brinks}, {Elmegreen}, {Schruba}, {Walter}, {Rupen}, {Young}, {Simpson}, {Johnson}, {Herrmann}, {Ficut-Vicas}, {Cigan}, {Heesen}, {Ashley}, \& {Zhang}}]{2015AJ....149..180O}
{Oh}, S.-H., {Hunter}, D.~A., {Brinks}, E., {et~al.} 2015, \aj, 149, 180

\bibitem[{{Oman} {et~al.}(2015){Oman}, {Navarro}, {Fattahi}, {Frenk}, {Sawala}, {White}, {Bower}, {Crain}, {Furlong}, {Schaller}, {Schaye}, \& {Theuns}}]{2015MNRAS.452.3650O}
{Oman}, K.~A., {Navarro}, J.~F., {Fattahi}, A., {et~al.} 2015, \mnras, 452, 3650

\bibitem[{{Pence} \& {Blackman}(1984)}]{1984MNRAS.210..547P}
{Pence}, W.~D. \& {Blackman}, C.~P. 1984, \mnras, 210, 547

\bibitem[{{Pontzen} \& {Governato}(2012)}]{2012MNRAS.421.3464P}
{Pontzen}, A. \& {Governato}, F. 2012, \mnras, 421, 3464

\bibitem[{{Read} {et~al.}(2016{\natexlab{a}}){Read}, {Agertz}, \& {Collins}}]{2016MNRAS.459.2573R}
{Read}, J.~I., {Agertz}, O., \& {Collins}, M.~L.~M. 2016{\natexlab{a}}, \mnras, 459, 2573

\bibitem[{{Read} \& {Gilmore}(2005)}]{2005MNRAS.356..107R}
{Read}, J.~I. \& {Gilmore}, G. 2005, \mnras, 356, 107

\bibitem[{{Read} {et~al.}(2016{\natexlab{b}}){Read}, {Iorio}, {Agertz}, \& {Fraternali}}]{2016MNRAS.462.3628R}
{Read}, J.~I., {Iorio}, G., {Agertz}, O., \& {Fraternali}, F. 2016{\natexlab{b}}, \mnras, 462, 3628

\bibitem[{{Read} {et~al.}(2019){Read}, {Walker}, \& {Steger}}]{2019MNRAS.484.1401R}
{Read}, J.~I., {Walker}, M.~G., \& {Steger}, P. 2019, \mnras, 484, 1401

\bibitem[{{Rocha} {et~al.}(2013){Rocha}, {Peter}, {Bullock}, {Kaplinghat}, {Garrison-Kimmel}, {O{\~n}orbe}, \& {Moustakas}}]{2013MNRAS.430...81R}
{Rocha}, M., {Peter}, A. H.~G., {Bullock}, J.~S., {et~al.} 2013, \mnras, 430, 81

\bibitem[{{Romano} {et~al.}(2023){Romano}, {Nanni}, {Donevski}, {Ginolfi}, {Jones}, {Shivaei}, {Junais}, {Salak}, \& {Sawant}}]{2023A&A...677A..44R}
{Romano}, M., {Nanni}, A., {Donevski}, D., {et~al.} 2023, \aap, 677, A44

\bibitem[{{Romeo} \& {Falstad}(2013)}]{2013MNRAS.433.1389R}
{Romeo}, A.~B. \& {Falstad}, N. 2013, \mnras, 433, 1389

\bibitem[{{Ryan-Weber} {et~al.}(2002){Ryan-Weber}, {Koribalski}, {Staveley-Smith}, {Jerjen}, {Kraan-Korteweg}, {Ryder}, {Barnes}, {de Blok}, {Kilborn}, {Bhathal}, {Boyce}, {Disney}, {Drinkwater}, {Ekers}, {Freeman}, {Gibson}, {Green}, {Haynes}, {Henning}, {Juraszek}, {Kesteven}, {Knezek}, {Mader}, {Marquarding}, {Meyer}, {Minchin}, {Mould}, {O'Brien}, {Oosterloo}, {Price}, {Putman}, {Sadler}, {Schr{\"o}der}, {Stewart}, {Stootman}, {Waugh}, {Webster}, {Wright}, \& {Zwaan}}]{Ryan-Weber2002}
{Ryan-Weber}, E., {Koribalski}, B.~S., {Staveley-Smith}, L., {et~al.} 2002, \aj, 124, 1954

\bibitem[{{Sales} {et~al.}(2022){Sales}, {Wetzel}, \& {Fattahi}}]{2022NatAs...6..897S}
{Sales}, L.~V., {Wetzel}, A., \& {Fattahi}, A. 2022, Nature Astronomy, 6, 897

\bibitem[{{S{\'a}nchez-Salcedo}(2006)}]{2006MNRAS.365..555S}
{S{\'a}nchez-Salcedo}, F.~J. 2006, \mnras, 365, 555

\bibitem[{{Santos-Santos} {et~al.}(2020){Santos-Santos}, {Navarro}, {Robertson}, {Ben{\'\i}tez-Llambay}, {Oman}, {Lovell}, {Frenk}, {Ludlow}, {Fattahi}, \& {Ritz}}]{2020MNRAS.495...58S}
{Santos-Santos}, I. M.~E., {Navarro}, J.~F., {Robertson}, A., {et~al.} 2020, \mnras, 495, 58

\bibitem[{{Sardone} {et~al.}(2021){Sardone}, {Pisano}, {Pingel}, {Sorgho}, {Carignan}, \& {de Blok}}]{Sardone2021}
{Sardone}, A., {Pisano}, D.~J., {Pingel}, N.~M., {et~al.} 2021, \apj, 910, 69

\bibitem[{{Schruba} {et~al.}(2012){Schruba}, {Leroy}, {Walter}, {Bigiel}, {Brinks}, {de Blok}, {Kramer}, {Rosolowsky}, {Sandstrom}, {Schuster}, {Usero}, {Weiss}, \& {Wiesemeyer}}]{2012AJ....143..138S}
{Schruba}, A., {Leroy}, A.~K., {Walter}, F., {et~al.} 2012, \aj, 143, 138

\bibitem[{{Sorgho} {et~al.}(2019){Sorgho}, {Carignan}, {Pisano}, {Oosterloo}, {de Blok}, {Korsaga}, {Pingel}, {Sardone}, {Goedhart}, {Passmoor}, {Dikgale}, \& {Sirothia}}]{2019MNRAS.482.1248S}
{Sorgho}, A., {Carignan}, C., {Pisano}, D.~J., {et~al.} 2019, \mnras, 482, 1248

\bibitem[{{Spergel} \& {Steinhardt}(2000)}]{2000PhRvL..84.3760S}
{Spergel}, D.~N. \& {Steinhardt}, P.~J. 2000, \prl, 84, 3760

\bibitem[{{Swaters} {et~al.}(2003){Swaters}, {Madore}, {van den Bosch}, \& {Balcells}}]{2003ApJ...583..732S}
{Swaters}, R.~A., {Madore}, B.~F., {van den Bosch}, F.~C., \& {Balcells}, M. 2003, \apj, 583, 732

\bibitem[{Swaters {et~al.}(2009)Swaters, Sancisi, van Albada, \& van~der Hulst}]{Swaters2009}
Swaters, R.~A., Sancisi, R., van Albada, T.~S., \& van~der Hulst, J.~M. 2009, Astronomy \& Astrophysics, 493, 871

\bibitem[{{Tollet} {et~al.}(2016){Tollet}, {Macci{\`o}}, {Dutton}, {Stinson}, {Wang}, {Penzo}, {Gutcke}, {Buck}, {Kang}, {Brook}, {Di Cintio}, {Keller}, \& {Wadsley}}]{2016MNRAS.456.3542T}
{Tollet}, E., {Macci{\`o}}, A.~V., {Dutton}, A.~A., {et~al.} 2016, \mnras, 456, 3542

\bibitem[{{Toomre}(1964)}]{1964ApJ...139.1217T}
{Toomre}, A. 1964, \apj, 139, 1217

\bibitem[{{Tulin} \& {Yu}(2018)}]{2018PhR...730....1T}
{Tulin}, S. \& {Yu}, H.-B. 2018, \physrep, 730, 1

\bibitem[{{Tully} {et~al.}(2016){Tully}, {Courtois}, \& {Sorce}}]{2016AJ....152...50T}
{Tully}, R.~B., {Courtois}, H.~M., \& {Sorce}, J.~G. 2016, \aj, 152, 50

\bibitem[{{Veronese} {et~al.}(2025){Veronese}, {de Blok}, {Healy}, {Kleiner}, {Marasco}, {Maccagni}, {Kamphuis}, {Brinks}, {Holwerda}, {Zabel}, {Chemin}, {Adams}, {Kurapati}, {Sorgho}, {Spekkens}, {Combes}, {Pisano}, {Walter}, {Amram}, {Bigiel}, {Wong}, \& {Athanassoula}}]{2025A&A...693A..97V}
{Veronese}, S., {de Blok}, W.~J.~G., {Healy}, J., {et~al.} 2025, \aap, 693, A97

\bibitem[{{Walter} {et~al.}(2008){Walter}, {Brinks}, {de Blok}, {Bigiel}, {Kennicutt}, {Thornley}, \& {Leroy}}]{2008AJ....136.2563W}
{Walter}, F., {Brinks}, E., {de Blok}, W.~J.~G., {et~al.} 2008, \aj, 136, 2563

\bibitem[{{Whiting} {et~al.}(2002){Whiting}, {Hau}, \& {Irwin}}]{Whting2002}
{Whiting}, A.~B., {Hau}, G. K.~T., \& {Irwin}, M. 2002, \apjs, 141, 123

\bibitem[{{Whiting} {et~al.}(2007){Whiting}, {Hau}, {Irwin}, \& {Verdugo}}]{2007AJ....133..715W}
{Whiting}, A.~B., {Hau}, G. K.~T., {Irwin}, M., \& {Verdugo}, M. 2007, \aj, 133, 715

\bibitem[{{Wright} {et~al.}(2010){Wright}, {Eisenhardt}, {Mainzer}, {Ressler}, {Cutri}, {Jarrett}, {Kirkpatrick}, {Padgett}, {McMillan}, {Skrutskie}, {Stanford}, {Cohen}, {Walker}, {Mather}, {Leisawitz}, {Gautier}, {McLean}, {Benford}, {Lonsdale}, {Blain}, {Mendez}, {Irace}, {Duval}, {Liu}, {Royer}, {Heinrichsen}, {Howard}, {Shannon}, {Kendall}, {Walsh}, {Larsen}, {Cardon}, {Schick}, {Schwalm}, {Abid}, {Fabinsky}, {Naes}, \& {Tsai}}]{2010AJ....140.1868W}
{Wright}, E.~L., {Eisenhardt}, P. R.~M., {Mainzer}, A.~K., {et~al.} 2010, \aj, 140, 1868

\end{thebibliography}
\end{document}